\documentclass{article}
\pdfoutput=1

\usepackage[template=plain,defineTheorems=true]{arxiv_plain/charles_preamble}
\graphicspath{ {figures/} } 
\setcitestyle{round}  

\title{Matching of Users and Creators in Two-Sided Markets with Departures}

\usepackage{authblk}  
\author[1]{Daniel Huttenlocher}
\author[2]{Hannah Li}
\author[1]{Liang Lyu}
\author[1]{Asuman Ozdaglar}
\author[3]{James Siderius}
\affil[1]{Massachusetts Institute of Technology, 77 Massachusetts Ave, Cambridge, Massachusetts, United States 02139}
\affil[2]{Columbia Business School, 665 West 130th Street, New York, New York, United States 10027}
\affil[3]{Tuck School of Business, Dartmouth College, 100 Tuck Hall, Hanover, New Hampshire, United States 03755}
\date{}

 \bibpunct[, ]{(}{)}{,}{a}{}{,}%

\begin{document}

\maketitle

\begin{abstract}
    
    Many online platforms of today, including social media sites, are two-sided markets bridging content creators and users. Most of the existing literature on platform recommendation algorithms largely focuses on user preferences and decisions, and does not simultaneously address creator incentives. We propose a model of content recommendation that explicitly focuses on the dynamics of user-content matching, with the novel property that both users and creators may leave the platform permanently if they do not experience sufficient engagement. In our model, each player decides to participate at each time step based on utilities derived from the current match: users based on alignment of the recommended content with their preferences, and creators based on their audience size. 
    We show that a user-centric greedy algorithm that does not consider creator departures can result in arbitrarily poor total engagement, relative to an algorithm that maximizes total engagement while accounting for two-sided departures. Moreover, in stark contrast to the case where only users or only creators leave the platform, we prove that with two-sided departures, approximating maximum total engagement within any constant factor is NP-hard. We present two practical algorithms, one with performance guarantees under mild assumptions on user preferences, and another that tends to outperform algorithms that ignore two-sided departures in practice.
\end{abstract}

\section{Introduction}
Many platforms today, including most social media platforms, are in essence two-sided markets bringing content creators and users together. These platforms highly value intense engagement of users, because they generate revenue by selling digital ads to third parties or sometimes products and services to users, which become more profitable when users are more engaged. User engagement in turn depends on both the quality of the content created and how successfully this content is matched to the interests and preferences of users, making content recommendation algorithms a centerpiece of the operation of these platforms. Even though these algorithms vary between platforms such as Spotify, YouTube and Etsy to social media sites such as Facebook, X (formerly Twitter) and TikTok, all of these platforms rely on high-quality content and matching of this content to interested users. Content recommendations not only affect user behavior, but also production decisions of creators who aim to attract more users, either to more successfully monetize their offered content or because of intrinsic incentives (such as to get a bigger fan base). %

There is a growing literature exploring how platforms can best provide users with content aligned with their preferences, as well as a smaller body of work that looks at incentives of content creators. However, a common assumption among these is that agents on one side of the market are taken ``as given,'' and content recommenations affect either user \textit{or} creator behavior, but not both simultaneously. In our paper, we endogenize both user and creator behaviors through platform participation constraints, where creators require sufficient viewership and users need access to the types of content they desire. Recommendations that do not meet these constraints will induce some users to leave, some creators to leave, or both.  

For example, Neil Young, a prominent singer and songwriter who once had 2.4 million followers and 6 million monthly listeners on Spotify, left the platform in January 2022 due to external reasons. While his departure was unrelated to the shortcomings of the recommendation algorithm, it nonetheless serves as a powerful illustration of the significant consequences of a cascade of departures. Almost immediately, Spotify lost \$4 billion in market value from loss in user engagement, but it did not stop there: many indie artists who had relied on patronage from Neil Young listeners also saw a drop in engagement, making it less appealing to produce new music. It is not hard to believe that, in turn, faithful listeners of these indie artists became disengaged, further exacerbating the spiral of user departures from Spotify. \cite{Qian2022ThePower} observe similar phenomena on the popular streaming platform Twitch: when a prominent ``star creator'' leaves the platform, the platform experiences a drop in both content creation and consumption in categories that are immediately relevant to the departing creator, and smaller creators producing similar content also reduce their effort in content creation.

We propose a model of platform recommendations as a sequential decision problem that incorporates these dynamics of user-content matching. Both users and content creators derive utility from this match -- for users, on the basis of the similarity between their preferences and the content they are assigned to, and for creators, on the basis of how many users are consuming their product. Crucially, both users and content creators leave the platform if their utilities fall below a certain threshold, hence the set of users and content creators dynamically evolve over time.

More specifically, in our model, each user and creator has a type given by a high-dimensional vector, which represents user preferences and characteristics of creator content along different attributes. We assume all types are fixed and known to the platform. At each time step, the platform chooses a fixed number of creators to assign to each user. We model the dynamic departure decisions of both users and creators via their “participation constraints”. This means, in particular, that users leave the platform if their recommended content is too dissimilar to their own interests, measured by cosine similarity between types, and creators will leave if they do not receive enough viewership. 

Within this setup, we study two different platform recommendation schemes. The first is a ``myopic'' \UC\ (UC) algorithm which maximizes the engagement of users without taking into account content creator participation constraints. This is similar to the algorithms that many platforms such as Facebook and Twitter currently use \citep{Lada2021HowDoes, Twitter2023Recommendation}. The second is a \FL\ (FL) algorithm that chooses recommendations to maximize long-term user engagement, taking into account possible departures on both sides of the market. While this latter algorithm focuses on long-run metrics, the myopic \UC\ algorithm greedily assigns each user to their favorite creators and does not consider the possibility of creators leaving, which can then trigger a cascade of creator and user departures on the platform. 

Our first set of results shows that the recommendations given by the \FL\ algorithm over time converge to the solution of a generalized matching problem, which finds a maximum stable set that maximizes user engagement subject to participation constraints of users and creators in this set. We prove that if we ignore user participation constraints, i.e., assume that the user population is fixed, the objective function of this matching problem is submodular in the set of content creators. We show there exists a polynomial-time constant factor approximation algorithm for this case, by noting that the feasibility of satisfying the participation constraints of any subset of creators is entirely captured by a knapsack constraint, where each creator is an item whose cost is her audience threshold, and the budget is the total number of recommendations given to all users. %
This emits a greedy algorithm that incrementally adds creators to the matching. 

We then consider the more general two-sided participation constraints where users are allowed to depart dynamically if their engagement falls below a certain level. We show that computing the optimal solution of this two-sided formulation as well as its constant-factor approximation are NP-hard. Our proof considers a reduction from the maximum independent set problem to finding the maximum stable set in a user-creator instance, where vertices are translated to creators and edges to users. User and creator types are constructed to capture the underlying structure of the graph, i.e., a creator meets a user’s engagement threshold if and only if their corresponding vertex and edge are incident. We show that under this construction, any stable set can be mapped to an independent set in the graph.

On the other hand, we turn to understanding the performance of the myopic \IE\ algorithm. We show that in the worst case, the ratio of the long-term total engagement of the \UC\ algorithm to that of the \FL\ algorithm can be zero, demonstrating the highly suboptimal results that matching on the basis of myopic metrics can produce. The suboptimality may still apply even when user and creator types are randomly generated.

Motivated by the computational complexity of the \FL\ matching problem and the arbitrary suboptimality of the \IE\ algorithm in terms of long-term engagement, we design two practical algorithms that run in polynomial time, one based on clustering and the other based on a careful prioritization of content creator matchings. The \CA\ (LC) algorithm matches sets of users and creators in local neighborhoods, under a density assumption that requires each user to have a sufficiently dense neighborhood of users and creators around her. This assumption may hold when the platform consists of communities of users and creators with similar interests, preferences and identities. When this assumption holds, the algorithm achieves a constant fraction of total engagement relative to the \FL\ algorithm when the dimensionality of user and creator types is fixed. The second algorithm is the \AM\ (CR) algorithm, which examines all creators in increasing order of the size of their ``potential audience'', and prioritizes creators whose constraints are the hardest to satisfy by assigning as many users as possible to them. Simulations corroborate that \CRshort\ achieves good practical performance in terms of total engagement relative to \FL, especially in situations where the \IE\ algorithm is suboptimal (achieves low long-term engagement).

\subsection{Literature Review}
Our work relates to a growing literature on content recommender systems, largely motivated by algorithmic recommendations in social media. Much of this work focuses on matching of fixed content to user preferences (we refer to \cite{Roy2022ASystematicReview} for a review).  There are a few recent works that consider decisions on the supply side of these platforms and simultaneously study user and creator objectives. Among these, \cite{Zhan2021Towards} and \cite{zhu2023online} focus on long term optimization of aggregate utilities of users and creators through sequential decisions and without knowledge of their payoffs. \citeauthor{Zhan2021Towards} in particular presents a reinforcement learning formulation and a simulation environment to model changing user preferences and creator content production decisions. \citeauthor{zhu2023online}, on the other hand, focuses on a two-sided platform linking users and creators, and models the platform’s interaction with creators through contracts and users through recommendations. Their paper uses an online learning setup to analyze the regret achieved under different contracts and their impact on total user payoff. Our work differs from these two papers by assuming fixed user preferences and content attributes, represented by high-dimensional types and known by the platform. It instead focuses on the dynamics of user-creator matching, specifically accounting for departures on both sides of the platform.

The most closely related paper to our work is \cite{mladenov2020optimizing}, which studies maximizing user payoffs while accounting for creator decisions to stay on the platform. \citeauthor{mladenov2020optimizing} assumes user population is fixed without departures, and notes that myopic user-centric policies in this setting can be suboptimal in the long term. Unlike \citeauthor{mladenov2020optimizing}, our model considers both user and creator departures and allows assignment of each user to content from multiple creators at each point in time. We show that considering departures of both sides of the market leads to a computationally much harder matching problem, and provide various practical algorithmic solutions to this computational challenge. More detailed on \citeauthor{mladenov2020optimizing}'s model and comparisons to our work can be found in Section~\ref{sec:users_never_leave} and Appendix~\ref{sec:Mladenov_appendix}.

There are also a number of papers that focus on supply-side competition and equilibria with content creators shaping their content to maximize viewership \citep{Basat2017AGameTheoretic, benporat2019recommendation, benporat2019convergence, BenPorat2020ContentProvider, BenPorat2018AGA, Bhargava2022CreatorEconomy, hron2023modeling, jagadeesan2023supplyside, Qian2023Digital, Yao2023HowBad}. Of these, \cite{jagadeesan2023supplyside} and \cite{hron2023modeling} are closest to our model in their focus on the platform having full information on heterogeneous and continuous user and creator types. These papers assume creators derive utility from one shot matchings where each user is assigned to their favorite content. Our paper shows such user-centric assignments can have arbitrarily bad performance for the platform in the long run. Recently, \cite{Yao2023HowBad} shows the interesting result that a user-centric algorithm with users assigned their top $K$ recommendations could achieve small loss in user welfare with competition among creators. While they assume creator utilities are perfectly aligned with users and depend on the quality of recommendations (user engagement), our model considers asymmetric incentives where creators only care about the cardinality of their audience, which may explain the contrasting results.

Our model also relates to the extensive literature on matching and search (see \cite{abdulkadiroğlu_sönmez_2013} and \cite{Chade2017Sorting} for comprehensive reviews). Much of the literature on platform design for search and matching originates from the classic stable marriage problem first proposed by \cite{Gale1962College} and studied further by works such as \cite{Lones2006Marriage} and \cite{Lauermann2014Stable}. Similar models were later developed for a variety of one-sided and two-sided matching problems, such as housing allocation \citep{Hylland1979Efficient}, school choice \citep{Abdulkadiroğlu2003SchoolChoice} and kidney exchange \citep{Roth2004Kidney}. Notably, the stable marriage problem and similar models typically assume that agents have ordinal preferences over other agents, whereas in our model, agent payoffs are captured by engagement they derive from a given matching.

More recent works on directed search study models where platforms facilitate matching by choosing meeting rates between finitely many agents \citep{immorlica2022designing, Kanoria2021Facilitating, Rios2023Improving, Banerjee2017Segmenting}. These papers solve for stationary equilibria given meeting rates, and design approximation algorithms for the platform’s problem of choosing meeting rates to maximize social welfare. Our work differs from these papers by considering asymmetric objectives for two sides (engagement for users and audience size for creators). More importantly, satisfying participation constraints for users and creators lead to a subset selection problem on both sides and over time, creating computational challenges.

Finally, we would like to note that the generalized matching problem we study as the “solution” of the \FL\ problem is related to the canonical combinatorial optimization problem of generalized assignment problem, which in turn is connected to facility location problems \citep{Ross1977Modeling}. Several variants of the assignment problem have been studied, which involve assigning agents to tasks to maximize total utility from all matches subject to constraints defined on agents and tasks \citep{Fleischer2011TightApproximation, Gavish1991Algorithms, LITVINCHEV20101115, ZHU201630}. These works typically use Lagrangian heuristics and branch-and-bound methods for their approximate solution, without performance guarantees. Our problem is distinguished from most variants of generalized assignment problem in its consideration of subset selection on both sides of the market.

\section{Our Model of Content Recommendations}  \label{sec:model}
We consider a temporal setup with an infinite time horizon. There are two types of \textit{players} on the platform: \textit{users} and \textit{content creators}. Initially, when $t=0$, we denote the initial set of users as $\userset_0=\{1,2,\dots,U\}$ and content creators as $\creatorset_0=\{1,2,\dots,C\}$, where $U\geq 1$ and $C\geq 1$. At each time step, the platform assigns personalized content recommendations, which we model as the content produced by a subset of content creators that the platform chooses for each user. Users then consume the recommended content, and users and creators make decisions on whether to stay on the platform or leave.

\subsection{User and Creator Types}
Each user $i$ has a non-negative \textit{type} $u_i\in \R^D_{\geq 0}$, and each creator $j$ has a non-negative type $c_j\in \R^D_{\geq 0}$.\footnote{For convenience, we may use the notation $u_i$ to refer to both the user and her type vector. When the context is clear, we may also drop the index $i$ and use $u$ to refer to the user and her type. Similarly, we may use either $c$ or $c_j$ to refer to individual creators and their types.} We assume all types are unit vectors in the $L_2$ norm: $\left\lVert u_i \right\rVert_2 = \left\lVert c_j \right\rVert_2 = 1$, for all $i\in \userset_0$ and $j\in \creatorset_0$.\footnote{We show in Appendix~\ref{sec:unit_vector_wlog} that the unit vector assumption holds without loss of generality.} When user $i$ consumes content from creator $j$ in a single time step, she obtains \textit{engagement} $u_i^T c_j$.

The dimensionality constant $D>1$ represents the number of different content attributes or the richness of content: for example, they can include different topics (e.g., politics, sports, entertainment) and attributes (e.g., veracity, political leaning, importance). Each element of the user type measures her preference for that specific content attribute, and each element of the creator type measures the suitability of her content along that dimension. 

Using high-dimensional vectors to represent types and cosine similarity for engagement is common in the literature \citep{hron2023modeling, jagadeesan2023supplyside, mladenov2020optimizing, Yao2023HowBad}. We assume the platform has complete knowledge of all $u_i$'s and $c_j$'s. The platform may obtain knowledge of user and creator types from their past activities or surveys.

\subsection{Recommendations, Utilities and Platform Dynamics}
We assume that through interactions on the platform, the set of users and content creators decrease monotonically over time due to departures of players, and refer to these sets at time $t$ as $\userset_t$ and $\creatorset_t$.\footnote{While our model focuses on endogeneous departure of users and creators, one can also incorporate exogeneous arrivals and departures such that the platform is not always shrinking. We show in Appendix~\ref{sec:user_arrivals_appendix} that a slow arrival rate of users does not change our main results meaningfully, most notably Theorem~\ref{thm:approx_ratio} which shows suboptimality of the \IE\ algorithm.} The timing of events during each $t\geq 0$ is as follows:

\begin{itemize}
    \item The platform uses a \textit{recommendation algorithm} which assigns each user to $K$ content creators, where $K$ is a fixed constant that describes capacity of users. For user $i$, the platform assigns her to a set of creators $\rect(i) \subseteq \creatorset_t$, such that $|\rect(i)| = K$ for all $i\in \userset_t$.\footnote{In some cases, the recommendation algorithm may assign a user to fewer than $K$ creators, i.e., $|\rect(i)| < K$. Most notably, this necessarily happens if there are fewer than $K$ creators remaining on the platform. When this happens, we consider the recommendation to violate user $i$'s participation constraints.} 
    \item Each user $i\in \userset_t$ consumes content from all assigned creators in $\rect(i)$. Her \textit{user engagement} during this time step is the sum of engagements she receives from each assigned creator:
    \begin{align*}
        e_{i,t} &:= \sum_{j\in \rect(i)} u_i^T c_j.
    \end{align*}
    \item Each creator $j\in \creatorset_t$ obtains \textit{audience size} equal to the number of users assigned to her (regardless of their user engagements). Formally, creator $j$'s audience size at time $t$ is
        $$a_{j,t} := \left| \left\{ i \in \userset_t: j\in \rect(i) \right\} \right|.$$
    This can be interpreted as the number of views, likes and other forms of engagement that the creator receives.
    \item All users and creators evaluate their \textit{participation constraints} (defined in the next section), and leave the platform if their respective participation constraints are violated. The remaining sets of users and creators, $\userset_{t+1}\subseteq \userset_t$ and $\creatorset_{t+1}\subseteq \creatorset_t$, are those whose constraints are satisfied at time $t$. These players will be considered for time $t+1$.
\end{itemize}
Formally, a \textit{recommendation algorithm} $\recseq$ describes an infinite sequence of recommendations, $\recseq = \left\{ \rec_0, \rec_1, \dots \right\}$, such that each $\rect$ is a function from $\userset_t$ to subsets of $\creatorset_t$ with size $K$. (We say any such $\rect$ is a set of recommendations \textit{between} users in $\userset_t$ and creators in $\creatorset_t$.) The sequence $\recseq$ induces sets $\userset_1, \userset_2, \dots$ and $\creatorset_1, \creatorset_2, \dots$, the sets of remaining users and creators at each time step. Note that a recommendation algorithm can also be viewed as a ``matching algorithm'', as each $\rect$ describes a many-to-many matching between users and creators, where each user is matched with $K$ creators.

\subsection{User and Creator Participation Constraints}  \label{sec:participation_constraints}
At the end of each time step $t$, users and creators evaluate their participation constraints based on the recommendations $\rect$ that they were provided with at this time step.
\\~\\
\textbf{User constraints.} Each user $i$ requires that the engagement she receives from each assigned creator must be above a lower bound $\ebar$. Formally, $u_i^T c_j \geq \ebar$ for all $j\in \rect(i)$. This is a requirement on the \textit{quality} of recommendations: content produced by each assigned creator must be similar enough to the user's own type, so that this creator provides sufficient engagement and relevant content for the user. We say user $i$ is \textit{happy} with creator $j$ if $u_i^T c_j \geq \ebar$.
\\\\
\textbf{Creator constraints.} Each creator $j$ requires her audience size to be no smaller than a lower bound $\abar$. Formally, $a_{j,t} \geq \abar$. This means creators care about \textit{quantity}: they must receive sustained attention from users (views, likes etc) in order to keep producing content, but they do not care about how much engagement users get from them.
\\\\
Each player stays on the platform after time $t$ if the recommendation $\rect$ \textit{satisfies} her participation constraint, and leaves the platform permanently otherwise (i.e., if $\rect$ \textit{violates} her constraint). Formally,
\begin{align*}
    \userset_{t+1} &:= \left\{ i\in \userset_t: \left| \rect(i) \right| = K \ \text{and}\ u_i^T c_j \geq \ebar \ \forall j\in \rect(i) \right\},  \\
    \creatorset_{t+1} &:= \left\{ j\in \creatorset_t: a_{j,t} = \left| \left\{ i \in \userset_t: j\in \rect(i) \right\} \right| \geq \abar \right\}.
\end{align*}
The user constraint value $\ebar \in \left[0,1\right]$ and creator constraint value $\abar \in \mathbb{N}_{0}$ are fixed input parameters that are known to the platform, and apply to all users and creators. We show in Appendix~\ref{sec:homogeneous_creators_wlog} that homogeneneity of creator constraints holds without loss of generality. In addition, we claim that homogeneity in user constraints is a reasonable assumption: allowing a user to have a lower threshold on similarity to her assigned creators can be achieved by modifying her type vector $u_i$, which also measures her similarity to creator types.

The platform's choice of the sequence of recommendations $\recseq$ induces the sets of users and creators over time, $\left\{ \left( \userset_t, \creatorset_t \right) \right\}$, due to participation constraints and departures. This is a nested sequence, as $\userset_{t+1} \subseteq \userset_t$ and $\creatorset_{t+1} \subseteq \creatorset_t$ for all $t\geq 0$; in other words, the sets of users and creators ``shrink'' over time. Even though participation constraints do not restrict the choice of $\rect$ directly, the platform may consider downstream effects of these participation constraints when choosing $\rect$.

\subsection{Platform Objectives and Optimal Algorithms}
The platform's goal is to assign recommendations that \textit{maximize long-term total user engagement}. For any sets of users $\userset \subseteq \userset_0$, creators $\creatorset \subseteq \creatorset_0$ and recommendations $\rec$ between them, we define the total engagement as the sum of each user's individual engagements:
\begin{align*}
    \engagement\left(\userset, \creatorset, \rec\right) := \sum_{i\in \userset} \sum_{j\in \rec(i)} u_i^T c_j.  %
\end{align*}
In particular, we use $\engagement_t := \engagement\left(\userset_t, \creatorset_t, \rect \right)$ to denote the total engagement at time $t$ following a sequence of recommendations $\recseq = \left\{ \rec_0, \rec_1, \dots \right\}$, where $\userset_t$ and $\creatorset_t$ result from the recommendation algorithm $\recseq$. Note that $\engagement_t$ includes all users $\userset_t$ at the start of time $t$, even if they leave at the end of this time step after recommendations $\rect$ violate their participation constraints. This is because prior to making their decisions on departures, these users have already consumed the content and thus contributed to platform engagement.  \possiblefootnote

We assume the platform aims to maximize total engagement over an infinite time horizon. In particular, when the platform uses a recommendation algorithm $\recseq$, we define its objective, the \textit{long-term total engagement}, as the average total engagement over $T$ time steps as $T\rightarrow \infty$:
\begin{align}
    \longtermengagement \left( \recseq \right) := \lim_{T\rightarrow \infty} \frac{1}{T} \sum_{t=0}^{T-1} \engagement_t = \lim_{T\rightarrow \infty} \frac{1}{T} \sum_{t=0}^{T-1} \engagement\left(\userset_t, \creatorset_t, \rect \right).  \label{eqn:def_long_term_engagement}
\end{align}
When the context is clear, we may simply denote the long-term engagement as $\longtermengagement$. 
Note that the limit in \eqref{eqn:def_long_term_engagement} is well-defined, because $\engagement\left( \userset_t, \creatorset_t, \rect \right)$ is bounded at all times $t$ since $\userset_t$ and $\creatorset_t$ shrink over time.  \possiblefootnote

\subsection{Example}  \label{sec:examples}

\begin{example}  \label{example:model_simple}
    Consider the instance shown in Figure~\ref{fig:example_model_simple} with 2 creators and 6 users. Here, $D=2$ (hence user and creator types are non-negative unit vectors in $\mathbb{R}^2$), $K=1$ (each user is assigned to one creator), $\abar = 3$ (each creator needs 3 users), and $\ebar=\cos (\pi/3)$ (for a user to be happy with a creator, their types must be within an angular distance of $\pi/3$). Creator $c_1$ and users $u_1$ and $u_2$ have type $(0,1)$, creator $c_2$ and users $u_4$, $u_5$ and $u_6$ have type $(1,0)$, and user $u_3$ has type $\left( \cos (\pi/6), \sin (\pi/6) \right)$, as illustrated in Figure~\ref{fig:example_model_simple_vectors}.
    
    \begin{figure}[h]
        \centering
        \begin{subfigure}[c]{0.45\textwidth}
            \centering
            \includegraphics[width=0.7\textwidth]{tikz/ex_simple_vectors.tikz}
            \caption{User and creator types in $\mathbb{R}^2$. Solid dots are creators, and open dots are users. The points in the top left and bottom right clusters are all located at $(0,1)$ and $(1,0)$ respectively. User constraints are $\ebar = \cos (\pi/3)$, so $u_3$ is happy with both $c_1$ and $c_2$.}
            \label{fig:example_model_simple_vectors}
        \end{subfigure}
        \hfill%
        \begin{minipage}[c]{0.45\textwidth}
            \begin{subfigure}{\textwidth}
                \centering
                \includegraphics{tikz/ex_simple_bipartite.tikz}
                \caption{An alternative representation of Example~\ref{example:model_simple} as a bipartite graph. An edge exists if the user is happy with the creator. Edge weights are engagement values (dot products).}
                \label{fig:example_model_simple_bipartite}
            \end{subfigure}
            \medskip
            \begin{subfigure}{\textwidth}
                \centering
                \includegraphics{tikz/ex_simple_fl.tikz}
                \caption{A recommendation that satisfies all user and creator constraints, with total engagement $5.5$.}
                \label{fig:example_model_simple_fl}
            \end{subfigure}
        \end{minipage}
        \par\bigskip
        \begin{subfigure}{\textwidth}
            \centering
            \begin{subfigure}[t]{0.3\textwidth}
                \centering
                \includegraphics{tikz/ex_simple_ie0.tikz}
                \caption*{$t=0$, engagement $5+\sqrt{3}/2$}
            \end{subfigure}
            \hfill
            \begin{subfigure}[t]{0.3\textwidth}
                \centering
                \includegraphics{tikz/ex_simple_ie1.tikz}
                \caption*{$t=1$, engagement $3+\sqrt{3}/2$. Creator $c_1$ leaves due to insufficient audience size.}
            \end{subfigure}
            \hfill
            \begin{subfigure}[t]{0.3\textwidth}
                \centering
                \includegraphics{tikz/ex_simple_ie2.tikz}
                \caption*{$t=2$, engagement $3+\sqrt{3}/2$. Users $u_1$ and $u_2$ leave due to insufficient engagement.}
            \end{subfigure}
            \caption{A sequence of recommendations that achieves higher engagement than $5.5$ at $t=0$, but lower engagement at later times.}
            \label{fig:example_model_simple_ie}
        \end{subfigure}
        \caption{Illustration of Example~\ref{example:model_simple}}
        \label{fig:example_model_simple}
    \end{figure}

    Users $u_1$ and $u_2$ must be assigned to creator $c_1$ in order to stay on the platform, and likewise, $u_4$, $u_5$ and $u_6$ must be assigned to $c_2$. User $u_3$ is happy with both creators, but receives higher engagement from $c_2$.

    Suppose the platform greedily assigns each user to her favorite creator (the one that gives her the highest engagement). Such a recommendation scheme and its evolution over time is shown in Figure~\ref{fig:example_model_simple_ie}. At $t=0$, user $u_3$ is assigned to $c_2$ which she prefers, and the platform receives immediate total engagement $5+\sqrt{3}/2$. However, creator $c_1$ only receives 2 users, so her participation constraint of an audience size of at least $\abar=3$ is violated, and she leaves the platform after $t=0$. At the next time step $t=1$, $u_1$ and $u_2$ have no choice but to be assigned to $c_2$, the only remaining creator. However, their user constraints are violated, as they receive engagement 0 from $c_2$ which is below $\ebar$. From $t=2$ onwards, only $u_3$, $u_4$, $u_5$, $u_6$ and $c_2$ stay on the platform. The long-term engagement is $3+\sqrt{3}/2$, which is below the initial engagement at $t=0$.
    
    An alternative recommendation algorithm is shown in Figure~\ref{fig:example_model_simple_fl}: at all times $t$, assign users $u_1$, $u_2$ and $u_3$ to creator $c_1$, and users $u_4$, $u_5$ and $u_6$ to creator $c_2$. This recommendation satisfies all user and creator constraints, and therefore keeps all players on the platform at all times. In fact, this is the recommendation that achieves maximum long-term total engagement of $5.5$. Note that $u_3$ is not assigned to the creator that she prefers the most; however, doing so allows $c_1$ to receive sufficient audience, so that both $c_1$ and users $u_1$ and $u_2$ stay on the platform. Essentially, the platform sacrifices engagements of users with more flexible constraints ($u_3$) to ensure sustainability of a greater number of creators \textit{and} users ($c_1$, $u_1$ and $u_2$).
\end{example}

\begin{remark}
    The recommendation in Figure~\ref{fig:example_model_simple_fl} corresponds to the outcome of the \FL\ algorithm (maximizing total engagement subject to both user and creator participation constraints), whereas the sequence of recommendations in Figure~\ref{fig:example_model_simple_ie} is the outcome of the \IE\ algorithm (maximizing total engagement while only accounting for user constraints but not creators). We will describe these two baseline recommendation algorithms in more detail in Sections~\ref{sec:fl} and \ref{sec:uc}. %

\end{remark}

\section{The \FL\ Algorithm}  \label{sec:fl}
In this section, we define the \FL\ (FL) recommendation algorithm, the first-best solution that maximizes long-term total engagement. We first show that under the \FL\ algorithm, the sets of users and creators must converge to a \textit{maximum stable set}, which is a subset of users and creators associated with recommendations that achieve maximum engagement while ensuring all their participation constraints are satisfied. We then formulate the maximum stable set problem as an integer linear program. As a special case, we show that if user participation constraints are ignored (i.e., users are always satisfied with the quality of recommendations and never leave), the resulting matching problem with one-sided creator constraints can be approximated efficiently. Finally, we show that in the more general case with two-sided participation constraints, obtaining an exact or approximate solution is computationally hard.

\subsection{Definition}  \label{sec:def_fl}

\begin{definition}[\FL]  \label{def:fl}
The \textit{\FL} (FL) recommendation algorithm is one that maximizes long-term total user engagement.  Formally, $\FLseq = \left\{ FL_0, FL_1, \dots \right\}$ is a sequence of recommendations such that
\begin{align*}
    \FLseq &\in \argmax_{\recseq = \left\{ \rec_0, \rec_1, \dots \right\} } \longtermengagement\left(\recseq\right) = \argmax_{\recseq = \left\{ \rec_0, \rec_1, \dots \right\} } \lim_{T\rightarrow \infty} \frac{1}{T} \sum_{t=0}^{T-1} \engagement\left(\userset_t, \creatorset_t, \rect \right).  %
\end{align*}
\end{definition}

Note that the sequence of recommendations $\FLseq = \left\{ FL_0, FL_1, \dots \right\}$ that achieves this maximum is not necessarily unique; however, all such sequences give the same long-term total engagement by definition.

\subsection{Stable Set and its Relation to the \FL\ Algorithm}  \label{sec:stable_set}
We now develop a deeper understanding of the \FL\ algorithm by examining its long-term behavior. Observe that for any sequence of recommendations $\FLseq$ given by the algorithm, as $t\rightarrow \infty$, the sets of users and creators on the platform induced by $\FLseq$ shrink and therefore converge to a smaller subset (possibly empty), and all participation constraints of players in this subset are satisfied. We refer to such a set as a \textit{stable set} of users and creators, and show that the sets of users and creators induced by any recommendation sequence $\FLseq$ must converge to a \textit{maximum stable set} that maximizes engagement among all stable sets. This means our first-best \FL\ objective can be solved using an integer linear program that computes the maximum stable set.

\begin{definition} [Stable set] \label{def:stable_set}
    Given a subset of users $\userset \subseteq \userset_0$, a subset of creators $\creatorset \subseteq \creatorset_0$ and recommendations $\rec$ between them, $\left(\userset, \creatorset \right)$ is a \textit{stable set (associated) with recommendations} $\rec$, if $\rec$ satisfies all participation constraints of users in $\userset$ and creators in $\creatorset$. In other words, $\rec(i) \subseteq \creatorset$ for all $i\in \userset$, such that 
    \begin{align*}
        \left| \rec(i) \right| &= K, \quad \forall i\in \userset,  \\
        u_i^T c_j &\geq \ebar, \quad \forall i\in \userset, j\in \rec(i),  \\
        \left| \left\{ i \in \mathcal{U}: j\in \rec(i) \right\} \right| &\geq \abar, \quad \forall j\in \creatorset.
    \end{align*}

    In particular, a stable set $\left(\userset^*, \creatorset^* \right)$ is a \textit{maximum stable set (associated) with recommendations} $\rec^*$ if it achieves maximum total engagement among all stable sets: $\left(\userset^*, \creatorset^*, \rec^* \right) \in \argmax_{\left( \userset, \creatorset, \rec  \right)} \engagement\left( \userset, \creatorset, \rec \right).$ 
    
    (In some contexts, we may refer to $\left(\userset, \creatorset \right)$ as a stable set while omitting the associated recommendations $\rec$, as long as such an $\rec$ exists. Likewise, we may simply call $\left(\userset^*, \creatorset^* \right)$ a maximum stable set as long as there exists $\rec^*$ that achieves maximum engagement out of all stable sets.)  \possiblefootnote
\end{definition}
There can be multiple different maximum stable sets in any instance, and multiple recommendations associated with each, as long as they have the same total engagement. Note that $(\emptyset, \emptyset)$ is a stable set and may be a maximum stable set.  \possiblefootnote %

Recall from Section~\ref{sec:participation_constraints} that under any sequence of recommendations $\recseq$, the induced sets of users and creators $\left\{ \left( \userset_t, \creatorset_t \right) \right\}$ shrink over time. This means it converges to a limit, i.e., there exists $t_0$ such that for all $t\geq t_0$, $\userset_t = \userset_{t_0}$ and $\creatorset_t = \creatorset_{t_0}$. Moreover, since the sets stop shrinking after $t_0$, this means no users or creators leave at such times $t$, so the participation constraints of all users and creators in the limit are satisfied by $\rect$. Therefore, the sets of users and creators \textit{converge to a stable set} $\left( \userset_{t_0}, \creatorset_{t_0} \right)$ as $t\rightarrow \infty$.\footnote{There may be multiple recommendations associated with the stable set $\left( \userset_{t_0}, \creatorset_{t_0} \right)$. In particular, the recommendation sequence $\recseq$ may assign different recommendations at different time steps even after convergence: it is possible that $\rec_t \neq \rec_{t_0}$ for some $t\geq t_0$, but both $\rec_t$ and $\rec_{t_0}$ satisfy all participation constraints of $\userset_{t_0}$ and $\creatorset_{t_0}$, and are therefore valid recommendations associated with the stable set $\left( \userset_{t_0}, \creatorset_{t_0} \right)$.}

Next, we show that the \FL\ algorithm induces sets of users and creators that converge to a maximum stable set:

\begin{proposition}  \label{prop:fl_mss}
    Let $\FLseq=\left\{ FL_0, FL_1, \dots \right\}$ be any sequence of recommendations given by the \FL\ algorithm, $\left(\userset^*, \creatorset^* \right)$ be any maximum stable set with recommendations $\rec^*$ (that may not relate to $\FLseq$), and $\mathbf{ALG}=\left\{ ALG_0, ALG_1, \dots \right\}$ be an infinite sequence of recommendations such that $ALG_t = \rec^*$ at all times $t$.\footnote{Technically, recommendations $ALG_0(i)$ still needs to be defined for any user in $\userset_0 \backslash \userset^*$, i.e., not in the maximum stable set. This can be done by setting $ALG_0(i)$ to be an arbitrary subset of $\creatorset^*$ with size $K$, and letting these users leave after $t=0$.} Then, $\longtermengagement\left(\FLseq\right) = \longtermengagement\left( \mathbf{ALG} \right)$.
\end{proposition}
Proposition~\ref{prop:fl_mss} shows that the long-term engagement achieved by $\FLseq$ can be obtained with recommendations induced by any maximum stable set. Therefore, we can focus on finding a maximum stable set to characterize the engagement given by the \FL\ algorithm. We will refer to a maximum stable set as the \textit{\FL\ solution}. We present the proof of Proposition~\ref{prop:fl_mss} in Appendix~\ref{sec:fl_mss_proof}, where we also show additional statements such as convergence of any sequence $\FLseq$ to a maximum stable set.

Finally, we present an integer linear program that numerically solves for the maximum stable set and thus the \FL\ solution. Given a subset of users $\userset \subseteq \userset_0$ and creators $\creatorset \subseteq \creatorset_0$, with recommendations $\rec$ between them, $\left( \userset, \creatorset \right)$ is a maximum stable set with recommendations $\rec$ if and only if it gives a solution to the following integer linear program:
\begin{alignat*}{4}
    & \max_{x_{ij}, y_i, z_j} & \sum_{i=1}^U \sum_{j=1}^C x_{ij} u_i^T c_j &&&&& \quad \text{(total engagement)} \\
    & \text{subject to} \quad & x_{ij} \left( u_i^T c_j - \ebar \right) & \geq 0, \quad && \forall 1\leq i\leq U, 1\leq j\leq C, && \quad \text{(user constraints)} \\
    && \sum_{i=1}^U x_{ij} &\geq z_j \abar, \quad && \forall 1\leq j\leq C, && \quad \text{(creator constraints)}  \\
    && \sum_{j=1}^C x_{ij} &= K, && \forall 1\leq i\leq U, && \quad \text{(number of recommendations)}  \\
    && x_{ij} &\leq y_i, \quad && \forall 1\leq i\leq U, 1\leq j\leq C,  && \\
    && x_{ij} &\leq z_j, \quad && \forall 1\leq i\leq U, 1\leq j\leq C,  && \\
    && x_{ij}, y_i, z_j &\in \{0,1\}, \quad && \forall 1\leq i\leq U, 1\leq j\leq C,  &&
\end{alignat*}
where $x_{ij}$ indicates whether user $i$ is matched to creator $j$, $y_i\in \{0,1\}$ indicates whether user $i$ is in $\userset$, and $z_j\in \{0,1\}$ indicates whether creator $j$ is in $\creatorset$. More formally, $x_{ij} = \mathbf{1}_{R(i)}(j), y_i = \mathbf{1}_{\userset}(i)$, and $z_j = \mathbf{1}_{\creatorset}(j)$, where $\mathbf{1}_S(a)$ is an indicator function whose value is 1 if $a\in S$ and 0 otherwise.

\subsection{One-Sided Creator Participation Constraints}  \label{sec:users_never_leave}
We first consider the case where all user constraints are ignored; equivalently, $\ebar = 0$. This means users never leave the platform as long as they are assigned to any $K$ creators, regardless of whether their content match the user's preferences. In this case, the model only has one-sided creator participation constraints. This is a generalization of the baseline model in \cite{mladenov2020optimizing}, who also consider creator departures due to insufficient audience but without departure of users. The main modeling difference is that they assume each user is only assigned to one creator ($K=1$), whereas we allow arbitrary values of $K$ such that users can be assigned to multiple creators. (This difference affects the analysis but not the results.)

We show that for user-creator matching with one-sided creator constraints, the first-best solution can be approximated efficiently:
\begin{theorem}[Approximability without user constraints]  \label{thm:approx_bigK}
    If users never leave (equivalently, $\ebar=0$), \MSS\ can be approximated up to factor $\left( 1/e-\epsilon \right)$ in polynomial time.
\end{theorem}
Theorem~\ref{thm:approx_bigK} extends the analogous result in \citeauthor{mladenov2020optimizing} on approximation to $K>1$.\footnote{We also remark that while our model defines the engagement each user derives from each creator as the dot product of their types, Theorem~\ref{thm:approx_bigK} and Proposition~\ref{prop:submodular_fixU} still hold under a slight generalization that allows engagement of each user-creator pair to be arbitrarily defined.  \possiblefootnote}

The proof of Theorem~\ref{thm:approx_bigK} relies on two key observations. First, when the set of users is exogenously fixed (for example, by assuming $\ebar=0$ and $\userset = \userset_0$), the objective function is submodular in the set of creators:
\begin{proposition}[Submodularity when users are fixed]  \label{prop:submodular_fixU}    
    Suppose $\ebar = 0$.\footnote{This assumption allows Proposition~\ref{prop:submodular_fixU} to be applied to $\userset = \userset_0$.}
    Let $f(\userset, \creatorset)$ denote the maximum engagement of any stable set $\left(\userset, \creatorset \right)$, i.e.,
    $$ f\left( \userset, \creatorset \right) := \max_{\rec} \engagement \left( \userset, \creatorset, \rec \right), $$
    where $\rec$ is subject to participation constraints of all players in $\userset$ and $\creatorset$, as long as $\left\lvert \creatorset \right\rvert \geq K$.\footnote{When $\left\lvert \creatorset \right\rvert < K$, we instead define $f(\userset, \creatorset)$ as the maximum engagement in a slightly modified instance where each user receives $\left\lvert \creatorset \right\rvert$ recommendations instead of $K$. More formal definitions and explanations are in Appendix~\ref{sec:Mladenov_appendix}.}
    Then, for any $\userset \subseteq \userset_0$, $c_0, c_1 \in \creatorset_0$ and $\creatorset \subseteq \creatorset_0 \backslash \left\{ c_0, c_1 \right\}$,
    \begin{align}
        f\left( \userset, \creatorset \cup \left\{ c_0, c_1 \right\} \right) - f\left( \userset, \creatorset \cup \left\{ c_1 \right\} \right) &\leq f\left( \userset, \creatorset \cup \left\{ c_0 \right\} \right) - f\left( \userset, \creatorset \right),  \label{eqn:submodular_fixU}
    \end{align}
    as long as all four terms are well-defined. This applies to all values of $K$.
\end{proposition}
Second, when $\ebar=0$, the ``feasibility region'' of whether $f\left( \userset_0, \creatorset \right)$ is well-defined for any set of creators $\creatorset \subseteq \creatorset_0$ (i.e., whether $\left( \userset_0, \creatorset \right)$ is a stable set) is characterized by a single knapsack constraint: $\sum_{j\in \creatorset} \abar \leq UK$, or more succinctly, $\left| \creatorset \right| \abar \leq UK$. Intuitively, treating each creator as an item with its ``cost'' being her audience requirement $\abar$, selecting all creators in $\creatorset$ is feasible if and only if the total cost does not exist the ``budget'' of the total number of recommendations $UK$ given to all users. Combining submodularity and the knapsack constraint means a constant factor approximation can be computer in polynomial time using the continuous greedy algorithm given by \cite{Feldman2011AUnified}, thus showing Theorem~\ref{thm:approx_bigK}.

\begin{figure}[h!]
    \centering
    \begin{tikzpicture}[node distance={20mm}, %
            main/.style = {draw, circle}, 
            every edge quotes/.style = {auto, font=\footnotesize}] 
        \begin{scope}[nodes=main]
            \node [ellipse, align=center, anchor=center] (R) at (0,0) {$\rec$: Optimal \\$f\left(\userset, \creatorset\right)$};
            \node [ellipse, align=center, anchor=center] (R1) at (6.5,1.5) {$\rec^1$: Optimal \\$f\left(\userset, \creatorset \cup \left\{ c_1 \right\} \right)$};
            \node [ellipse, align=center, anchor=center] (R01) at (13,0) {$\rec^{0,1}$: Optimal \\$f\left(\userset, \creatorset \cup \left\{ c_0, c_1 \right\} \right)$};
            \node [ellipse, align=center, anchor=center] (R0) at (6.5,-1.5) {$\rec^0$: Feasible \\$\leq f\left(\userset, \creatorset \cup \left\{ c_0 \right\} \right)$};
        \end{scope}
        
        \draw [->, blue, thick] (R) -- (R1) node [midway, sloped, above] {$G^1$: add $c_1$} node [midway, sloped, below] {\footnotesize paths that end at $c_1$};
        \draw [->, red, thick] (R1) -- (R01) node [midway, sloped, above] {$G^{0,1}$: add $c_0$} node [midway, sloped, below] {\footnotesize paths that end at $c_0$};
        \draw [->, violet, thick] (R) -- (R0) node [midway, sloped, above] {$G^0$: add $c_0$} node [midway, sloped, align=center, anchor=north] {{\footnotesize \textcolor{red}{tails of all red paths}} \\ {\footnotesize \textcolor{blue}{heads of some blue paths}}};
    \end{tikzpicture} 
    \caption[Outline of the proof sketch of Proposition~\ref{prop:submodular_fixU}]{Outline of the proof sketch of Proposition~\ref{prop:submodular_fixU}. Vertices in this figure indicate matchings between the specified sets of users and creators that satisfy all constraints. Edges in this figure indicate relocation graphs that capture changes from one matching to another: for example, $G^1$ represents changes from $R$ to $R^1$. The left hand side of \eqref{eqn:submodular_fixU} is the sum of edge weights of $G^{0,1}$ (red edge), and the right hand side is lower bounded by that of $G^0$ (violet edge).}
    \label{fig:submod_proof_outline}
\end{figure}
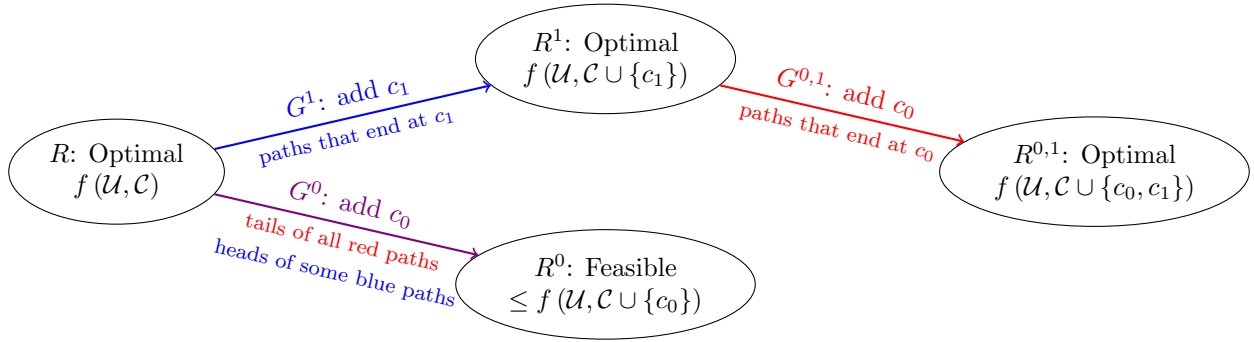

Below, we provide a sketch of the proof of Proposition~\ref{prop:submodular_fixU}. %
The proof constructs a ``relocation graph'' $G^1$ that represents changes in user assignments when a new creator $c_1$ is added to the optimal recommendations between $\userset$ and $\creatorset$, and another graph $G^{0,1}$ when another new creator $c_0$ is then added. The weights of edges in each graph are constructed such that they represent the change in engagement with the new assignments. We then show that there exists a feasible $G^0$ that adds creator $c_0$ but not $c_1$, which can be constructed by combining some edges from both $G^{0,1}$ and $G^1$. We show that $G^0$ gives a higher difference in engagement than $G^{0,1}$, by proving that replacing the unselected edges in $G^{0,1}$ with the selected edges in $G^1$ must not reduce the sum of edge weights, due to optimality of matchings used to construct $G^1$. This gives the submodularity property. A high-level picture of relocation graphs and their associated matchings is shown in Figure~\ref{fig:submod_proof_outline}.

More specifically, consider the two terms on the left hand side of \eqref{eqn:submodular_fixU}. We explicitly write down the optimal matchings between the sets of users and creators specified, i.e., the recommendations that achieves maximum engagement: $\rec^{0,1} \in \argmax_{\hat{\rec}} \engagement\left( \userset, \creatorset \cup \left\{ c_0, c_1 \right\}, \hat{\rec}\right)$ and $\rec^{1} \in \argmax_{\hat{\rec}} \engagement\left( \userset, \creatorset \cup \left\{ c_1 \right\}, \hat{\rec}\right)$. Furthermore, we construct a \textit{relocation graph} $G^{0,1}$ that expresses all changes in each user's recommendations from $\rec^{1} \left( u \right)$ to $\rec^{0,1} \left( u \right)$ when the new creator $c_0$ is added. The relocation graph is a directed weighted graph whose vertices are creators and edges are labeled with users, such that an edge from $c$ to $c'$ with label $u$ can only exist if $c \in \rec^1 \left( u \right)$ and $c'\in \rec^{0,1} \left( u \right)$ (i.e., $c$ is ``replaced'' with $c'$). Edge weights are defined as the difference in engagement that user $u$ receives between the two creators, i.e., $u^T c' - u^T c$. This means the left hand side of \eqref{eqn:submodular_fixU} equals to the total engagement of $\rec^{0,1}$ minus that of $\rec^1$, which is the sum of weights of all edges in $G^{0,1}$. We say all edges in $G^{0,1}$ are colored \textit{red}.

To analyze the right hand side of \eqref{eqn:submodular_fixU}, we let $\rec \in \argmax_{\hat{\rec}} \engagement\left( \userset, \creatorset, \hat{\rec}\right)$ be the optimal matching for the second term $f\left( \userset, \creatorset \right)$, but construct a feasible matching $\rec^0$ between $\userset$ and $\creatorset \cup \left\{ c_0 \right\}$ that satisfies all participation constraints but may not give maximum engagement. Thus, $\engagement \left( \userset, \creatorset \cup \left\{ c_0 \right\}, \rec^0 \right) \leq f\left( \userset, \creatorset \cup \left\{ c_0 \right\} \right)$, and we aim to show that \eqref{eqn:submodular_fixU} holds when replacing the right hand side with $\engagement \left( \userset, \creatorset \cup \left\{ c_0 \right\}, \rec^0 \right) - f\left( \userset, \creatorset \right)$.

Ideally, we would like to construct $\rec^0$ by ``duplicating'' all changes in matchings that are represented by the red graph $G^{0,1}$ (which collectively assign users to $c_0$), but with the initial matching being $\rec$ instead of $\rec^1$. However, some of these changes cannot be applied when the starting point is $\rec$, as it is possible that $\rec\left(u\right) \neq \rec^1 \left(u\right)$ for some users $u$ (for example, $\rec^1 \left(u\right)$ may contain $c_1$, but $\rec\left(u\right)$ cannot). To reason about these cases, we consider another relocation graph $G^1$ that expresses changes from $\rec$ to $\rec^1$, whose edges are colored \textit{blue}.

We show that both $G^1$ and $G^{0,1}$ are directed acyclic graphs that can each be decomposed into edge-disjoint paths, such that all paths in $G^1$ end at $c_1$ and all paths in $G^{0,1}$ end at $c_0$. Intuitively, when a new creator is added, all changes between the optimal matchings only serve the fundamental purpose of reassigning at least $\abar$ users to the new creator, but the reassignments may be done indirectly (e.g., user $u_1$ is reassigned from $c$ to $c'$, $u_2$ reassigned from $c'$ to $c''$, etc.) We further characterize all possible cases that a red edge in $G^{0,1}$ (a change from $\rec^1$ to $\rec^{0,1}$) cannot be applied to $\rec$ without first performing relocations given by a blue edge in $G^1$ (a change from $\rec$ to $\rec^1$), which we call \textit{junctions}. In these cases, the blue edge is a prerequisite for the red edge. 

Finally, we present an algorithm that constructs a new feasible graph $G^0$ by selecting a subset of red edges and a subset of blue edges, and construct $\rec^0$ to be the matching induced by $G^0$. The selected edges in $G^0$ have the property that 
if a selected red edge has a prerequisite blue edge, then the blue edge is also selected,
thereby ``resolving'' the junctions and ensuring $\rec^0$ is well-defined and satisfies all constraints.
Furthermore, we show that the sum of weights of all selected edges in $G^0$ are at least the sum of weights of all red edges in $G^{0,1}$, i.e., replacing the unselected red edges with the selected blue edges cannot reduce the total edge weights, because otherwise the optimality of either $\rec$ or $\rec^1$ is violated. The second property shows the inequality in \eqref{eqn:submodular_fixU}.

Our proof builds on that of the analogous result in \citeauthor{mladenov2020optimizing}. In particular, we use the same high-level idea of capturing changes between matchings as relocation graphs that decompose into paths (Figure~\ref{fig:submod_proof_outline}), but the most significant difference is in the construction of $G^0$ and its analysis. Full proofs of Proposition~\ref{prop:submodular_fixU} and Theorem~\ref{thm:approx_bigK}, as well as more details on the comparison to \citeauthor{mladenov2020optimizing}, can be found in Appendix~\ref{sec:Mladenov_appendix}.

\subsection{Hardness of Approximation with User Participation Constraints}  \label{sec:MSS_hard}
We now consider the more general case where user participation constraints can be non-trivial, i.e., $\ebar \geq 0$. This allows users to leave the platform due to insufficient engagement from recommendations. In this setting, even obtaining an approximate solution to the maximum stable set problem is computationally hard, unlike the case in Section~\ref{sec:users_never_leave} with only creator constraints.

\begin{theorem}[Inapproximability of the general model] \label{thm:hard}
    Given any $D\geq 6$ that does not depend on $U$ and $C$, finding a maximum stable set for an arbitrary instance with dimensionality $D$ is NP-hard. In addition, for any constant $c\geq 1$, finding a stable set whose total engagement is a $c$-approximation of the maximum stable set is NP-hard.
\end{theorem}

The proof of Theorem~\ref{thm:hard} uses a reduction from the NP-hard maximum independent set (MIS) problem \citep{Karp1972Reducibility}. 
The hardness of approximation follows naturally from the hardness of approximation of \MIS, as \cite{Zuckerman2007Linear} showed that approximating the equivalent maximum clique problem within a factor of $n^{1-\epsilon}$ is NP-hard.

Comparing the inapproximability result of Theorem~\ref{thm:hard} to the approximability result of Theorem~\ref{thm:approx_bigK} demonstrates the additional complexity due to two-sided participation constraints. When both users and creators can leave the platform, computing the first-best \FL\ solution is harder than the case with only one-sided creator participation constraints.

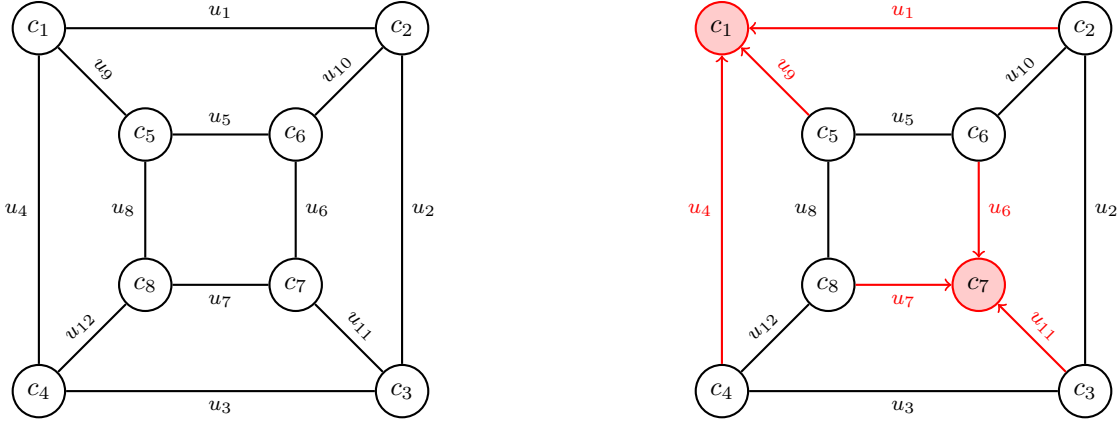
\begin{figure}[h!]
    \centering
    \begin{subfigure}[t]{0.45\textwidth}
        \centering
        \begin{tikzpicture}[node distance={20mm}, thick, 
                main/.style = {draw, circle}, 
                every edge quotes/.style = {auto, font=\footnotesize}] 
            \begin{scope}[nodes=main]
                \node (5) {$c_5$};
                \node (6) [right of=5] {$c_6$};
                \node (7) [below of=6] {$c_7$};
                \node (8) [left of=7] {$c_8$};
                \node (1) [above left of=5] {$c_1$};
                \node (2) [above right of=6] {$c_2$};
                \node (3) [below right of=7] {$c_3$};
                \node (4) [below left of=8] {$c_4$};
            \end{scope}
            
            \draw (1) edge["$u_1$"] (2)
            (2) edge["$u_2$"] (3)
            (3) edge["$u_3$"] (4)
            (4) edge["$u_4$"] (1)
            (5) edge["$u_5$"] (6)
            (6) edge["$u_6$"] (7)
            (7) edge["$u_7$"] (8)
            (8) edge["$u_8$"] (5)
            (1) edge["$u_9$", sloped] (5)
            (2) edge["$u_{10}$", sloped] (6)
            (3) edge["$u_{11}$", sloped] (7)
            (4) edge["$u_{12}$", sloped] (8);
        \end{tikzpicture} 
        \caption{A 3-regular graph $G$ with vertices labeled as creators and edges labeled as users. Each user is only happy with the two adjacent creators. Each user receives $K=1$ recommendation and each creator needs $\abar=3$ users.}
        \label{fig:reduction_regular_G}
    \end{subfigure}
    \hfill
    \begin{subfigure}[t]{0.45\textwidth}
        \centering
        \begin{tikzpicture}[node distance={20mm}, thick, 
                main/.style = {draw, circle}, 
                set/.style = {draw=red, circle, fill=red!20}, 
                every edge quotes/.style = {auto, font=\footnotesize}] 
            \begin{scope}[nodes=main]
                \node (5) {$c_5$};
                \node (6) [right of=5] {$c_6$};
                \node [set] (7) [below of=6] {$c_7$};
                \node (8) [left of=7] {$c_8$};
                \node [set] (1) [above left of=5] {$c_1$};
                \node (2) [above right of=6] {$c_2$};
                \node (3) [below right of=7] {$c_3$};
                \node (4) [below left of=8] {$c_4$};
            \end{scope}
            
            \draw (1) edge[<-, red, "$u_1$"] (2)
            (2) edge["$u_2$"] (3)
            (3) edge["$u_3$"] (4)
            (4) edge[->, red, "$u_4$"] (1)
            (5) edge["$u_5$"] (6)
            (6) edge[->, red, "$u_6$"] (7)
            (7) edge[<-, red, "$u_7$"] (8)
            (8) edge["$u_8$"] (5)
            (1) edge[<-, red, "$u_9$", sloped] (5)
            (2) edge["$u_{10}$", sloped] (6)
            (3) edge[->, red, "$u_{11}$", sloped] (7)
            (4) edge["$u_{12}$", sloped] (8);
        \end{tikzpicture} 
        \caption{Stable set that corresponds to an independent set with original vertices $\left\{ v_1, v_7 \right\}$ in $G$. Users and creators in the stable set are colored red, and arrows point to the creator that the user is assigned to. This matching satisfies all constraints of red players, and captures the independence condition in $G$ that vertices $v_1$ and $v_7$ are not adjacent, as no users are assigned to two creators.}
        \label{fig:reduction_regular_G_independent_set}
    \end{subfigure}
    
    \caption{Reduction of a possible input graph to a user-creator instance in Theorem~\ref{thm:hard}, and a stable set that corresponds to an independent set}
    \label{fig:reduction_regular_example}
\end{figure}

We give a sketch of the proof on regular graphs (where every vertex has the same degree). 
Suppose we want to find the maximum independent set on an arbitrary $\Delta$-regular graph $G=\left(V,E\right)$ with $n$ vertices and $m$ edges. We construct a user-creator instance with $C=n$ creators and $U=m$ users, such that each creator corresponds to a vertex and each user corresponds to an edge. In addition, we construct type vectors for users and creators in $\R^6$, and choose a corresponding user constraint value $\ebar$, such that for all $1\leq i\leq m, 1\leq j\leq n$,
\begin{align}
    u_i^T c_j &= \ebar, \quad \text{if edge $i$ is incident on vertex $j$,}  \nonumber \\ %
    u_i^T c_j &< \ebar, \quad \text{otherwise.}  \nonumber %
\end{align}
This implies that user $i$ is happy with creator $j$ if and only if they are incident in $G$, and all happy user-creator pairs have the same engagement $\ebar$. Essentially, we use type vectors in metric space to capture the structure of the graph (whether each pair of vertices is connected with an edge). Furthermore, we set $K=1$ (each user is only assigned to one creator), and $\abar=\Delta$ (each creator requires an audience size of $\Delta$). 
Figure~\ref{fig:reduction_regular_G} shows reduction of a possible input graph to a user-creator instance, with user engagement constraints corresponding to incidence of edges and vertices through the construction of type vectors.

We claim that the choice of parameters and participation constraint values, namely $K=1$ and $\abar=\Delta$, encode the independence condition in $G$: any two creators cannot coexist in a stable set if their corresponding vertices in $G$ are adjacent. The key observation is that in any stable set, each creator (vertex) must have all her $\Delta$ incident users assigned to her in order for her audience constraint of $\Delta$ to be satisfied; however, each user (edge) can only be assigned to $K=1$ creator, so her two incident creators cannot both be in the stable set. Conversely, any independent set in $G$ induces a stable set with the corresponding creators, which uniquely determines the users and recommendations in the stable set. Figure~\ref{fig:reduction_regular_G_independent_set} shows such a stable set that corresponds to an independent set in the input graph, with a recommendation that satisfies the participation constraints of all selected players.

Therefore, there exists a 1-1 correspondence between independent sets in $G$ and stable sets in our instance. We further observe that the engagement of the stable set is proportional to the size of the independent set, as each creator receives $\Delta$ users each with engagement $\ebar$. This implies the problem of finding a maximum independent set in $G$ can be solved by finding a maximum stable set in our instance, which completes the reduction and proof.

The full proof of Theorem~\ref{thm:hard} is presented in Appendix~\ref{sec:proof_MSS_hard}. In the appendix, we also show that the hardness result holds for non-regular graphs, and when restricted to a particular value of $K$ as long as $K\geq 3$.

\section{The \IE\ Algorithm}  \label{sec:uc}
Sections~\ref{sec:users_never_leave} and \ref{sec:MSS_hard} showed that while finding the \FL\ solution with two-sided constraints is computationally challenging, the case with one-sided \textit{creator} participation constraints reduces the complexity and allows efficient approximations. In this section, we consider a similar algorithm used in practice that only accounts for one-sided \textit{user} participation constraints, which we refer to as the \UC\ (UC) recommendation algorithm.

We first formally define the \IE\ algorithm as a myopic algorithm that ignores creator constraints and greedily assigns each user to her favorite creators. Many social media platforms in the real world use such a recommendation scheme to maximize immediate user engagement without considering the dynamics of content creators. We show that when both user and creator departures are taken into account, the performance of the \UC\ algorithm can be arbitrarily bad on both deterministic instances and random instances.

\subsection{Definition}  \label{sec:def_ie}
\begin{definition}[\IE]  \label{def:ie}
The \textit{\IE} (UC) recommendation algorithm, at each time step, assigns each user to the $K$ content creators that she receives greatest engagement from (or all remaining creators on the platform if there are fewer than $K$ of them). Formally, $\UCseq = \left\{ UC_0, UC_1, \dots \right\}$ is a sequence of recommendations such that
    \begin{align*}  
        UC_t(i) &= \argmax_{\creatorset\subseteq \creatorset_t, |\creatorset| \leq K} \sum_{j\in \creatorset} u_i^T c_j, \quad \text{for all } t\geq 0, \text{for all } i\in \userset_t. %
    \end{align*}
\end{definition}

\IE\ is a \textit{one-sided}, \textit{myopic} algorithm that \textit{greedily} assigns each user the best possible recommendations for her, without considering departures of creators who may not obtain sufficient viewership from these myopic assignments.\footnote{Note that according to our model, at time $t$, the platform first makes the recommendations $UC_t$, and then users and creators decide whether to leave the platform based on their utilities derived from $UC_t$. The subset of users and creators who choose to stay will then progress onto time $t+1$, when they will receive a new matching $UC_{t+1}$ from the platform. Definition~\ref{def:ie} implies that $UC_t$ is generated without considering which users and creators will leave following the recommendations.}
Note that without creator departures (i.e., $\abar = 0$), \UC\ maximizes total engagement: in this case, recommendations given to any user are independent of any other user due to the lack of creator audience constraints that bind them together (a user does not need ``help'' from other users to support her favorite creators), so maximizing total engagement reduces to separate problems of maximizing each user's individual engagement, achieved with $UC_t\left(i\right)$.

Many real-world social media platforms, such as Facebook and X (formerly Twitter), deploy recommendation algorithms that are entirely user-centric \citep{Lada2021HowDoes, Twitter2023Recommendation}. Most of the literature that studies incentives and strategic behaviors of content creators \citep{hron2023modeling, jagadeesan2023supplyside} also assumes that the platform uses the \UC\ algorithm or a stochastic variant of it.

\subsection{Performance of \IE\ on Deterministic Instances}  \label{sec:IE_approx_ratio}
We show that the \IE\ algorithm's approximation ratio in long-term engagement relative to the \FL\ is zero. In other words, there exist instances where a myopic platform achieves zero engagement in the long run, due to every user and creator leaving the platform.

\begin{theorem} \label{thm:approx_ratio}
    
    For any $n\geq 1$ and $m\geq 1$, there exists instances of initial users and creators $\left(\userset_0, \creatorset_0 \right)$ where $\left| \userset_0 \right| \geq n$, $\left| \creatorset_0 \right| \geq m$, and
    $$ \frac{\longtermengagement\left(\UCseq\right)}{\longtermengagement\left(\FLseq\right)} = 0. $$
\end{theorem}
Theorem~\ref{thm:approx_ratio} states that a platform that only focuses on users may achieve arbitrarily low long-term engagement, compared to one that considers both user and creator constraints, regardless of its size. This illustrates the importance of considering both user and creator incentives when providing recommendations.

We remark that Theorem~\ref{thm:approx_ratio} applies even with all of the following restrictions on the instance: $(\mathcal{U}_0, \mathcal{C}_0)$ is a stable set (i.e., \FL\ can retain all users and creators); $D$ is a small constant; and a ``market balance'' condition that $UK=C\abar$ holds, which we elaborate on in Section~\ref{sec:increase_users_condensed}.  %

\begin{proofCharles}[Proof of Theorem~\ref{thm:approx_ratio}]
    We present the following example where $\longtermengagement\left(\UCseq\right) = 0$ at $t=\infty$. In fact, \UC\ converges to an empty set within two time steps, due to too many creators leaving at $t=0$. 
    \begin{example}  \label{example:model_megacrown}
        Given $n$ and $m$, assume $m\geq 3$ without loss of generality. Choose $K=m-1$, $D=2$, and creator constraint value $\abar$ to be an arbitrary integer that satisfies $\abar \geq \max \left\{ n/2+1, 3 \right\}$. Consider the instance shown in Figure~\ref{fig:example_model_megacrown}, with $m$ creators and $2\abar+2$ users. Creator $c_1$ and $\abar-2$ users have type $(0,1)$, creators $c_2, \dots, c_{m-1}$ have type $\left( \cos (\pi/4), \sin (\pi/4) \right)$, user $u_1$ has type $\left( \cos (\pi/3), \sin (\pi/3) \right)$, user $u_2$ has type $\left( \cos (\pi/6), \sin (\pi/6) \right)$, and creator $c_m$ and $\abar-2$ users have type $(1,0)$. The user constraint value is $\ebar=\cos \left(\pi/3\right)$, which means $u_1$ and $u_2$ are happy with all $m$ creators but prefer creators $c_2,\dots,c_{m-1}$, while the other $2\abar-4$ users are not happy with the creator in the opposite corner.
    
        \begin{figure}[h]
            \centering
            \begin{subfigure}[c]{0.45\textwidth}
                \centering
                \includegraphics[width=0.7\textwidth]{tikz/ex_megacrown_vectors.tikz}
                \caption{User and creator types in $\mathbb{R}^2$. Solid dots are creators, and open dots are users. The points in the top left, center and bottom right of the curve are all located at $(0,1)$, $\left( \sqrt{2}/2, \sqrt{2}/2 \right)$ and $(1,0)$ respectively. Parameters are $K=m-1$ and $\ebar = \cos \left(\pi/3\right)$, so any user-creator pair is happy if and only if their types are not in opposite corners.}
                \label{fig:example_model_megacrown_vectors}
            \end{subfigure}
            \hfill%
            \begin{minipage}[c]{0.45\textwidth}
                \begin{subfigure}{\textwidth}
                    \centering
                    \includegraphics{tikz/ex_megacrown_bipartite.tikz}
                    \caption{Bipartite graph representation of Example~\ref{example:model_megacrown}}
                    \label{fig:example_model_megacrown_bipartite}
                \end{subfigure}
                \par\bigskip
                \begin{subfigure}{\textwidth}
                    \centering
                    \includegraphics{tikz/ex_megacrown_fl.tikz}
                    \caption{The \FL\ recommendation, which satisfies all user and creator constraints}
                    \label{fig:example_model_megacrown_fl}
                \end{subfigure}
            \end{minipage}
            \par\bigskip
            \begin{subfigure}{\textwidth}
                \centering
                \begin{subfigure}[t]{0.3\textwidth}
                    \centering
                    \includegraphics[width=\textwidth]{tikz/ex_megacrown_ie0.tikz}
                    \caption*{$t=0$ (dashed red lines indicate each user is only assigned to $m-2$ of the $m-1$ center creators)}
                \end{subfigure}
                \hfill
                \begin{subfigure}[t]{0.3\textwidth}
                    \centering
                    \includegraphics[width=\textwidth]{tikz/ex_megacrown_ie1.tikz}
                    \caption*{$t=1$, creators $c_1$ and $c_m$ leave due to insufficient audience size}
                \end{subfigure}
                \hfill
                \begin{subfigure}[t]{0.3\textwidth}
                    \centering
                    \includegraphics[width=\textwidth]{tikz/ex_megacrown_ie2.tikz}
                    \caption*{$t=2$, engagement $0$. All users leave due to insufficient number of recommendations.}
                \end{subfigure}
                \caption{The \IE\ recommendations, which receive zero engagement as $t\rightarrow \infty$ due to losing all users and creators}
                \label{fig:example_model_megacrown_ie}
            \end{subfigure}
            \caption{Illustration of Example~\ref{example:model_megacrown}}
            \label{fig:example_model_megacrown}
        \end{figure}

        In this instance, the \FL\ algorithm keeps all users and creators by satisfying all participation constraints, as shown in Figure~\ref{fig:example_model_megacrown_fl}. Thus, $\longtermengagement\left(\FLseq\right) > 0$. On the other hand, the \IE\ algorithm makes recommendations as shown in Figure~\ref{fig:example_model_megacrown_ie}. Creators $c_1$ and $c_m$ leave after $t=0$ due to insufficient audience (expected $\abar$ users, got $\abar-1$), so no users can receive $m-1$ recommendations at $t=1$ and they all leave afterwards. Therefore, $\longtermengagement\left(\UCseq\right) = 0$ and $\frac{\longtermengagement\left(\UCseq\right)}{\longtermengagement\left(\FLseq\right)} = 0$, which proves Theorem~\ref{thm:approx_ratio}.
    \end{example}
    
    \end{proofCharles}

\begin{remark}
    While the fact that \IE\ only retains fewer than $K$ creators at $t=1$ may appear to be a degenerate case, it turns out that such cases are not uncommon even in randomly generated instances with appropriate choices of parameters. This will be elaborated in Theorem~\ref{thm:increase users}, whose proof uses a similar underlying structure as Example~\ref{example:model_megacrown}.
\end{remark}

We also present Example~\ref{example:model_cascade} below that illustrates an alternative scenario in which \IE\ can receive arbitrarily low long-term engagement. While this example does not have $\longtermengagement\left( \UCseq \right) = 0$, it constructs a family of instances parameterized by $n$ such that \UC's approximation ratio converges to zero as $n\rightarrow \infty$.

\begin{example}  \label{example:model_cascade}
    Let $n\geq 2$ be an arbitrary integer and $\theta=\pi/6(2n-1)$. Consider the instance shown in Figure~\ref{fig:example_model_cascade} with $2n$ creators and $2n$ users. Creator types $c_1, \dots, c_{2n}$ are spaced out evenly along the arc from $(0,1)$ to $(1,0)$. For $1\leq i\leq 2n-1$, user $u_i$ is placed between $c_i$ and $c_{i+1}$, but closer to $c_i$; the last user $u_{2n}$ is between the last two creators $c_{2n-1}$ and $c_{2n}$. Notably, for any $2\leq i\leq 2n-1$, user $u_i$ is happy with creators $c_{i-1}$, $c_i$ and $c_{i+1}$, with the latter two being her two most preferred creators. However, $u_i$ is not happy with $c_{i+2}$ for any $1\leq i\leq 2n-2$.
    
    \begin{figure}[h!]
        \centering
        \begin{subfigure}[t]{0.45\textwidth}
            \centering
            \includegraphics[width=0.7\textwidth]{tikz/ex_cascade_vectors.tikz}
            \caption{User and creator types in $\mathbb{R}^2$. Solid dots are creators, and open dots are users. Parameters are $\ebar = \cos 4\theta$, $K=2$ and $\abar = 2.$}
            \label{fig:example_model_cascade_vectors}
        \end{subfigure}
        \hfill%
        \begin{subfigure}[t]{0.45\textwidth}
            \centering
            \includegraphics[width=0.6\textwidth]{tikz/ex_cascade_fl.tikz}
            \caption{The \FL\ recommendation which satisfies all user and creator constraints. Red edges are recommendations for odd-numbered users, and blue edges are for even-numbered users.}
            \label{fig:example_model_cascade_fl}
        \end{subfigure}
        \par\bigskip
        \begin{subfigure}{\textwidth}
            \centering
            \begin{subfigure}[t]{0.3\textwidth}
                \centering
                \includegraphics[width=0.8\textwidth]{tikz/ex_cascade_ie0.tikz}
                \caption*{$t=0$}
            \end{subfigure}
            \hfill
            \begin{subfigure}[t]{0.3\textwidth}
                \centering
                \includegraphics[width=0.8\textwidth]{tikz/ex_cascade_ie1.tikz}
                \caption*{$t=1$}
            \end{subfigure}
            \hfill
            \begin{subfigure}[t]{0.3\textwidth}
                \centering
                \includegraphics[width=0.8\textwidth]{tikz/ex_cascade_ie2.tikz}
                \caption*{$t=2$}
            \end{subfigure}
            \par\medskip
            \begin{subfigure}[t]{0.3\textwidth}
                \centering
                \includegraphics[width=0.8\textwidth]{tikz/ex_cascade_ie3.tikz}
                \caption*{$t=3$}
            \end{subfigure}
            \hfill
            \begin{subfigure}[t]{0.3\textwidth}
                \centering
                \includegraphics[width=0.8\textwidth]{tikz/ex_cascade_ie4.tikz}
                \caption*{$t=4$}
            \end{subfigure}
            \hfill
            \begin{subfigure}[t]{0.3\textwidth}
                \centering
                \includegraphics[width=0.8\textwidth]{tikz/ex_cascade_ie5.tikz}
                \caption*{$t\geq 2(2n-2)$}
            \end{subfigure}
            \caption{The \UC\ recommendations, which lose all but two users and creators one by one every time step}
            \label{fig:example_model_cascade_ie}
        \end{subfigure}
        \caption{Illustration of Example~\ref{example:model_cascade}}
        \label{fig:example_model_cascade}
    \end{figure}

    The \UC\ algorithm gradually loses a creator after every even time step and a user after every odd time step, until only the last two users and creators ($2n-1$ and $2n$) are left, as shown in Figure~\ref{fig:example_model_cascade_ie}. On the other hand, the \FL\ algorithm satisfies the participation constraints of all users and creators by pairing up users and creators $2i-1$ and $2i$ for each $1\leq i\leq n$, as shown in Figure~\ref{fig:example_model_cascade_fl}. Therefore, $\frac{\longtermengagement\left( \UCseq \right)}{\longtermengagement\left( \FLseq \right)} \approx \frac{2}{2n} \rightarrow 0$ as $n\rightarrow \infty$.
\end{example}

\subsection{Performance of \IE\ on Random Instances}  \label{sec:random_instances}
One may wonder whether the examples in Section~\ref{sec:IE_approx_ratio} are outliers with carefully placed user and creator types, and whether any guarantees in expectation can be achieved when they are randomly generated. We give a negative answer in this section, and show that even when user and creator type vectors are i.i.d. random variables from prior distributions, the \textit{average} performance of the \IE\ algorithm can still be upper bounded, and decays rapidly as the size of the instance grows.

Let $\distribution$ be a uniform prior distribution over $\left\{ x\in \R^D_{\geq 0}: \left\lVert x \right\rVert = 1 \right\}$, the space of non-negative unit vectors in $\R^D$. Suppose parameters such as $U$, $C$, $K$, $\ebar$ and $\abar$ are fixed, but user and creator types are generated i.i.d. from the uniform prior: $u_i \sim \distribution$ for all $1\leq i\leq U$, and $c_j \sim \distribution$ for all $1\leq j\leq C$. We then calculate the expected approximation ratio of \UC\ over \FL\ in long-term engagement, defined as
$$ \mathbb{E}_{u_i, c_j \sim \distribution} \left[ \frac{\longtermengagement\left(\UCseq\right)}{\longtermengagement\left(\FLseq\right)} \, \middle\vert \, \longtermengagement\left(\FLseq\right)>0 \right]. $$
In the next two subsections, we present two scenarios in which this expected approximation ratio can be shown to be low or converge to zero. We also discuss main takeaways from our simulations that support the two theoretical results while also generalizing them to broader setups empirically; full simulation results are in Appendix~\ref{sec:appendix_sims}.

\subsubsection{When \texorpdfstring{$C$}{C} Increases}  \label{sec:increase_creators_condensed}
\begin{proposition}  \label{prop:increase_creators}
    Fix $U$, $K$, $\ebar$, $\abar$ such that $U\geq \abar$, $\ebar<1$, and assume $u_i, c_j \sim \distribution$ i.i.d. for all $1\leq i\leq U$ and $1\leq j\leq C$. When $C\rightarrow \infty$, %
    \begin{align*}
        \lim_{C\rightarrow \infty} \mathbb{E}\left[ \frac{\longtermengagement\left(\UCseq\right)}{\longtermengagement\left(\FLseq\right)} \, \middle| \, \longtermengagement\left(\FLseq\right)>0 \right] &= 0.
    \end{align*}
\end{proposition}
Proposition~\ref{prop:increase_creators} suggests that when the creators side of the market increases to infinity but the users side is held constant (i.e., the number of users and the sizes of their recommendations stay the same), \UC's performance degrades and converges to zero, but \FL's performance is unaffected (in fact, it improves empirically).

As a proof sketch, we first show $\lim_{C\rightarrow \infty} \mathbb{E}\left[ \longtermengagement\left(\UCseq\right) \right] = 0$. This is because each creator's expected audience size converges to 0 regardless of its type vector. To see this, we note that for each user $u$ and each potential creator type vector $c$ such that $c\neq u$, as $C\rightarrow \infty$, with probability one, there will eventually be an infinite number of other creators that are closer to $u$ than $c$ is (i.e., whose types give a higher cosine similarity score). Thus, $c$ eventually falls out of $u$'s recommendations under \IE. To complete the proof, we also show $\lim_{C\rightarrow \infty} \mathbb{E}\left[ \longtermengagement\left(\FLseq\right) \right] > 0$ by noting that at least some realizations of user and creator types will give non-empty stable sets. We present the full proof of Proposition~\ref{prop:increase_creators} in Appendix~\ref{sec:proof_increase_creators}.  %

\subsubsection{When Both \texorpdfstring{$U$}{U} and \texorpdfstring{$\abar$}{a} Increase Proportionally}  \label{sec:increase_users_condensed}
Proposition~\ref{prop:increase_creators} considers an imbalanced market where creators collectively demand more audience ($C\abar$) than the platform is able to provide them via the recommendations given to all users ($UK$). However, even when the parameters satisfy a ``market balance'' condition where these two quantities are equal ($UK=C\abar$), \IE\ may still perform badly on average:

\begin{theorem}  \label{thm:increase users}
    Fix constant integers $r$, $C$ and $K$ such that $K \in \left( C/2, C \right)$, and let $U=rC$, $\abar=rK$, $D=2$, and $\ebar \in [0,1]$ be arbitrary. Assume $u_i, c_j \sim \distribution$ i.i.d. for all $1\leq i\leq U, 1\leq j\leq C$, 
    and define $\epsilon := \Pr \left( \longtermengagement\left(\FLseq\right)=0 \right)$ given by the parameters $U, C, K, \ebar$ and $\abar$. 
    Then, when $r\rightarrow \infty$, 
    \begin{align}
        &\phantom{{}={}} \lim_{r\rightarrow \infty} \mathbb{E}\left[ \frac{\longtermengagement\left(\UCseq\right)}{\longtermengagement\left(\FLseq\right)} \, \middle| \, \longtermengagement\left(\FLseq\right)>0 \right]  \nonumber \\
        &\leq  \frac{1}{1-\epsilon} \left[ \sum_{i=1}^{C-K+1} \Pr \left( \frac{X_i + X_{K+i}}{2} \geq \frac{K}{C} \text{ and } \frac{X_{i-1} + X_{K+i-1}}{2} \leq 1-\frac{K}{C} \right) \right],  \label{eqn:thm_increase_users}
    \end{align}
    where $X_1, \dots, X_C$ are sorted from $C$ i.i.d. random variables drawn from $\text{Uniform}(0,1)$, such that $X_1 \leq X_2 \leq \dots \leq X_C$, and we define $X_0 := -\infty$ and $X_{C+1} := +\infty$ deterministically. In addition, the right hand side of \eqref{eqn:thm_increase_users} evaluates to
    
    \begin{align}
        \eqref{eqn:thm_increase_users} &= \frac{1}{1-\epsilon} \left[ \sum_{i=1}^{C-K+1} \frac{C!}{\left(i-2\right)! \left(K-2\right)! \left(C-K-i+1\right)!} \int_{0}^{\min\left\{ 1-K/C, 3-4K/C \right\}} \int_{\max\left\{ a, 2K/C-1 \right\}}^{2-2K/C-a} \int_b^{2-2K/C-a} \right.  \nonumber \\
        &\phantom{{}={}} \left.  a^{i-2} \left(c-b\right)^{K-2} \left(1-\max\left\{ c, 2\frac{K}{C}-b \right\} \right)^{C-K-i+1} \ dc \ db \ da \right].  \label{eqn:thm_increase_users_bound}
    \end{align}
\end{theorem}
Table~\ref{table:bound_results} shows numerical results of the terms in square brackets in the upper bound \eqref{eqn:thm_increase_users_bound}, or equivalently its value when $\epsilon = 0$, for some values of $C$ and $K$. Simulations in Appendix~\ref{sec:appendix_sims} show that $\epsilon$ is close to 0 for most choices of parameters and constraint values, i.e., $\longtermengagement\left(\FLseq\right)>0$ holds with high probability. Therefore, the upper bound on \UC's approximation ratio is very small for most choices of $C$ and $K$ as long as $K>C/2$, especially when $K>\left\lceil \frac{C+1}{2}\right\rceil$.

\begin{table}[h!]
    \centering
        \begin{tabular}{|c|c|c||c|c|c||c|c|c|} 
         \hline
         \multicolumn{3}{|c||}{$K=\left\lceil \frac{C+1}{2}\right\rceil$} & \multicolumn{3}{c||}{$K=\left\lceil \frac{C+1}{2}\right\rceil+1$} & \multicolumn{3}{c|}{$K=C-1$}   \\
         \hline
         $C$ & $K$ & Bound & $C$ & $K$ & Bound & $C$ & $K$ & Bound  \\
         \hline
         6 & 4 & 0.0453 & 6 & 5 & $1.37\times 10^{-3}$ & 6 & 5 & $1.37\times 10^{-3}$   \\
         7 & 4 & 0.232 & 7 & 5 & $9.95\times 10^{-3}$ & 7 & 6 & $1.55\times 10^{-4}$  \\
         8 & 5 & 0.0328 & 8 & 6 & $1.95\times 10^{-3}$ & 8 & 7 & $1.53\times 10^{-5}$  \\
         9 & 5 & 0.239 & 9 & 6 & $6.61\times 10^{-3}$ & 9 & 8 & $1.32\times 10^{-6}$  \\
         10 & 6 & 0.0301 & 10 & 7 & $1.51\times 10^{-3}$ & 10 & 9 & $1.02\times 10^{-7}$  \\
         15 & 8 & 0.252 & 15 & 9 & $3.24\times 10^{-3}$ & 15 & 14 & $7.48\times 10^{-14}$  \\
         20 & 11 & 0.0403 & 20 & 12 & $3.21\times 10^{-4}$ & 20 & 19 & $1.00\times 10^{-20}$  \\
         \hline
        \end{tabular}
    
    \caption{Numerical results of upper bound \eqref{eqn:thm_increase_users_bound} in Theorem~\ref{thm:increase users} for selected values of $C$ and $K$ when $\epsilon=0$}
    \label{table:bound_results}
\end{table}

Theorem~\ref{thm:increase users} means that when $K>C/2$, the creator constraint $\abar$ satisfies the aforementioned ``market balance'' condition, and the other mild assumptions in Theorem~\ref{thm:increase users} hold, the \IE\ algorithm performs very poorly even in expectation and has a high probability of achieving zero long-term engagement. This applies even when all user and creator types are uniformly at random, and even when the first-best \FL\ solution is non-zero.

The proof relies on showing that the expected approximation ratio is upper bounded by the probability that at least $K$ creators stay on the platform after the first time step. (This shares a similar flavor as Example~\ref{example:model_megacrown}, in that we consider instances where fewer than $K$ creators stay after $t=0$.) We then divide the prior space into segments that characterize possible recommendations that \IE\ gives, and consequently show that when $K>C/2$, at most $K$ creators can stay after the first time step and they must be adjacent. We define random variables for the proportion of users in each segment, and show that when $r\rightarrow \infty$, they converge to the sizes of the segments. This allows us to express the probability of $K$ creators staying in the form of \eqref{eqn:thm_increase_users}. Full proof of Theorem~\ref{thm:increase users} is in Appendix~\ref{sec:proof_increase_users}.

\subsubsection{Takeaways from Simulations}  \label{sec:random_instances_simulation_takeaways}
While both Proposition~\ref{prop:increase_creators} and Theorem~\ref{thm:increase users} are asymptotic statements on scenarios where $U$ or $C$ approach infinity, we have run additional simulations to show empirically that their key insights often generalize to non-asymptotic setups with fewer assumptions. The simulations also suggest additional scenarios where \IE's expected approximation ratio is poor relative to \FL.

While we defer full details on the simulations to Appendix~\ref{sec:appendix_sims}, here is a list of main observations regarding the \IE\ algorithm:
\begin{itemize}
    \item \IE's performance degrades when $U$ and $\abar$ both increase proportionally, subject to the market balance condition. This generalizes Theorem~\ref{thm:increase users}, while also relaxing its requirements that $K>C/2$ and $D=2$.
    \item \IE's performance degrades when $C$ increases. This generalizes Proposition~\ref{prop:increase_creators}.
    \item \IE's performance is poor when the ratio of $K/C$ is high and the market balance condition holds.
\end{itemize}

\section{Practical Two-Sided Algorithms}  \label{sec:heuristic_algs}
We have shown that the \IE\ algorithm can perform poorly because it ignores the participation constraints of the entire creator side of the market. On the other hand, the optimal \FL\ algorithm that does consider participation constraints of both sides is NP-hard to approximate.

This section develops two practical algorithms that consider both sides of the market, but still run in polynomial time. 
The first is the \CA\ algorithm (LC), which aims to create local neighborhoods in which constraints of both creators and users are both satisfied. Under fixed dimensionality and a density assumption on users and creators, we show that \LC\ achieves a constant approximation guarantee. 
The second is the \AM\ (CR) algorithm that first computes the potential audience of each creator to incorporate user constraints, then repeatedly matches a creator with the smallest potential audience size to her maximum audience to satisfy her creator constraints.
While we do not have theoretical guarantees for the \CRshort algorithm, simulations in Section~\ref{sec:appendix_sims} suggest that it performs well in a variety of scenarios.

\subsection{The \CA\ Algorithm (LC)}  %

The \CA\ algorithm takes into account the participation constraints on both sides of the market, but only considers smaller subsets of users and creators at a time rather than the entire instance. This algorithm leverages the intuition that, when creating recommendations for a user $u$, it is more important to consider users and creators whose types are close to $u$ than participants farther away. The \CA\ algorithm thus decomposes the recommendation problem into several smaller and localized problems, which become much easier under a density assumption that we will define in Section~\ref{sec:clustering_guarantee_density}.

\subsubsection{The Algorithm}  \label{sec:clustering_alg_def}
Before we present the algorithm, we first give an alternative, geometric interpretation of the user engagement constraint: $u_i^T c_j \geq \ebar$ is equivalent to $\left\lVert u_i - c_j \right\rVert_2 \leq d$, where $d := 2 \sin \left( \frac{\cos^{-1} \ebar}{2} \right)$ is an upper bound on the distance between the types of user $i$ and creator $j$. This is due to all types being non-negative unit vectors. We call $d$ the \textit{happy distance}.

For each user $i\in \userset$, we define the \textit{neighborhood ball centered at user $i$}, denoted $B\left(u_i\right)$, to be a ball with ``effective diameter'' $d$: that is, the intersection of its boundary with the unit-vector type space $\left\{ x\in \R^D_{\geq 0}: \left\lVert x \right\rVert_2 = 1 \right\}$ is a sphere with diameter $d$.\footnote{The actual diameter of the ball is greater than $d$ due to convexity of the unit-vector type space. This is elaborated and defined more formally in Appendix~\ref{sec:proof_clustering}.} This means that any user (not necessarily $i$) whose type is in $B\left(u_i\right)$ is happy with any creator in the same ball.

Now we proceed to define the algorithm.
The \CA\ algorithm, detailed in Algorithm~\ref{alg:neighbor_apx}, scans all users and their associated neighborhood balls in an arbitrary order.
For each user $u_i$, the algorithm examines the set of all creators and users within the neighborhood ball $B(u_i)$. If the number of unassigned users in $B(u_i)$ (that had not been assigned any recommendations) is at least $\abar$, and the number of creators in this ball is at least $K$, then the algorithm forms a stable set for these users and (a subset of) the creators, by assigning all unassigned users to $K$ creators in the ball chosen at random. If there are either not enough users or not enough creators in the ball, then the algorithm skips this user's neighborhood ball and moves on to the next one. 

This algorithm leverages the fact that if the subsets of users and creators are sufficiently close (pairwise distance at most $d$), then all user participation constraints are automatically satisfied, and the platform only needs to ensure that creator constraints are satisfied and that users receive a sufficient number of recommendations.

\begin{algorithm}[h]
\caption{\CA\ algorithm} 
\label{alg:neighbor_apx}
\begin{algorithmic} 
    \State $\userset_0 = \{1, \ldots, U\} \gets$ set of users, in arbitrary order
    \State $\creatorset_0 = \{1, \ldots, C\} \gets$ set of creators
    \State $\rec:\userset_0 \mapsto P(\creatorset_0) \gets$ recommendations (initially empty, i.e., $\rec(i)=\emptyset$ for all $ i\in \userset_0$)
    \For{$i \in \userset_0$}
        \If{$\rec(i) \neq \emptyset$}  
            \State Skip user $i$   \Comment{Already assigned recommendations from another neighborhood ball}
        \Else
            \State $\userset_i \gets \left\{ j\in \userset_0: u_j \in B(u_i) \text{ and } R(j) = \emptyset \right\}$  \Comment{Unassigned users in $B(u_i)$}
            \State $\creatorset_i \gets \left\{ k\in \creatorset_0: c_k \in B(u_i) \right\}$  \Comment{Creators in $B(u_i)$}
            \If{$\left|\userset_i\right| \geq \abar$ and $\left| \creatorset_i \right| \geq K$}
                \State $\creatorset'_i \gets $ arbitrary subset of $\creatorset_i$ of size $K$
                \ForAll{$j\in \userset_i$}
                    \State $\rec(j) \gets \creatorset'_i$  \Comment{Assign all unassigned users and $K$ creators in $B(u_i)$ to one another}
                \EndFor
            \Else
                \State Skip user $i$  \Comment{$B(u_i)$ failed (user $i$ may be assigned later due to another ball)}
            \EndIf
        \EndIf
    \EndFor
    \State \Return{$\rec$}  %
\end{algorithmic}
\end{algorithm}

\subsubsection{Performance guarantee under density assumption}  \label{sec:clustering_guarantee_density}

We first note that the \CA\ algorithm runs in polynomial time, since the algorithm inspects at most $U$ neighborhood balls, and the assignment within each neighborhood ball is polynomial in the number of users and creators. Now, we show that with fixed dimensionality, the algorithm admits a constant approximation guarantee under a density assumption that we define below. 

\begin{assumption}[Density assumption]  \label{assumption:density}
   For every user $i\in \userset_0$, the neighborhood ball $B(u_i)$ contains at least $\abar$ users and at least $K$ creators.
\end{assumption}
Assumption~\ref{assumption:density} implies that for any user $i$, the platform can give a recommendation that satisfies the constraints of all users and $K$ creators in $B(u_i)$, by choosing $K$ arbitrary creators within the neighborhood ball and then assigning each user in the ball to these $K$ creators. However, the platform may not be able to do this to every ball concurrently, as some users may be in multiple neighborhood balls.

Intuitively, this assumption requires that each user is located in a ``self-sustainable'' neighborhood, which has enough users of similar types and enough creators that these users are mutually happy with. This approximates the structure of many real-world social media platforms, where users and creators may congregate into communities that can form based on their shared interests, preferences and identities. On the other hand, in regions where no users exist, the density assumption places no restrictions. This means the overall type space can be sparse, yet allow small, dense communities to scatter across the space.

When the dimensionality $D$ is fixed, the algorithm guarantees a constant factor approximation of the \FL\ algorithm in long-term engagement: %
\begin{theorem}[Approximation of \CA]  \label{thm:neighbor_apx_ratio_3D}
    There exists function $r\left( \ebar, D \right)$ which only depends on user constraint value $\ebar$ and number of dimensions $D$, such that on all deterministic instances with dimensionality $D$ that satisfy the density Assumption \ref{assumption:density}, 
    $$ \frac{\longtermengagement\left( \CAseq \right)}{\longtermengagement\left( \FLseq \right)} \geq r\left( \ebar, D \right).$$
    In addition, when $D=3$, a lower bound for the approximation ratio is given by $r\left( \ebar, 3 \right) \geq \frac{\ebar}{37}$.
\end{theorem}
Theorem~\ref{thm:neighbor_apx_ratio_3D} shows that the \CA\ algorithm achieves \textit{some} constant-factor approximation guarantee in long-term engagement compared to \FL, and gives a lower bound for $D=3$ (note that it may not be tight). 

We detail the full proof of Theorem~\ref{thm:neighbor_apx_ratio_3D} in Appendix~\ref{sec:proof_clustering}.
As a proof sketch, we first associate each unassigned user with another user whose neighborhood ball is included entirely in the final matching. This allows us to define the concept of \textit{large balls}, which extend from neighborhood balls that are chosen in the matching, so that every user is within a large ball. Thus, the approximation ratio is lower bounded by the number of users in all chosen neighborhood balls divided by the number of users in all large balls, i.e., the density of ``chosen regions'' of large balls. Finally, we upper bound the maximum density of ``unchosen regions'' in large balls that are not in the corresponding neighborhood balls, which we call \textit{rings}. 

We conjecture that for dimensions $D>3$, an analogous lower bound for Theorem \ref{thm:neighbor_apx_ratio_3D} is 
$ r\left( \ebar, D \right) \geq \frac{\ebar}{1+\kappa_D (\kappa_D + 1)}$.
The number $\kappa_D$ is the \textit{kissing number} in $\mathbb{R}^D$, a concept from geometry defined as the maximum number of non-overlapping unit spheres in $\mathbb{R}^D$ that can exist such that they all touch a common unit sphere. Note that the kissing number is conjectured to grow exponentially with $D$. Appendix~\ref{sec:proof_clustering} provides intuition for the relationship between the \CA\ algorithm and the kissing number.

\subsection{The \AM\ Algorithm (CR)}  \label{sec:AM}
The \AM\ (CR) algorithm is a polynomial-time algorithm that attempts to satisfy both user and creator participation constraints by repeatedly executing two steps. The first step computes the potential audience of each creator, i.e., the set of all users that are happy with the creator and have capacity for more recommendations. The second step chooses the creator with the smallest potential audience size (whose participation constraint is the hardest to satisfy), and attempts to satisfy her constraint by assigning maximum audience to her.
This procedure is repeated until no more creators can be added to the matching.  %

We present two versions of the \CRshort\ algorithm, CR1 and CR2. The latter is an improvement over the former due to a more efficient way to assign maximum audience to each creator, and generally achieves better performance in simulations. While most of our empirical analysis will focus on CR2 for this reason, we first introduce CR1 to offer a more intuitive understanding of the algorithm.

\subsubsection{The CR1 Algorithm}
\begin{algorithm}  
\caption{\CR\ 1 (CR1) algorithm for time $t$}
 \label{alg:am1}
\begin{algorithmic}[1]
    \State $\userset_t \gets$ current set of users, in arbitrary order
    \State $\creatorset \gets \creatorset_t$, current set of creators  \Comment{$\creatorset$ is set of all remaining creators to be examined}
    \State $\rect:\userset_t \mapsto P(\creatorset_t) \gets$ recommendations (initially empty, i.e., $\rect(i)=\emptyset$ for all $i\in \userset_t$)
    \While{$\creatorset \neq \emptyset$}
        \State $a(j) := \left\{ i \in \userset_t: u_i^T c_j \geq \ebar \text{ and } \left| \rect(i) \right| < K \right\}$, for all $j \in \creatorset$  \Comment{Potential audience for creator $j$}  \label{algline:am1_potential_audience}
        \State $j \gets $ creator in $\creatorset_t$ with smallest value of $\left| a(j) \right|$  \Comment{Smallest potential audience size}  \label{algline:am1_lowest_creator}
        \State $\creatorset \gets \creatorset \backslash \left\{j\right\}$
        \If{$\left| a(j) \right| \geq \abar$}  \Comment{Creator $j$ has enough potential audience}  \label{algline:am1_creator_audience_compare}
            \For{user $i \in a(j)$}
                \State $\rect(i) \gets \rect(i) \cup \left\{j\right\}$  \Comment{Add $j$ and assign all her potential audience to her}
            \EndFor
        \EndIf  \label{algline:am1_creator_endif}
    \EndWhile
    \State \Return{$\rect$}  \Comment{Recommendation for time $t$ (note that some users may leave)}
\end{algorithmic}
\end{algorithm}
\AM\ 1 (CR1), shown in Algorithm~\ref{alg:am1}, is the more basic of the two \CRshort\ algorithms. At the core of the algorithm is the concept of \textit{potential audience}:

\begin{definition}[Potential audience]  \label{def:potential_audience}
    Given a set of (current) recommendations $\rect$, for any creator $j\in \creatorset_t$, the \textit{potential audience} of $j$ is
    $$ a(j) := \left\{ i \in \userset_t: u_i^T c_j \geq \ebar \text{ and } \left| \rect(i) \right| < K  \right\}. $$
    In other words, it is the set of users that are happy with creator $j$ and have capacity for more recommendations (i.e., can be assigned to $j$ without violating their user constraints).
\end{definition}
The potential audience of $j$ is the maximum set of users that $j$ can receive with the current state of recommendations being $\rect$. Note that $a(j)$ is a dynamic function whose values change with $\rect$ and are re-evaluated each time it is used. 
The potential audience function is updated as the first step of each iteration on Line~\ref{algline:am1_potential_audience}, thus accounting for user constraints.

After the potential audience sets have been computed, the algorithm chooses a creator $j$ with the smallest potential audience size $\left| a(j) \right|$ from the set of remaining creators $\creatorset$ that have not been examined yet. For each such creator $j$, if her current potential audience size meets her participation constraint, add creator $j$ to the current recommendations $\rec_t$ by assigning her entire potential audience to her. If the potential audience size is insufficient, $j$ is removed from the set of creators and will not be considered again (since $j$'s constraints can no longer be satisfied without changing existing recommendations). This is the second step of each iteration (Lines~\ref{algline:am1_lowest_creator}-\ref{algline:am1_creator_endif}) which focuses on creator constraints. Both steps are repeated until all creators have been examined.

Intuitively, CR1 aims to produce a matching that prioritizes creators whose constraints are the hardest to satisfy due to a smaller potential audience. We do so under the intuition that creators with a larger potential audience can afford to lose some of these users early on, and still have enough users for their creator constraints to be satisfied in later iterations. %

However, we note that the resulting recommendations $R_t$ from CR1 may not induce a stable set, since there may exist users whose number of recommendations at the end of the algorithm is fewer than $K$. Each of these users will need to be assigned to additional creators to ensure she has $K$ recommendations, but she may not be happy with all such creators. Therefore, after time step $t$, some users may leave. The algorithm will need to be executed again at time $t+1$ and all future times, until a stable set is reached. %

\subsubsection{The CR2 Algorithm}  \label{sec:am2_high_level}
While CR1 is intuitive and simple to implement, one drawback is that it often assigns an unnecessarily large number of users to creators who are examined earlier in the algorithm, and is unable to reassign them to other creators in later iterations who may need these users to meet their audience constraints. 

To get around this issue, we also define an \AM\ 2 (CR2) algorithm that proceeds similarly, but makes an improvement to Lines~\ref{algline:am1_creator_audience_compare}-\ref{algline:am1_creator_endif} which assign maximum audience to creator $j$. In addition to her potential audience set $a\left(j\right)$ (containing users with spare capacity), CR2 assigns additional users not in $a\left(j\right)$ to creator $j$ by changing one of their recommendations from another creator to $j$. This uses a concept that we refer to as \textit{augmenting path}, which allows us to reassign previously made recommendations without violating the constraints of any users or earlier creators. This algorithm is described in detail in Appendix~\ref{sec:appendix_am2}.

\begin{figure}[h]
    \centering
    \includegraphics[width=\textwidth]{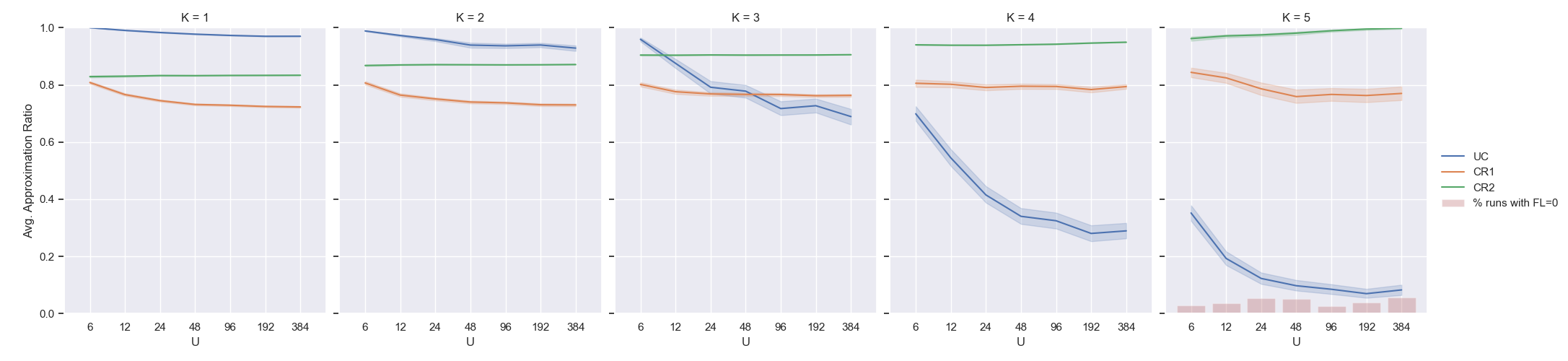}
    \caption[Average approximation ratios of UC, CR1 and CR2 when $U$ and $\abar$ increase proportionally]{Approximation ratios of UC, CR1 and CR2 relative to FL. Across subplots, $K$ increases from $1$ (left) to $5$ (right). Within each subplot, $U$ increases exponentially on the $x$-axis. In all plots, $C=6$, $\abar = UK/C$, $D=10$ and $\ebar$ is held constant. UC, CR1 and CR2 are in blue, red and green respectively, with 95\% confidence intervals.}
    \label{fig:sims_conj1_maintext_am2}
\end{figure}

We note that neither CR1 nor CR2 are guaranteed to return a stable set, due to users possibly receiving an insufficient number of recommendations. Regardless, simulations show that CR2 in particular performs well in many settings, especially when \IE\ performs poorly on long-term engagement as highlighted in Section~\ref{sec:random_instances_simulation_takeaways}. One such case is presented in Figure~\ref{fig:sims_conj1_maintext_am2}, which shows that \IE's performance degrades when either $K$ becomes large or both $U$ and $\abar$ become large, but CR2 still performs well in these cases. More simulation results are in Appendix~\ref{sec:appendix_sims}. 

\section{Conclusion and Future Work}  \label{sec:conclusion}
In this paper, we introduce a stylized model of content recommendations in two-sided markets, where we consider dynamic departures of both users and content creators due to insufficient utilities. We show that the platform's objective of maximizing long-term engagement (the \FL\ algorithm) is computationally hard to solve for or approximate. While the problem becomes more tractable in special cases with only one-sided departures of either users or creators (but not both), applying these algorithms such as the \UC\ algorithm to the more general model with two-sided departures fails to produce any performance guarantees. Finally, we present the \CA\ (LC) and \AM\ (CR) algorithms that account for incentives from both sides of the platform.

Below is a list of natural extensions of our work and future directions we hope to pursue.

\begin{itemize}
    \item Users and creators may have more general participation constraints than our model assumes. For example, users may accept a range of possible numbers of recommendations instead of a single integer $K$, creators may have an upper bound on attention (especially on e-commerce platforms like Etsy where creators may be overwhelmed with high demand), and feasible pairings of users and creators may also be determined by other factors beyond cosine similarity. While these settings are likely still NP-hard to approximate in general, it would be interesting to characterize scenarios in which the optimal matching can be found exactly or approximately.
    \item In the integer linear program presented in Section~\ref{sec:stable_set}, the constraints can be represented as a matrix whose rank is arbitrary. A special case to consider is when the rank of the matrix is low, which may happen if users can be categorized into a small number of groups (or ``profiles''), such that all users in each group are happy with the same sets of creators. This assumption may give rise to polynomial-time approximation schemes.
    \item The \CA\ algorithm requires a density assumption to produce theoretical guarantees. We may be able to improve the algorithm in cases where the density assumption does not hold, in order to either obtain approximations on these more general instances or achieve better empirical performance.
    \item While the CR2 algorithm performs well in simulations, it remains an open question whether the algorithm can be shown to guarantee a constant-factor approximation, or whether its empirical performance has an underlying intuition that is grounded in the optimization literature.
    \item Beyond making decisions on departures, there are other ways in which users and creators can respond to platform recommendations. Creators may change the characteristics of contents that they produce with hopes of receiving a bigger audience, user preferences may shift under the influence of the recommended content, and users may also vary the intensity at which they interact with the content and its creators. These may give rise to a model with a richer set of dynamics and behavior.
    \item We set aside central learning aspects by assuming the platform knows the user and creator types. An interesting future direction would be to consider platforms learning the types and constraints of users and creators through their responses to recommended content.
\end{itemize}

\clearpage
\appendix
\section{Additional Notes on the Model}  \label{sec:appendix_model}
In this section, we explain several design decisions behind our model, and show that most of them either hold without loss of generality or do not affect our main results.

\subsection{Unit Vector Types}  \label{sec:unit_vector_wlog}
We assume all user types $u_i$ and creator types $c_j$ are unit vectors: $\left\lVert u_i \right\rVert = \left\lVert c_j \right\rVert = 1$, for all $i\in \userset_0, j\in \creatorset_0$. This holds without loss of generality:

\begin{proposition}[Unit vector holds without loss of generality]  \label{prop:unit_vector_wlog}
    For any instance of users and creators $\left( \userset_0, \creatorset_0 \right)$ with dimensionality $D$ whose player types are not unit vectors (but still non-negative), another instance $\left( \userset'_0, \creatorset'_0 \right)$ with dimensionality $D+2$ and unit vector types can be constructed in polynomial time, such that all user and creator constraints are equivalent between the two instances, and the total engagement of the new instance is that of the original instance multiplied by a constant factor.
\end{proposition}
Proposition~\ref{prop:unit_vector_wlog} implies that any recommendation algorithm on the new instance has the same evolution over time steps as the corresponding recommendation on the old instance, and their engagements only differ by a scalar multiple. This means the two instances are effectively equivalent up to a constant factor.

\begin{proofCharles}
    We construct the new instance $\left( \userset'_0, \creatorset'_0 \right)$. Use $u_i$ and $c_j$ to denote original user and creator type vectors, and $u'_i$ and $c'_j$ to denote the new types.

    Let $l_u=\max_i \left\lVert u_i \right\rVert$ be the maximum $L_2$ norm of all user types, and $l_c=\max_j \left\lVert c_j \right\rVert$ be the maximum $L_2$ norm of all creator types. Define
    \begin{align*}
        u'_i &= \left[ \frac{\left( u_i \right)_1}{l_u}, \dots, \frac{\left( u_i \right)_D}{l_u}, \frac{\sqrt{l_u^2 - \left\lVert u_i \right\rVert^2}}{l_u}, 0 \right]^T, \quad \forall 1\leq i\leq U,  \\
        c'_j &= \left[ \frac{\left( c_j \right)_1}{l_c}, \dots, \frac{\left( c_j \right)_D}{l_c}, 0, \frac{\sqrt{l_c^2 - \left\lVert c_i \right\rVert^2}}{l_u} \right]^T, \quad \forall 1\leq j\leq C.
    \end{align*}
    In other words, the first $D$ dimensions of $u'_i$ and $c'_j$ are those of $u_i$ and $c_j$ scaled by $1/l_u$ and $1/l_c$ respectively; $u'_i$ uses the $(D+1)$-th dimension as a balancing term to make it a unit vector, and $c'_j$ similarly uses the $(D+2)$-th dimension to do so. Additionally, define the new user constraint value $\ebar'$ to be
    $$ \ebar' = \frac{\ebar}{l_u l_c}, \quad \forall 1\leq i\leq U, $$
    where $\ebar$ is the user constraint value in the original instance, and define other parameters ($K$, $\abar$) to be equal to the original instance.

    It's easy to see that both $u'_i$ and $c'_j$ are unit vectors. Additionally, for any $i,j$, 
    \begin{align}
        u_i^T c_j \leq \ebar  
        &\iff \left( \frac{u_i}{l_u} \right)^T \frac{c_j}{l_c} \leq \frac{\ebar}{l_u l_c}  \nonumber \\
        &\iff {u'}_i^T c'_j \leq \ebar',  \label{eqn:unit_vector_wlog_dotprod}
    \end{align}
    where \eqref{eqn:unit_vector_wlog_dotprod} uses the fact that ${u'}_i^T c'_j$ only possibly gets non-zero values from the first $D$ dimensions, which are simply scaled from $u_i$ and $c_j$. This shows that participation constraints in the two instances are equivalent.

    Finally, ${u'}_i^T c'_j = \frac{u_i^T c_j}{l_u l_c}$, which means each individual user engagement between user $i$ and creator $j$ in the new instance is a scalar multiple $1/l_u l_c$ of the original instance. Therefore, total engagement in the new instance under a particular set of recommendations is also $1/l_u l_c$ of the original instance.
\end{proofCharles}

\subsection{Homogeneous Creator Constraints}  \label{sec:homogeneous_creators_wlog}
We assume that all creators require the same audience size $\abar$ to stay on the platform. We can show that this assumption is without loss of generality: given any instance with dimensionality $D$ where creator $j$ requires audience size $\abar_j$, we can construct another instance with dimensionality $D+3$ where all creators have the same constraint $\abar'$, such that all user and creator constraints are equivalent between the two instances, and the total engagement of the new instance is that of the original instance multiplied by a constant factor.

As a proof sketch, for each creator $j$, replace her audience constraint $\abar_j$ with a homogeneous constraint $\abar' := \max_k \abar_k$, and create $\left(\max_k \abar_k\right)-\abar_j$ auxiliary users whose types are configured such that they get maximum engagement from $j$ among all creators and the engagement is exactly $\ebar$. This can be done by setting the user type's first $D$ dimensions to be $\ebar c_j$, then adding a new, $(D+1)$-th dimension to make it a unit vector. Such a construction ensures this auxiliary user is only happy with creator $j$ among the original creators. 

The remaining step is the find $K-1$ additional creators to assign the auxiliary users to. To do so, we scale all user types (original and auxiliary) by some constant $\lambda$ and add two new dimensions, so that the first $D+1$ dimensions of all users are $\lambda$ of their original value, the $(D+2)$-th dimension of all original users is $\sqrt{1-\lambda^2}$, and the $(D+3)$-th dimension of all auxiliary users is $\sqrt{1-\lambda^2}$. Set all user constraints to $\lambda \ebar$ to account for the scaling. We then add $K-1$ auxiliary creators whose types are all $(0,0,\dots,0,1)$ with the $(D+3)$-th dimension being $1$. Choose $\lambda$ such that $\sqrt{1-\lambda^2}\geq \lambda \ebar$, so that each auxiliary user is happy with all $K-1$ auxiliary creators and gets higher engagement than any original creators, while the original users do not have such an option.

Finally, we need to ensure the auxiliary creators have sufficient audience. If at least $\abar'$ auxiliary users were created, this is already done; if not, raise $\abar'$ by 1, which then gives additional auxiliary users for the auxiliary creators while only raising their constraint by 1. Repeat this until there are at least $\abar'$ auxiliary users.

\subsection{Arrival of Users}  \label{sec:user_arrivals_appendix}
Our model assumes that the sets of all possible users and creators are determined at the start. Given our focus on departures, it is natural to also consider arrivals of new players on the platform, and in particular, whether they can offset the departures that the \IE\ algorithm experiences.

In this section, we give a negative answer to this question when new users arrive at a sufficiently slow rate. We show this using two examples, which are extensions of the counter-examples we used to prove \IE's zero approximation ratio in Theorem~\ref{thm:approx_ratio}.

\subsubsection{When Users and Creators Leave Rapidly: Example~\ref{example:model_megacrown}}

Recall that in Example~\ref{example:model_megacrown}, which we presented in the proof of Theorem~\ref{thm:approx_ratio}, there are $m$ creators and $2\abar-2$ users each being given $K=m-1$ recommendations. Each creator requires $\abar$ users, but at $t=0$, the \IE\ algorithm only assigns $\abar-1$ users to creators $c_1$ and $c_m$. These two creators leave after the first time step, and subsequently all users leave after $t=1$ due to insufficient quantity of recommendations. Therefore, in this example, any user arrivals that take place after the conclusion of the first time step $t=0$ will not help \IE\ avoid having zero engagement in the long run, as they do not increase the audience size of the two departing creators.

One might wonder what if user arrivals take place before the first time step. However, an adversary can adjust the initial sets of users and their type vectors accordingly. In fact, the proof of Theorem~\ref{thm:increase users} generalizes Example~\ref{example:model_megacrown} and shows that even when all user types are drawn from a uniform distribution, \IE\ is still very likely to lose too many creators in the first time step when $K>C/2$, so additional random user arrivals will also not help.

\subsubsection{When Cascades of User and Creator Arrivals Happen Slowly: Generalization of Example~\ref{example:model_cascade}}
Recall that Example~\ref{example:model_cascade} shows a cascade of departures under \IE, where either one user or one creator leaves the platform at every time step until an arbitrary small set of them remain. On the other hand, \FL\ retains all users and creators.

\begin{figure}[h!]
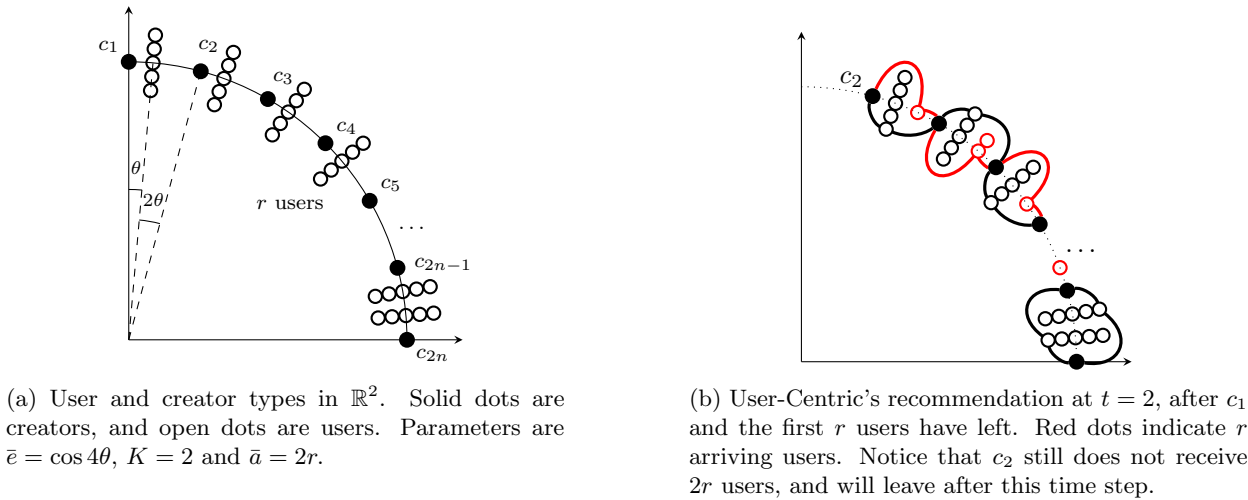

    \centering
    \begin{subfigure}[t]{0.45\textwidth}
        \centering
        \includegraphics[width=0.7\textwidth]{tikz/arr_cascadeR_vectors.tikz}
        \caption{User and creator types in $\mathbb{R}^2$. Solid dots are creators, and open dots are users. Parameters are $\ebar = \cos 4\theta$, $K=2$ and $\abar = 2r.$}
        \label{fig:arrival_cascadeR_vectors}
    \end{subfigure}
    \hfill%
    \begin{subfigure}[t]{0.45\textwidth}
        \centering
        \includegraphics[width=0.6\textwidth]{tikz/arr_cascadeR_ie.tikz}
        \caption{\IE's recommendation at $t=2$, after $c_1$ and the first $r$ users have left. Red dots indicate $r$ arriving users. Notice that $c_2$ still does not receive $2r$ users, and will leave after this time step.}
        \label{fig:arrival_cascadeR_ie}
    \end{subfigure}
    \caption{Illustration of a generalization of Example~\ref{example:model_cascade}}
    \label{fig:arrival_cascadeR}
\end{figure}

Consider a modification of the example as shown in Figure~\ref{fig:arrival_cascadeR}. We fix an arbitrary integer $r$, replace each individual user in Example~\ref{example:model_cascade} with $r$ identical users of the same type, and multiply the audience constraint $\abar$ by $r$. In addition, we allow $r$ users to arrive on the platform over some time horizon, whose types are drawn i.i.d. from a uniform distribution.

Notice that there are $2n-1$ intervals between each pair of adjacent creators, and \IE\ matches users within each interval to the two creators at its endpoints. Unless all $r$ arriving users fall within two consecutive intervals (which has low probability), users in each interval will not contribute enough audience to the creators at endpoints (such as $c_2$ in Figure~\ref{fig:arrival_cascadeR_ie}) to prevent them from leaving, after all previous users and creators have left. In these cases, \IE\ still only retains a $1/n$ fraction of users and creators as $t\rightarrow \infty$, even if the new users arrive as early as possible.

We remark that the result still holds when more than $r$ users arrive, as long as each creator does not receive $r$ arriving users (which include all users in the interval after it, and users in the last $1/3$ of the interval before it). When the distribution of user types is uniform, this still has low probability.

\section{Proofs on Maximum Stable Set and Its Hardness of Computation}  \label{sec:proof_MSS_hard}
In this section, we first prove Proposition~\ref{prop:fl_mss}, which formally establishes the relationship between the maximum stable set problem and the \FL\ algorithm. We then devote most of this appendix to the proof of Theorem~\ref{thm:hard}, which establishes the NP-hardness of the maximum stable set problem and its constant factor approximation, as well as proofs of additional results, particularly on instances where $K$ is fixed (Corollary~\ref{cor:hard_fixed_K}).

\subsection{Relationship between \FL\ and Maximum Stable Sets}  \label{sec:fl_mss_proof}
We present the full statement of Proposition~\ref{prop:fl_mss} with additional claims:
\begin{customTheorem}{Proposition~\ref{prop:fl_mss}}[With additional claims]
    Let $\FLseq=\left\{ FL_0, FL_1, \dots \right\}$ be any sequence of recommendations given by the \FL\ algorithm, $\left(\userset^*, \creatorset^* \right)$ be any maximum stable set with recommendations $\rec^*$ (that may not relate to $\FLseq$), and $\mathbf{ALG}=\left\{ ALG_0, ALG_1, \dots \right\}$ be an infinite sequence of recommendations such that $ALG_t = \rec^*$ at all times $t$. Then: 
    \begin{enumerate}
        \item There exists $t_0\in \mathbb{N}_0$ such that for all $t \geq t_0$ under $\FLseq$, $\userset_t = \userset_{t_0}, \creatorset_t = \creatorset_{t_0}$, and $\left( \userset_{t_0}, \creatorset_{t_0} \right)$ is a maximum stable set. (Note that the maximum stable set is not necessarily associated with $\rect$, and not necessarily the same as $\left(\userset^*, \creatorset^* \right)$.)  \label{enum:fl_mss_converge}
        \item $\longtermengagement\left( \FLseq \right) = \engagement\left(\userset^*, \creatorset^*, \rec^* \right)$.   \label{enum:fl_mss_engagement}
        \item $\longtermengagement\left( \FLseq \right) = \longtermengagement\left( \mathbf{ALG} \right)$.  \label{enum:fl_mss_achievable}
    \end{enumerate}
\end{customTheorem}
Proposition~\ref{prop:fl_mss} means that under any sequence of recommendations $\FLseq$ that the \FL\ algorithm may give, the sets of users and creators converge to a maximum stable set (Statement~\ref{enum:fl_mss_converge}), and its long-term engagement equals to the total engagement derived from any maximum stable set (Statement~\ref{enum:fl_mss_engagement}). In particular, we can construct one such sequence of recommendations by assigning the maximum stable set's recommendations at each time step (Statement~\ref{enum:fl_mss_achievable}). In other words, $\mathbf{ALG}$ is one of the sequences of recommendations that the \FL\ algorithm may give.

\begin{proofCharles}
    Let $\engagement^* := \engagement\left(\userset^*, \creatorset^*, \rec^* \right)$ be the engagement of the maximum stable set $\left(\userset^*, \creatorset^* \right)$, and by definition, it also equals to the engagement of any other maximum stable set. Since $ALG_t = \rec^*$ at all times $t$, this means $\mathbf{ALG}$ achieves engagement $\engagement^*$ at each time step, so $\longtermengagement\left(\mathbf{ALG}\right) = \engagement^*$. 
    
    On the other hand, 
    due to convergence of all sequences of recommendations, there exists $t_0 \in \mathbb{N}_0$ such that for all $t \geq t_0$ under $\FLseq$, $\userset_t = \userset_{t_0}, \creatorset_t = \creatorset_{t_0}$, and $\left( \userset_{t}, \creatorset_{t} \right)$ is a stable set with recommendations $\rect$. 
    
    Let $\rec^\dag$ be the recommendation between $\userset_{t_0}$ and $\creatorset_{t_0}$ that maximizes engagement while satisfying all participation constraints of $\userset_{t_0}$ and $\creatorset_{t_0}$. 
    Thus, for all $t\geq t_0$, $\engagement \left( \userset_{t_0}, \creatorset_{t_0}, \rec_{t} \right) \leq \engagement \left( \userset_{t_0}, \creatorset_{t_0}, \rec^\dag \right)$, since $\rec_{t}$ also satisfies all participation constraints of $\userset_{t_0}$ and $\creatorset_{t_0}$. On the other hand, since $\left( \userset_{t_0}, \creatorset_{t_0} \right)$ is a stable set with recommendations $\rec^\dag$, its engagement at most that of any maximum stable set, so $\engagement \left( \userset_{t_0}, \creatorset_{t_0}, \rec^\dag \right) \leq \engagement^*$.
    
    The long-term engagement of $\FLseq$ satisfies
    \begin{align}
        \longtermengagement\left( \FLseq \right) &= \lim_{T\rightarrow \infty} \frac{1}{T} \sum_{t=0}^{T-1} \engagement \left(\userset_{t}, \creatorset_{t}, \rec_{t} \right)  \nonumber \\
        &= \lim_{\substack{T\rightarrow \infty \\ T\geq t_0}} \frac{1}{T} \left( \engagement_{0}^{t_0-1} + \sum_{t=t_0}^{T-1} \engagement \left(\userset_{t_0}, \creatorset_{t_0}, \rec_{t} \right) \right), \quad \text{where } \engagement_{0}^{t_0-1} := \sum_{t=0}^{t_0-1} \engagement \left(\userset_{t}, \creatorset_{t}, \rec_{t} \right)  \nonumber \\
        &\leq \lim_{\substack{T\rightarrow \infty \\ T\geq t_0}} \frac{1}{T} \left( \engagement_{0}^{t_0-1} + \sum_{t=t_0}^{T-1} \engagement \left( \userset_{t_0}, \creatorset_{t_0}, \rec^\dag \right) \right)  \nonumber \\
        &= \lim_{\substack{T\rightarrow \infty \\ T\geq t_0}} \left( \frac{1}{T} \engagement_{0}^{t_0-1} + \frac{T-t_0}{T} \engagement \left( \userset_{t_0}, \creatorset_{t_0}, \rec^\dag \right) \right)  \nonumber \\
        &= \engagement \left( \userset_{t_0}, \creatorset_{t_0}, \rec^\dag \right)  \label{eqn:converge_all_remove_limit} \\
        &\leq \engagement^* = \longtermengagement\left( \mathbf{ALG} \right),  \label{eqn:converge_all_ineq}
    \end{align}
    where \eqref{eqn:converge_all_remove_limit} uses the fact that $\engagement_{0}^{t_0-1}$ is finite (since $t_0$ is finite and the engagement at any time step is finite) and does not depend on $T$.
    
    On the other hand, by maximality of the \FL\ algorithm, $\longtermengagement\left( \FLseq \right) \geq \longtermengagement\left( \mathbf{ALG} \right)$. So $\longtermengagement\left( \FLseq \right)= \longtermengagement\left( \mathbf{ALG} \right)$ which is Statement~\ref{enum:fl_mss_achievable}, and $\longtermengagement\left( \FLseq \right) = \engagement^*$ which is Statement~\ref{enum:fl_mss_engagement}. In particular, \eqref{eqn:converge_all_ineq} holds with equality, which means $\left( \userset_{t_0}, \creatorset_{t_0} \right)$ is a maximum stable set, thus shows Statement~\ref{enum:fl_mss_converge}.
\end{proofCharles}

\subsection{Proof of Theorem~\ref{thm:hard}}
Recall Theorem~\ref{thm:hard} from Section~\ref{sec:MSS_hard}:
\begin{customTheorem}{Theorem~\ref{thm:hard}}[Inapproximability of the general model]
    Given any $D\geq 6$ that does not depend on $U$ and $C$, finding a maximum stable set for an arbitrary instance with dimensionality $D$ is NP-hard. In addition, for any constant $c\geq 1$, finding a stable set whose total engagement is a $c$-approximation of the maximum stable set is NP-hard.
\end{customTheorem}
We will prove Theorem~\ref{thm:hard} using a reduction of the \MIS\ (MIS) problem, both on regular graphs and general graphs:

\begin{proposition}[MIS reduction on regular graphs]  \label{prop:hard_regular}
    For any positive integer $\Delta$, the \MIS\ problem on $\Delta$-regular simple graphs (i.e., graphs with every vertex having degree $\Delta$) can be reduced to the \MSS\ problem with $D=6$ in polynomial time.
\end{proposition}

\begin{corollary}[MIS reduction on general graphs]  \label{cor:hard_general}
    The \MIS\ problem on any simple graph can be reduced to the \MSS\ problem with $D=6$ in polynomial time. %
\end{corollary}

Proposition~\ref{prop:hard_regular} is strictly weaker than Corollary~\ref{cor:hard_general}, but the proof of the former is more illustrative of our techniques and builds the foundations for the latter. Both statements show the hardness of an exact solution in Theorem~\ref{thm:hard}, as the \MIS\ problem is NP-hard on $\Delta$-regular graphs for any $\Delta \geq 3$ \citep{Fricke1998Independence}, while Corollary~\ref{cor:hard_general} additionally shows the hardness of constant-factor approximation \citep{Zuckerman2007Linear}.

In the proofs below, we use ``MSS instance'' to refer to an instance of our model with users and creators, or our \MSS\ problem defined in Section \ref{sec:stable_set}; and ``MIS instance'' to refer to an instance of the \MIS\ problem with a graph. 

Note that Theroem~\ref{thm:hard} has no additional limitations on the MSS instance other than $D$, so it suffices to show that finding a maximum stable set on a class of MSS instances with constant $D$ is NP-hard. In the proof, we construct such a class with $K=1$ and $D=6$.

\subsubsection{Constructing Type Vectors and User Constraints}  \label{sec:appendix_reduction_types}
Given a graph, we first construct the type vectors of all users and creators as well as constraint values of all users, such that the happiness structure of user-creator pairs (whether each user is happy with each creator) captures the structure of the graph. This procedure applies to both regular graphs and general graphs.

\begin{lemma}[Type vectors and user constraints]  \label{lemma:reduction_types}
    For any simple graph $G=(V,E)$ with vertices $V=\left\{ v_1,\dots,v_n \right\}$ and edges $E=\left\{ e_1,\dots,e_m \right\}$, the type vectors of $m$ users and $n$ creators as well as user constraint $\ebar$ can be constructed in polynomial time, such that all types are unit vectors in $D=6$ dimensions, and user $i$ is happy with creator $j$ if and only if vertex $v_j$ is incident on edge $e_i$. In addition, for all such $i$ and $j$, the engagement that user $i$ gets from creator $j$ is $\ebar$.
\end{lemma}

\begin{proofCharles}
    We first provide an intuition for what the type vectors should achieve, and gradually work towards constructing their concrete values to satisfy all requirements in Lemma~\ref{lemma:reduction_types}.

    Consider $m$ users (edges) and $n$ creators (vertices). We need user types $u_1, \dots, u_m$, creator types $c_1, \dots, c_n$ and a constant $A$ such that
    \begin{align}
        \left( u_i^T c_j \geq A \right) &\iff \text{$e_i$ is incident on $v_j$}, \quad \text{for all } 1\leq i\leq m, 1\leq j\leq n.  \label{eqn:reduction_types_condition}
    \end{align}
    This allows us to set all user constraints as $\ebar = A$ to achieve the requirements in Lemma~\ref{lemma:reduction_types}.
    
    For each edge $e_i$, denote its two incident vertices as $v_p$ and $v_q$. Suppose we can find integer $D$, non-negative real value $A$ and functions $f: \left\{ 1,\dots,n \right\}^3 \mapsto \mathbb{R}$, $g: \left\{ 1,\dots,n \right\}^2 \mapsto \mathbb{R}^D$ and $h: \left\{ 1,\dots,n \right\} \mapsto \mathbb{R}^D$, that have the following properties:
    \begin{enumerate}[align=left, label={Property \arabic*.},ref=\arabic*]
        \item $f(p, q, j) = g\left(p,q\right)^T h\left(j\right)$, for all $1\leq p,q,j\leq n$;  \label{enum:reduction_fgh_decompose}
        \item $\argmax_j f\left(p,q,j\right) = \left\{ p, q \right\}$, for all $1\leq p,q\leq n$;  \label{enum:reduction_fgh_argmax}
        \item $\max_j f\left(p,q,j\right) = A$, for all $1\leq p,q\leq n$; \label{enum:reduction_fgh_max}
        \item  $g\left(p,q\right)\geq \mathbf{0}$ and $h\left(j \right) \geq \mathbf{0}$, for all $1\leq p,q,j\leq n$;  \label{enum:reduction_fgh_nonneg}
        \item  $g\left(p,q\right)$ and $h\left(j \right)$ are unit vectors, for all $1\leq p,q,j\leq n$,  \label{enum:reduction_fgh_unitvec}
    \end{enumerate}
    then \eqref{eqn:reduction_types_condition} can be satisfied by setting $u_i = g\left( p,q \right)$ and $c_j = h\left(j\right)$ for all $i,j$, which would imply $u_i^T c_j \geq A$ iff $f\left(p,q,j\right) \geq A$ (due to Property~\ref{enum:reduction_fgh_decompose}), which occurs iff $j\in \left\{ p, q \right\}$ (due to Properties~\ref{enum:reduction_fgh_argmax} and \ref{enum:reduction_fgh_max}). Properties~\ref{enum:reduction_fgh_nonneg} and \ref{enum:reduction_fgh_unitvec} ensure that $u_i$ and $c_i$ are valid type vectors.

    In the rest of the proof, we will construct $f$, $g$ and $h$ by satisfying all five properties one by one. Consider function $f'\left(p,q,j\right) := - \left( j - p \right)^2 \left( j - q \right)^2$, which satisfies Property~\ref{enum:reduction_fgh_argmax}. Note that
    \begin{align*}
        f'\left(p,q,j\right) &= - \left( j-p \right)^2 \left( j-q \right)^2  \\
        &= -j^4 + \left( p+q \right) \left( 2j^3 \right) + \left( p^2 + 4pq + q^2 \right) \left( -j^2 \right) + \left( p^2 q + q^2 p \right) \left( 2j \right) + \left( p^2 q^2 \right) \left(-1\right)  \\
        &= \left[ 1, p+q, p^2 + 4pq + q^2, p^2 q + q^2 p, p^2 q^2 \right]^T \left[ -j^4, 2j^3, -j^2, 2j, -1 \right],
    \end{align*}
    so we can set $g'\left( p, q \right) := \left[ 1, p+q, p^2 + 4pq + q^2, p^2 q + q^2 p, p^2 q^2 \right]$ and $h'\left(j\right) := \left[ -j^4, 2j^3, -j^2, 2j, -1 \right]$, such that $f'$, $g'$ and $h'$ satisfy Property~\ref{enum:reduction_fgh_decompose}. However, $h'$ violates Property~\ref{enum:reduction_fgh_nonneg} due to negative entries $-j^4$, $-j^2$ and $-1$. To satisfy this property, we add constant terms $n^4$, $n^2$ and $1$ respectively. Namely, for all $p, q, j$, define:
    \begin{align*}
        h''\left(j\right) &:= \left[ n^4 -j^4, 2j^3, n^2 -j^2, 2j, 0 \right],  \\ 
        g''\left(p,q\right) &:= g'\left(p,q\right), \quad f''\left(p,q,j\right) := g''\left(p,q\right)^T h''\left(j\right).
    \end{align*}
    These functions satisfy Property~\ref{enum:reduction_fgh_nonneg} while maintaining Property~\ref{enum:reduction_fgh_argmax}. To see this, note that $f''\left(p,q,j\right) = f'\left(p,q,j\right) + n^4 + n^2\left( p^2 + 4pq + q^2 \right) + p^2 q^2 $, so $\argmax_j f''\left(p,q,j\right)$ is unaffected by the additional terms which do not depend on $j$.

    To satisfy Property~\ref{enum:reduction_fgh_max}, note that $\max_j f''\left(p,q,j\right) = n^4 + n^2\left( p^2 + 4pq + q^2 \right) + p^2 q^2 > 0$. Thus, we define
    \begin{align*}
        g'''\left(p,q\right) &:= \frac{g''\left(p,q\right)}{n^4 + n^2\left( p^2 + 4pq + q^2 \right) + p^2 q^2},  \\ 
        h'''\left(j\right) &:= h''\left(j\right), \quad f'''\left(p,q,j\right) := g'''\left(p,q\right)^T h'''\left(j\right),  \\
        A' &:= 1.
    \end{align*}
    This makes $\max_j f'''\left(p,q,j\right) = A'$ for all $p,q$, so $f'''$, $g'''$, $h'''$ and $A'$ satisfy Property~\ref{enum:reduction_fgh_max}. Finally, we satisfy Property~\ref{enum:reduction_fgh_unitvec} by normalizing all function values using an additional dimension. Define constants
    \begin{align}
        G &:= \max \left\{ 1, \max_{1\leq p,q\leq n} \left\lVert g'''\left(p,q\right) \right\rVert \right\},  \label{eqn:reduction_types_G} \\
        H &:= \max \left\{ 1, \max_{1\leq j\leq n} \left\lVert h'''\left(j\right) \right\rVert \right\}.  \nonumber
    \end{align}
    Then, for all $p,q,j$, set
    \begin{align*}
        g\left(p,q\right) &:= \left[ \frac{g'''_1\left(p,q\right)}{G}, \dots, \frac{g'''_4\left(p,q\right)}{G}, \sqrt{1 - \frac{\left\lVert g'''_{1\text{-}4}\left(p,q\right) \right\rVert}{G^2}}, 0 \right],  \\
        h\left(j\right) &:= \left[ \frac{h'''_1\left(j\right)}{H}, \dots, \frac{h'''_4\left(j\right)}{H}, 0, \sqrt{1 - \frac{\left\lVert h'''\left(j\right) \right\rVert}{H^2}} \right],  \\
        f\left(p,q,j\right) &:= g\left(p,q\right)^T h\left(j\right),  \\
        A &:= \frac{1}{GH},
    \end{align*}
    where $g'''_{1\text{-}4}\left(p,q\right)$ denotes the first four components of the vector $g'''\left(p,q\right)$. This ensures $g\left(p,q\right)$ and $h\left(j\right)$ are non-negative unit vectors, and scales their dot product by $1/GH$. (Note that $h'''_5\left( j \right) = 0$, so $g'''\left(p,q\right)^T h'''\left(j\right)$ equals to the dot product of their first four components, which are preserved in $g$ and $h$ up to a constant scaling.) Therefore, $f$, $g$, $h$ and $A$ satisfy all five properties and thus \eqref{eqn:reduction_types_condition}. This completes the construction of user and creator types, $u_i = g\left( p,q \right)$ and $c_j = h\left(j\right)$ for all $1\leq i\leq m, 1\leq j\leq n$, as well as the user constraint value $\ebar = A$.
\end{proofCharles}

\subsubsection{Full Reduction from Regular Graphs}  \label{sec:appendix_reduction_regular}
Building upon Lemma~\ref{lemma:reduction_types}, we complete the MSS instance reduced from a $\Delta$-regular graph by filling in the two missing parameters:
\begin{itemize}
    \item $K=1$: Each user (edge) only receives one recommendation.
    \item $\abar=\Delta$: Each creator (vertex) requires $\Delta$ users to stay on the platform.
\end{itemize}
We can now prove the correctness of the reduction (Proposition~\ref{prop:hard_regular}). In doing so, we will establish a 1-1 correspondence between independent sets and stable sets with two lemmas.

\begin{lemma}[Stable sets $\implies$ independent sets for regular graphs]  \label{lemma:ss_to_is_regular}
    Given any $\Delta$-regular simple graph $G=(V,E)$ with vertices $V=\left\{ v_1,\dots,v_n \right\}$ and edges $E=\left\{ e_1,\dots,e_m \right\}$, construct an MSS instance with types and user constraints given by Lemma~\ref{lemma:reduction_types}, $K=1$, and $\abar=\Delta$. Given any stable set $\left( \userset, \creatorset \right)$ with recommendations $\rec$ in this MSS instance, define a corresponding set of vertices $S:=\left\{ v_j \ | \ j\in \creatorset \right\}$ in $G$. Then:
    \begin{itemize}
        \item $S$ is an independent set.
        \item User $i \in \userset$ if and only if $e_i$ is incident on exactly one vertex $v_j$ in $S$. In addition, user $i$ is assigned to creator $j$.
        \item $\engagement \left( \userset, \creatorset, \rec \right) = \ebar \Delta \left| S \right|$.
    \end{itemize}
\end{lemma}
Intuitively, Lemma~\ref{lemma:ss_to_is_regular} shows that stable sets in our MSS instance capture the same constraint as independent sets: no two adjacent creators (vertices) can co-exist. Note that the lemma also completely characterizes all stable sets by the set of creators, i.e., $\userset$ and $\rec$ are uniquely determined given $\creatorset$.

\begin{proofCharles}
    For any $j\in \creatorset$, $v_j \in S$ by definition. By Lemma~\ref{lemma:reduction_types}, the only users that are happy with creator $c_j$ correspond to the edges that are incident on $v_j$, and there are exactly $\Delta$ of them. Since $c_j$ has audience constraint $\abar = \Delta$, all these $\Delta$ users must be in $\userset$ and assigned to $c_j$ as their only recommendation (since $K=1$) in order to meet $c_j$'s constraint.

    Now consider another vertex $v_k$ that is adjacent to $v_j$ with edge $e_i$ connecting them. Since user $u_i$'s recommendation is $\rec(i)=\left\{ j \right\}$, $u_i$ cannot be assigned to $c_k$. As we have just shown that any creator (vertex) in $\creatorset$ needs to receive all its $\Delta$ incident edges (users), this implies $k\notin \creatorset$, so $v_k \notin S$. This means $S$ is an independent set.

    We now show the second statement that completely characterizes $\userset$ and $\rec$. Due to independence, any edge $e_i$ must be adjacent to either zero or one vertex in $S$. If $e_i$ is adjacent to one vertex $v_j$, it must hold that $i\in \userset$ and $\rec(i)=\left\{j\right\}$ as we have shown. If $e_i$ is not adjacent to any vertices in $S$, $u_i$ is not happy with any creator in $\creatorset$ (due to Lemma~\ref{lemma:reduction_types}), so $i\notin \userset$.

    Finally, we prove the third statement on engagement:
    \begin{align*}
        \engagement \left( \userset, \creatorset, \rec \right) &= \sum_{j\in \creatorset} \sum_{i: j\in \rec(i)} u_i^T c_j  \\
        &= \sum_{j\in \creatorset} \sum_{i: j\in \rec(i)} \ebar, \quad \text{since all happy pairs have engagement $\ebar$ by Lemma~\ref{lemma:reduction_types}}  \\
        &= \sum_{j\in \creatorset} \ebar \Delta, \quad \text{since $j$ gets $\Delta$ users}  \\
        &= \ebar \Delta \left| S \right|.
    \end{align*}
\end{proofCharles}

Lemma~\ref{lemma:ss_to_is_regular} show that any stable set corresponds to an independent set. Conversely, the following lemma shows that any independent set corresponds to a stable set:
\begin{lemma}[Independent sets $\implies$ stable sets for regular graphs]  \label{lemma:is_to_ss_regular}
    Given any $\Delta$-regular simple graph $G=(V,E)$ with vertices $V=\left\{ v_1,\dots,v_n \right\}$ and edges $E=\left\{ e_1,\dots,e_m \right\}$, construct an MSS instance with types and user constraints given by Lemma~\ref{lemma:reduction_types}, $K=1$, and $\abar=\Delta$ for all creators. For any independent set $S\subseteq V$, there exists a stable set $\left( \userset, \creatorset \right)$ with recommendations $\rec$ such that:
    \begin{itemize}
        \item $\creatorset = \left\{ j\in [1,n] \ | \ v_j \in S \right\}$.
        \item User $i \in \userset$ if and only if $e_i$ is incident on exactly one vertex in $S$. In addition, user $i$ is assigned to the corresponding creator.
        \item $\engagement \left( \userset, \creatorset, \rec \right) = \ebar \Delta \left| S \right|$.
    \end{itemize}
\end{lemma}
\begin{proofCharles}
    Construct $\userset, \creatorset$ and $\rec$ as stated in the lemma. We will show that $\left( \userset, \creatorset \right)$ is a stable set with recommendations $\rec$, i.e., all user and creator constraints are satisfied.

    For each user $i\in \userset$, its recommendation $\rec(i)=\left\{j \right\}$ is well-defined and has size 1: $j\in \creatorset$, and since $S$ is an independent set, $e_i$ cannot have both its neighbors in $S$. Also, since $e_i$ is incident on $v_j$, Lemma~\ref{lemma:reduction_types} suggests that $u_i$ is happy with $c_j$, so $\rec$ satisfies all user constraints.

    For each creator $j\in \creatorset$, all its incident edges have their corresponding users in $\userset$ and assigned to $c_j$ by construction. There are $\Delta$ such edges, so $\rec$ satisfies $c_j$'s audience constraint of $\Delta$. Therefore, $\left( \userset, \creatorset \right)$ is a stable set with recommendations $\rec$.

    It remains to show the third statement on engagement. This can be done similarly to Lemma~\ref{lemma:ss_to_is_regular}.
\end{proofCharles}
Finally, we prove Proposition~\ref{prop:hard_regular} by formally establishing the 1-1 correspondence between both sets and their sizes (engagement):
\begin{proofCharles}[Proof of Proposition~\ref{prop:hard_regular}]
    For any $\Delta$-regular graph $G$, construct a corresponding MSS instance following Lemmas~\ref{lemma:reduction_types} and \ref{lemma:ss_to_is_regular} in polynomial time, with dimensionality $D=6$.
    
    Lemma~\ref{lemma:ss_to_is_regular} maps each stable set in the MSS instance to a unique independent set in $G$ (note they are both uniquely identified by the sets of creators and vertices respectively). Lemma~\ref{lemma:is_to_ss_regular} maps each independent set to a unique stable set. Thus, there is a 1-1 correspondence between all independent sets in $G$ and all stable sets in the MSS instance.

    In addition, the two lemmas also establish a proportional relationship between sizes of independent sets and total engagement of stable sets. For an independent set $S$ and a corresponding stable set $\left( \userset, \creatorset \right)$ with recommendations $\rec$, $\engagement \left( \userset, \creatorset, \rec \right) = \ebar \Delta \left| S \right|$. This means finding a \textit{maximum} independent set in $G$ is equivalent to finding a \textit{maximum} stable set in the MSS instance, which completes the reduction.
\end{proofCharles}

\subsubsection{Reduction from General Graphs}  \label{sec:appendix_reduction_general}
In this section, we offer an alternative reduction procedure that proves Corollary~\ref{cor:hard_general} on general graphs, which is less restrictive and does not require all vertices to have the same degree. This is necessary for showing the inapproximability of the \MSS\ problem.

As a high-level summary, for any vertex (creator) whose degree is smaller than the maximum vertex degree $\Delta$ in the graph, we add auxiliary users who can only be assigned to this creator, so that its potential audience size becomes $\Delta$. These users can be seen as ``auxiliary edges''. The rest of the construction and proof are similar to Sections~\ref{sec:appendix_reduction_types} and \ref{sec:appendix_reduction_regular}, as none of the key properties are fundamentally changed.

We first present the construction of auxiliary users in the following lemma. This is an extension of Lemma~\ref{lemma:reduction_types}.

\begin{lemma}[Type vectors of auxiliary users]  \label{lemma:reduction_types_general}
    For any simple graph $G=(V,E)$ with vertices $V=\left\{ v_1,\dots,v_n \right\}$ and edges $E=\left\{ e_1,\dots,e_m \right\}$, let $\Delta$ be its maximum vertex degree. Consider some users and $n$ creators, where the types of all creators, $m$ of the users and the user constraints $\ebar$ are given by Lemma~\ref{lemma:reduction_types}. Then, the types of remaining auxiliary users in $D=6$ dimensions can be constructed in polynomial time, such that user $i$ is happy with creator $j$ if and only if one of the following is true:
    \begin{itemize}
        \item $i$ is not an auxiliary user, and vertex $v_j$ is incident on edge $e_i$, or
        \item $i$ is an auxiliary user associated with creator $j$.
    \end{itemize}
    In addition, for all such $i$ and $j$, the engagement that user $i$ gets from creator $j$ is $\ebar$.
\end{lemma}

\begin{proofCharles}
    Consider some users and $n$ creators (vertices), such that:
    \begin{itemize}
        \item The first $m$ users correspond to the $m$ edges in $G$ (or ``original users''). We index them with $1,\dots,m$.
        \item The remaining users are \textit{auxiliary} users. For each vertex $v_j$, we create $\left( \Delta-\deg(j) \right)$ users to be \textit{associated} with creator $j$, and denote this set of users as $\userset_j$. Each auxiliary user is associated with exactly one creator.
    \end{itemize}
    Set the types of $m$ original users and all creators as well as $\ebar$ according to Lemma~\ref{lemma:reduction_types}. In addition, for any creator $p \in \left\{ 1, \dots, n \right\}$, set the types of all users in $\userset_j$ to be $g\left(p,p\right)$, where $f, g, h$ are functions constructed in Lemma~\ref{lemma:reduction_types}.

    Note that this construction does not affect the five properties in Lemma~\ref{lemma:reduction_types} that $f$,$g$ and $h$ satisfy, and that all properties still hold when $p=q$. Thus, \eqref{eqn:reduction_types_condition} still holds for any original user $i\in [1,m]$ and any creator $j\in [1,n]$. 
    
    It remains to show that if $i$ is an auxiliary user, then $u_i^T c_j \geq A$ if and only if $i\in \userset_j$. To see this, note that if $i$ is associated with creator $p$, then $u_i^T c_j = g\left(p,p\right)^T h\left(j\right) = f\left( p,p,j \right)$, so $u_i^T c_j \geq A$ if and only if $p=j$ due to Properties~\ref{enum:reduction_fgh_argmax} and \ref{enum:reduction_fgh_max} in Lemma~\ref{lemma:reduction_types}.
\end{proofCharles}

We now complete the full reduction and prove its correctness:
\begin{proofCharles}[Proof of Corollary~\ref{cor:hard_general}]
    For any graph $G$, construct a corresponding MSS instance following Lemma~\ref{lemma:reduction_types_general}, and set:
    \begin{itemize}
        \item $K=1$: Each user (edge) only receives one recommendation. This is true for both original and auxiliary users.
        \item $\abar=\Delta$: Each creator (vertex) requires $\Delta$ users to stay on the platform, where $\Delta$ is the maximum vertex degree.
    \end{itemize}
    We can show similar statements as Lemma~\ref{lemma:ss_to_is_regular} and Lemma~\ref{lemma:is_to_ss_regular}, to show a 1-1 correspondence between all independent sets in $G$ and all stable sets in the MSS instance. In particular, any independent set $S$ corresponds to a stable set $\left( \userset, \creatorset \right)$ with recommendations $\rec$, such that:
    \begin{itemize}
        \item $\creatorset = \left\{ j\in [1,n] \ | \ v_j \in S \right\}$.
        \item User $i \in \userset$ if and only if one of the following is true:
        \begin{itemize}
            \item $i$ is an original user, and $e_i$ is incident on exactly one vertex in $S$. In addition, user $i$ is assigned to the corresponding creator.
            \item $i$ is an auxiliary user associated with creator $j$, such that $v_j \in S$. In addition, user $i$ is assigned to creator $j$.
        \end{itemize}
        \item $\engagement \left( \userset, \creatorset, \rec \right) = \ebar \Delta \left| S \right|$.
    \end{itemize}
    The detailed proofs are left as an exercise to the reader.

    Similarly to regular graphs, there is a proportional relationship between sizes of independent sets and total engagement of stable sets, given by the third statement above. This means finding a \textit{maximum} independent set in $G$ is equivalent to finding a \textit{maximum} stable set in the MSS instance, which completes the reduction.
\end{proofCharles}

\subsection{NP-Hardness for Arbitrary Values of \texorpdfstring{$K$}{K}}
While the proofs of NP-hardness for Theorem~\ref{thm:hard} construct instances where the number of recommendations per user is $K=1$, we will show that the hardness result still applies even when we only want to find maximum stable sets for instances whose $K$ is given as a specific fixed value, as long as $K\geq 3$. Formally:

\begin{corollary}[Hardness for fixed $K$]   \label{cor:hard_fixed_K}
    For any given value of $K$ such that $K\geq 3$, \MSS\ on the subset of instances with this value of $K$ is NP-hard, even when $D$ is a constant as long as $D\geq 7$.
\end{corollary}
To show this, we will reduce \MIS\ on $\Delta$-regular graphs (where $\Delta\geq 3$) to \MSS\ with the given $K$.\footnote{Note that \MIS\ on graphs with maximum degree at most 2 is solvable in polynomial time.} This builts upon Proposition~\ref{prop:hard_regular} which constructs MSS instances with $K=1$.

\subsubsection{Reduction Procedure}  \label{sec:hard_fixed_K_reduction}
We first list the full procedure of constructing a \MSS\ instance with the specified $K$, with dimensionality $D=7$.

Given any $\Delta$-regular simple graph $G=(V,E)$ with $\Delta \geq 3$, where $V=\left\{ v_1,\dots,v_n \right\}$ and edges $E=\left\{ e_1,\dots,e_m \right\}$, we first apply the procedure in Lemma~\ref{lemma:reduction_types} to obtain the types of $m$ users and $n$ creators in $\mathbb{R}^6$, user constraint $\ebar$ and creator constraint $\abar$. However, we make a small change: replace Equation~\eqref{eqn:reduction_types_G} with
\begin{align*}
    G &:= \max \left\{ \sqrt{2}, \max_{1\leq p,q\leq n} \left\lVert g'''\left(p,q\right) \right\rVert \right\}.
\end{align*}
This ensures $G\geq \sqrt{2}$ and thus $\ebar \leq 1/\sqrt{2}$.

Let $\userset_0$ be the set of these $m$ users, and $\creatorset_0$ be the set of these $n$ creators. Recall that in Lemma~\ref{lemma:reduction_types}, all users in $\userset_0$ have types of the form $g\left( p,q \right) \in \mathbb{R}^6$ and all creators in $\creatorset_0$ have types $h\left( j \right) \in \mathbb{R}^6$, where $f, g, h$ are functions defined in that lemma. We first embed all these type vectors in $\mathbb{R}^7$ by simply adding a trailing zero as the last component. Note that they also satisfy $\left( c_j \right)_5 = 0$ for all $j\in \creatorset_0$ and $\left( u_i \right)_6 = 0$ for all $i\in \userset_0$ by construction.

When $K>1$, each user $i\in \userset_0$ needs to be assigned to $K$ creators, but the previous reduction requires her to be assigned to exactly one original creator in $\creatorset_0$. Therefore, for each $i\in\left\{ 1, \dots, m \right\}$, we construct a set of $K-1$ creators, $\creatorset_i$, such that user $i$ is happy with any of them. In particular, all creators in $\creatorset_i$ have the same type:
\begin{align*}
    c_k &= \left[ \ebar \left( u_i \right)_1, \dots, \ebar \left( u_i \right)_5, \sqrt{1-\ebar^2}, 0 \right], \quad \text{for all } 1\leq i\leq m, k\in \creatorset_i.
\end{align*}
We still need to ensure that each creator in $\creatorset_i$ receives $\abar$ users. Thus, for each $i\in\left\{ 1, \dots, m \right\}$, we construct a set of $\abar-1$ users, $\userset_i$, such that each of them is happy with all creators in $\creatorset_i$. All users in $\userset_i$ have the same type:
\begin{align*}
    u_k &= \left[ \ebar^2 \left( u_i \right)_1, \dots, \ebar^2 \left( u_i \right)_5, \ebar \sqrt{1-\ebar^2}, \sqrt{1-\ebar^2} \right], \quad \text{for all } 1\leq i\leq m, k\in \userset_i.
\end{align*}
Finally, all $\left(\abar-1\right)m$ users in $\bigcup_{i=1}^m \userset_i$ still need one more creator in order to receive $K$ recommendations. To achieve this, we construct on single creator, called $X$, with type
\begin{align*}
    c_X &= \left[ 0, 0, 0, 0, 0, 0, 1 \right].
\end{align*}
All users have constraint $\ebar$, and all creators have constraint $\abar$. Note that $\abar \geq 3$ and $m\geq 4$ by our initial assumption that $\Delta\geq 3$.

\subsubsection{Proof of Correctness}
We now prove that the procedure in Section~\ref{sec:hard_fixed_K_reduction} gives a valid reduction. First, we will show properties of all happy user-creator pairs, analogously to Lemma~\ref{lemma:reduction_types}:

\begin{lemma}[Happy pairs]  \label{lemma:reduction_types_fixK}
    In the MSS instance constructed in Section~\ref{sec:hard_fixed_K_reduction} user $i$ is happy with creator $j$ if and only if one of the following is true:
    \begin{itemize}
        \item $i\in \userset_0$, $j\in \creatorset_0$, and vertex $v_j$ is incident on edge $e_i$;
        \item $i\in \userset_0$, and $j\in \creatorset_i$;
        \item $i\in \userset_k$ for some $k=\left\{ 1, \dots, m \right\}$, and $j\in \creatorset_k$;
        \item $i\in \userset_k$ for some $k=\left\{ 1, \dots, m \right\}$, and $j=X$.
    \end{itemize}
    In addition, for all such $i$ and $j$, the engagement that user $i$ gets from creator $j$ is $\ebar$, except for the last case where the engagement is $\sqrt{1-\ebar^2}$.
\end{lemma}
The main goal of the lemma is to show that for any users and creators that were added in Section~\ref{sec:hard_fixed_K_reduction}, the only valid assignments between them and any other players are those that we mentioned as the intuition for adding them.

\begin{proofCharles}
    Suppose user $i$ is happy with creator $j$. First, consider when $i\in \userset_0$. If additionally $j\in \creatorset_0$, Lemma~\ref{lemma:reduction_types} characterizes their happiness and thus implies the first case. If $j\in \creatorset_k$ for some $k>0$, notice that $u_i^T c_j = u_i^T \left( \ebar u_k \right) \leq \ebar \left\lVert u_i \right\rVert ^2 = \ebar$, so equality must hold. This happens if and only if $u_i = u_k$. Since $u_i$ is determined by $g\left( p,q \right)$ where $v_p, v_q$ are vertices incident on edge $e_i$, and $G$ is a simple graph, this means all such $u_i$'s are distinct and thus $i=k$. If $j=X$, $u_i^T c_j = 0$.

    Now consider when $i\in \userset_k$ for some $k\in \left\{ 1, \dots, m \right\}$. Notice that the first six components of its type, $\left(u_i\right)_{1\text{-}6}$, satisfies $\left\lVert \left(u_i\right)_{1\text{-}6} \right\rVert = \ebar$. Also, the types of all creators other than $X$ have the seventh component being zero. Thus, $u_i^T c_j \leq \ebar$ if $j\neq X$, and equality holds if and only if the unit vector $\left(c_j\right)_{1\text{-}6}$ is a scalar multiple of $\left(u_i\right)_{1\text{-}6}$. By construction, this can only happen if $j\in \creatorset_k$, since all other sets $\creatorset_l$ have different creator types from $\creatorset_k$ (including $l=0$). Finally, if $j=X$, $u_i^T c_j = \sqrt{1-\ebar^2} \geq \ebar$ since $\ebar \leq 1/\sqrt{2}$.
\end{proofCharles}
We will now complete the reduction by showing a 1-1 correspondence between all independent sets in $G$ and all stable sets in the MSS instance, with a proportional relationship on their sizes (total engagement), analogously to Lemmas~\ref{lemma:ss_to_is_regular} and \ref{lemma:is_to_ss_regular}.

\begin{proofCharles}[Proof of Corollary~\ref{cor:hard_fixed_K}]
    Consider the MSS instance constructed in Section~\ref{sec:hard_fixed_K_reduction}. For any stable set $\left( \userset, \creatorset \right)$ with recommendations $\rec$ in this MSS instance, define a corresponding set of vertices $S:=\left\{ v_j \ | \ j\in \creatorset \cap \creatorset_0 \right\}$ in $G$. If $S\neq \emptyset$, we make the following claims:
    \begin{enumerate}[label=(\alph*)]%
        \item $S$ is an independent set.  \label{enum:hard_K_claim_indep}
        \item For any user $i \in \userset_0$, $i \in \userset$ if and only if $e_i$ is incident on exactly one vertex $v_j$ in $S$. In addition, $i \in \userset$ if and only if $R(i) = \left\{j \right\} \cup \creatorset_i$.  \label{enum:hard_K_claim_U0}
        \item For any creator $j \in \creatorset_k$ for some $k\in \left\{ 1, \dots, m \right\}$, $j\in \creatorset$ if and only if $k \in \userset$.  \label{enum:hard_K_claim_Ck}
        \item For any user $i \in \userset_k$ for some $k\in \left\{ 1, \dots, m \right\}$, $i\in \userset$ if and only if $k \in \userset$. In addition, $i\in \userset$ if and only if $R(i) = \creatorset_k \cup \left\{X \right\}$.  \label{enum:hard_K_claim_Uk}
        \item $X \in \creatorset$.  \label{enum:hard_K_claim_X}
        \item $\engagement \left( \userset, \creatorset, \rec \right)$ is proportional to $\left| S \right|$.  \label{enum:hard_K_claim_engagement}
    \end{enumerate}
    Note that these claims imply the value of $\creatorset \cap \creatorset_0$ completely characterizes the rest of the stable set and its recommendations. To prove these claims, we show the following facts:
    \begin{enumerate}
        \item For each original creator $j\in \creatorset_0$, her potential audience consists entirely of $\abar$ original users in $\userset_0$, which correspond to the $\Delta=\abar$ edges that are incident on vertex $v_j$ (similar to Lemma~\ref{lemma:ss_to_is_regular}). Thus, $j\in \creatorset$ if and only if all these $\abar$ users are in $\userset$ and assigned to creator $j$.  \label{enum:hard_K_point1}
        \item Each original user $i\in \userset_0$ is only happy with $K+1$ creators: two creators $j, k \in \creatorset_0$ (where $v_j, v_k$ are the two endpoints of edge $e_i$), and the $K-1$ creators in $\creatorset_i$. This means $i\in \userset$ if and only if at least one of $j, k$ and at least $K-2$ creators in $\creatorset_i$ are in $\creatorset$, and user $i$ is assigned to them.  \label{enum:hard_K_point2}
        \item For any creator $j\in \creatorset_k$ for some $k\in \left\{ 1, \dots, m \right\}$, her potential audience consists of $\abar$ users: $\left\{ k \right\} \cup \userset_k$. Thus, $j\in \creatorset$ if and only if both of the following are true: $k$ and all users in $\userset_k$ are in $\userset$, and all of them are assigned to creator $j$.   \label{enum:hard_K_point3}
        \item Any user $i\in \userset_k$ for some $k\in \left\{ 1, \dots, m \right\}$ is only happy with $K$ creators: $\creatorset_k \cup \left\{ X \right\}$. Thus, $i\in \userset$ if and only if both of the following are true: both $X$ and all creators in $\creatorset_k$ are in $\creatorset$, and $R(i) = \creatorset_k \cup \left\{ X \right\}$.  \label{enum:hard_K_point4}
        \item Let $k\in \userset \cap \userset_0$ be an original user in the stable set. By Fact~\ref{enum:hard_K_point2} and the assumption that $K\geq 3$, there exists $j\in \creatorset_k$ such that creator $j$ is in the stable set and user $k$ is assigned to her. Applying Fact~\ref{enum:hard_K_point3} to creator $j$ gives $\userset_k \subseteq \userset$. Applying Fact~\ref{enum:hard_K_point4} to any user $i\in \userset_k$ gives $\creatorset_k \subseteq \creatorset$. Applying Fact~\ref{enum:hard_K_point3} again to all creators in $\creatorset_k$ gives $\creatorset_k \subseteq R(k)$, i.e., user $k$ must be assigned to all these $K-1$ creators. This means user $k$ is only assigned to exactly one original creator (one of its endpoints) in $\creatorset_0$.  \label{enum:hard_K_point5}
    \end{enumerate}
    Notice that Facts~\ref{enum:hard_K_point1} and \ref{enum:hard_K_point5} reestablish the key constraints used in Lemma~\ref{lemma:ss_to_is_regular} for original users and creators (each user only gets one creator, and each creator needs all $\Delta=\abar$ users). Thus, we can similarly show Claim~\ref{enum:hard_K_claim_indep}. 

    The $\left( \implies \right)$ directions of Claims~\ref{enum:hard_K_claim_U0}, \ref{enum:hard_K_claim_Ck} and \ref{enum:hard_K_claim_Uk} are given respectively by Facts~\ref{enum:hard_K_point5}, \ref{enum:hard_K_point3}, and a combination of Facts \ref{enum:hard_K_point4} and \ref{enum:hard_K_point3}. The $\left( \impliedby \right)$ directions of Claims~\ref{enum:hard_K_claim_Ck} and \ref{enum:hard_K_claim_Uk} are given respectively by Fact~\ref{enum:hard_K_point5} and a combination of Facts \ref{enum:hard_K_point5} and \ref{enum:hard_K_point3}. The $\left( \impliedby \right)$ direction of Claim~\ref{enum:hard_K_claim_U0} is a consequence of Fact~\ref{enum:hard_K_point1} by noting that if $e_i$ is incident on $v_j\in S$, then $j\in \creatorset_0 \cap \creatorset$. Claim~\ref{enum:hard_K_claim_X} is given by a combination of $S\neq \emptyset$ and Facts~\ref{enum:hard_K_point1}, \ref{enum:hard_K_point5}, \ref{enum:hard_K_point3} and \ref{enum:hard_K_point4}.

    For Claim~\ref{enum:hard_K_claim_engagement}, consider the contribution of each original creator in the stable set on total engagement. Each original creator in $\creatorset \cap \creatorset_0$ receives $\abar$ original users, and each such user is included concurrently with $\left(\abar-1\right)$ auxiliary users in $\mathcal{U}_k$. Hence, the stable set contains $\abar^2$ users with $K$ recommendations each. With the exception of $\abar \left(\abar-1\right)$ recommendations to creator $X$ from all auxiliary users, which have engagement $\sqrt{1-\ebar^2}$ each, all other $K \abar^2 - \abar \left(\abar-1\right)$ recommendations have engagement $\ebar$ each. Thus, the inclusion of each original creator contributes engagement
    \begin{align*}
        \left( K \abar^2 - \abar \left(\abar-1\right) \right) \ebar + \abar \left(\abar-1\right) \sqrt{1-\ebar^2}.
    \end{align*}
    This is a constant for each creator in $\creatorset \cap \creatorset_0$ and thus for each vertex in $S$, so total engagement is proportional to $\left| S \right|$ which shows ~\ref{enum:hard_K_claim_engagement}.

    Claims~\ref{enum:hard_K_claim_indep}-\ref{enum:hard_K_claim_engagement} establish a mapping from stable sets to independent sets. One can similarly show that any independent set $S$ maps to a stable set where $\creatorset \cap \creatorset_0 = \left\{ j \ | \ v_j \in S \right\} $, and the choice of other creators, $\userset$ and $\rec$ are uniquely determined in order to satisfy Claims~\ref{enum:hard_K_claim_U0}-\ref{enum:hard_K_claim_X}. In addition, the total engagement satisfies Claim~\ref{enum:hard_K_claim_engagement}. Details are omitted.

    The 1-1 correspondence between independent sets and stable sets with proportional sizes means solving \MIS\ on $G$ is equivalent to solving \MSS\ on the constructed instance, which completes the proof.
\end{proofCharles}

\section{Proofs and Discussions of One-Sided Creator Participation Constraints, and Comparison to \texorpdfstring{\cite{mladenov2020optimizing}}{Mladenov et al. (2020)}}  \label{sec:Mladenov_appendix}
In Section~\ref{sec:users_never_leave}, we discussed the case when $\ebar = 0$, where users do not have any participation constraints and never leave the platform as long as they receive $K$ recommendations. This case is a generalization of the baseline model of \cite{mladenov2020optimizing}. In this appendix, we first offer a more detailed summary of the model proposed by \citeauthor{mladenov2020optimizing} and compare it to our model, and then prove Proposition~\ref{prop:submodular_fixU} and Theorem~\ref{thm:approx_bigK} which generalize their analogous result. Finally, we discuss why the approximability result in this special case does not extend to our more general model where $\ebar \geq 0$.

\subsection{Summary of \texorpdfstring{\cite{mladenov2020optimizing}}{Mladenov et al. (2020)}'s Model}
\citeauthor{mladenov2020optimizing} propose a very similar model that formulates the task of content recommendation as a matching problem with creator participation constraints. The key differences from our model are that they do not allow departure of users, and that each user is only assigned to one creator per time step.  %

In \citeauthor{mladenov2020optimizing}'s most general model, the system progresses in terms of \textit{epochs} with a fixed number of time steps $T$. Within each epoch, each user issues a random \textit{query} at every time step $t$. In response to each query, the platform gives the user a fractional (or stochastic) recommendation which is a distribution over all creators that sums to 1. Users then derive immediate rewards from recommendations. At the end of an epoch, each creator evaluates her cumulative audience over the epoch, and leaves the platform (or becomes ``unviable'' in their terminology) if it is below her threshold. Similarly to our model, the platform's objective is to maximize long-term average total engagement of all users (which they denote as user utility).

The case with additive user utilities in Section 3.1 of \citeauthor{mladenov2020optimizing} shares the most similarities with our model. In this case, each user's utility per epoch is a weighted sum of her utilities at each time step, and thus maximizing long-term engagement can be formulated as a mixed-integer linear program (4) in their paper. In their Appendix A.1, they show that (4) can then be reduced to an equivalent integer linear program (6), where each epoch has a single time step ($T=1$, thus equating users with queries), and matchings between users and creators are 0-1 variables (no fractional assignments). Therefore, their problem (6) is almost a special case of our model, in which each user only receives one recommendation ($K=1$) and users never leave ($\ebar=0$), with a slight generalization that the engagement that each user derives from each creator can be arbitrary and not limited to a dot product of their types.

\subsection{Approximation Results from \texorpdfstring{\cite{mladenov2020optimizing}}{Mladenov et al. (2020)}}
With the two additional restrictions of $K=1$ and $\ebar=0$, \citeauthor{mladenov2020optimizing} show that their problem allows an efficient constant-factor approximation:
\begin{theorem}[Paraphrased and corrected from Theorem~1 in \cite{mladenov2020optimizing}]  \label{thm:mladenov_constant_approx}
    When $K=1$ and users never leave (equivalently, $\ebar=0$), \MSS\ (equivalently, problem (6) in \citeauthor{mladenov2020optimizing}) can be approximated up to factor $\left( 1/e-\epsilon \right)$ in polynomial time. This also applies if the engagement each user derives from each creator is an arbitrary value instead of the dot product of their types.
\end{theorem}

The constant factor approximation in Theorem~\ref{thm:mladenov_constant_approx} crucially relies on two properties. For one, the objective function is submodular in the set of creators that stay on the platform:
\begin{theorem}[Submodularity when $K=1$, paraphrased from Theorem~2 in \cite{mladenov2020optimizing}]  \label{thm:mladenov_submodular}
    Let $g(\creatorset)$ denote the maximum engagement of any stable set in which the set of creators is exactly $\creatorset$: that is,
    $$ g\left( \creatorset \right) := \max_{\userset \subseteq \userset_0, \rec} \engagement \left( \userset, \creatorset, \rec \right), $$
    where $\rec$ is subject to all user and creator participation constraints in $\userset$ and $\creatorset$. In particular, $g(\creatorset)$ is undefined if no such stable sets exist. Then, when $K=1$ and $\ebar=0$, for any $c_0, c_1 \in \creatorset_0$ and $\creatorset \subset \creatorset_0 \backslash \left\{ c_0, c_1 \right\}$,
    \begin{align}
        g\left( \creatorset \cup \left\{ c_0, c_1 \right\} \right) - g\left( \creatorset \cup \left\{ c_1 \right\} \right) &\leq g\left( \creatorset \cup \left\{ c_0 \right\} \right) - g\left( \creatorset \right).  \label{eqn:Mladenov_submodular}
    \end{align}
    This also applies if the engagement each user derives from each creator is an arbitrary value instead of the dot product of their types.
\end{theorem}
The other property is that feasibility of choosing creators $\creatorset$ (i.e., the existence of a stable set $\left( \userset_0, \creatorset \right)$) is characterized by a single knapsack constraint:
\begin{proposition}[Knapsack constraint for \cite{mladenov2020optimizing}]  \label{prop:mladenov_knapsack}
    When $K=1$ and $\ebar=0$, $g(\creatorset)$ is well-defined if and only if $\sum_{j\in \creatorset} \abar \leq UK$, or more succinctly, $\left| \creatorset \right|\abar \leq UK$. In addition, under a variation of the model where each creator $j$ has her own audience threshold $\abar_j$ that can be heterogeneous, $g(\creatorset)$ is well-defined if and only if $ \sum_{j\in \creatorset} \abar_j \leq UK$.  %
\end{proposition}
This is a knapsack constraint as we can interpret $\abar_j$ as the ``cost'' of choosing creator $j$, and $UK=U=\left| \userset_0 \right|$ as the ``budget''. The proof of Proposition~\ref{prop:mladenov_knapsack} is trivial: necessity of $\left| \creatorset \right|\abar \leq UK$ is given by creator constraints, while sufficiency can be shown by construction, as a feasible matching between $\userset$ and $\creatorset$ that satisfies all their constraints (regardless of total engagement) can be found by arbitrarily assigning each user to $K=1$ creator as long as all creator constraints are satisfied.

When Theorem~\ref{thm:mladenov_submodular} and Proposition~\ref{prop:mladenov_knapsack} hold, \cite{Feldman2011AUnified} show that there exists a continuous greedy algorithm that achieves a $\left( 1/e-\epsilon \right)$-approximation, thus proving Theorem~\ref{thm:mladenov_constant_approx}.\footnote{We believe the original proof of Theorem~\ref{thm:mladenov_constant_approx} in \cite{mladenov2020optimizing} is incorrect. In their paper, the sole justification is Theorem~\ref{thm:mladenov_submodular}, and they claim that submodularity alone suggests a greedy approximation algorithm by adding creators until no more can be added. However, this simple algorithm requires the objective function to be monotone, whereas $g\left(\creatorset\right)$ is not necessarily monotone as forcing more creators to stay can hurt total engagement. Their overall statement of constant approximation in Theorem~\ref{thm:mladenov_constant_approx} is still true with a difference of $\epsilon$, but it requires a different, more complicated greedy algorithm as suggested in \cite{Feldman2011AUnified}, and it relies on the knapsack feasibility constraint (Proposition~\ref{prop:mladenov_knapsack}) which \citeauthor{mladenov2020optimizing} do not show.}

\subsection{Generalizing \texorpdfstring{\cite{mladenov2020optimizing}}{Mladenov et al. (2020)} to Our Model When \texorpdfstring{$\ebar = 0$}{e=0}: Proofs of Proposition~\ref{prop:submodular_fixU} and Theorem~\ref{thm:approx_bigK}}  \label{sec:submodular_fixU}
We first relate the aforementioned results from \citeauthor{mladenov2020optimizing} to our model, which allows $K\geq 1$, and prove our results. First, we show that a slight variation of Theorem~\ref{thm:mladenov_submodular} holds when $K\geq 1$, but requires the set of users in the stable set to be exogenously fixed. This is our Proposition~\ref{prop:submodular_fixU} that we introduced in Section~\ref{sec:users_never_leave}:

\begin{customTheorem}{Proposition~\ref{prop:submodular_fixU}}[Submodularity when users are fixed]
    Suppose $\ebar = 0$. Given subsets of users $\userset \subseteq \userset_0$ and creators $\creatorset \subseteq \creatorset_0$, define function $f(\userset, \creatorset)$ as follows. When $\left| \creatorset \right| \geq K$, $f(\userset, \creatorset)$ is the maximum engagement of any stable set $\left(\userset, \creatorset \right)$, i.e.,
    \begin{align}
        f\left( \userset, \creatorset \right) := \max_{\rec} \engagement \left( \userset, \creatorset, \rec \right),  \label{eqn:submodular_fixU_func_def}
    \end{align}
    where $\rec$ is subject to participation constraints of all players in $\userset$ and $\creatorset$. When $\left| \creatorset \right| < K$, $f(\userset, \creatorset)$ is the maximum engagement of any stable set $\left(\userset, \creatorset \right)$ on a modified instance where the number of recommendations per user is replaced with $\left| \creatorset \right|$ instead of $K$.
    Then, for any $\userset \subseteq \userset_0$, $c_0, c_1 \in \creatorset_0$ and $\creatorset \subseteq \creatorset_0 \backslash \left\{ c_0, c_1 \right\}$,
    \begin{align}
        f\left( \userset, \creatorset \cup \left\{ c_0, c_1 \right\} \right) - f\left( \userset, \creatorset \cup \left\{ c_1 \right\} \right) &\leq f\left( \userset, \creatorset \cup \left\{ c_0 \right\} \right) - f\left( \userset, \creatorset \right),  \label{eqn:submodular_fixU_appendix}
    \end{align}
    as long as all four terms are well-defined. This applies to all values of $K$. %
\end{customTheorem}
We remark that Proposition~\ref{prop:submodular_fixU} still holds under the variation where engagements of user-creator pairs can be arbitrarily defined and not limited to dot products of their types, as is the case with Theorem~\ref{thm:mladenov_submodular}. On the other hand, we define the objective function as $f\left( \userset, \creatorset \right)$ in Proposition~\ref{prop:submodular_fixU}, instead of $g\left( \creatorset \right)$ in Theorem~\ref{thm:mladenov_submodular} from \citeauthor{mladenov2020optimizing}, in order to emphasize that the set of users $\userset$ being exogenously fixed and not chosen by the platform as part of the optimization problem. However, when $\ebar = 0$ (which is also the setting for Theorem~\ref{thm:mladenov_submodular}), choosing $\userset = \userset_0$ is always a feasible and optimal choice for both $f\left( \userset, \creatorset \right)$ and $g\left( \creatorset \right)$, so the set of users is implicitly fixed and $g\left( \creatorset \right) = f\left( \userset_0, \creatorset \right)$.\footnote{When $\ebar > 0$, most of our proof of Proposition~\ref{prop:submodular_fixU} still applies when $\left| \creatorset \right| \geq K$, but no longer holds for the edge case where $\left| \creatorset \right|<K$. It remains an open question whether the proof can be adapted to $\left| \creatorset \right|<K$ when $\ebar>0$.}

Secondly, when $K\geq 1$ but $\ebar=0$, Proposition~\ref{prop:mladenov_knapsack} still holds, i.e., feasibility of $g\left(\creatorset \right) = f\left( \userset_0, \creatorset \right)$ for any $\creatorset \subseteq \creatorset_0$ is still characterized by a knapsack constraint. Therefore, combining Propositions~\ref{prop:submodular_fixU} and \ref{prop:mladenov_knapsack} gives our main theorem on approximation, as we discussed in Section~\ref{sec:users_never_leave}:

\begin{customTheorem}{Theorem~\ref{thm:approx_bigK}}[Approximability without user constraints]
    If users never leave (equivalently, $\ebar=0$), \MSS\ can be approximated up to factor $\left( 1/e-\epsilon \right)$ in polynomial time.
\end{customTheorem}

The following subsections mostly focus on the proof of Proposition~\ref{prop:submodular_fixU}. The proof draws many similarities from Theorem~\ref{thm:mladenov_submodular} in \cite{mladenov2020optimizing}, but with significant modification to their algorithm that constructs the matching between $\userset$ and $\creatorset \cup \left\{ c_0 \right\}$, as well as changes to notations and definitions to account for user constraints.

\subsubsection{Alternative View of \texorpdfstring{$f\left(\userset, \creatorset \right)$}{f(U,C)} as an Integer Linear Program}
Before presenting the proof of Proposition~\ref{prop:submodular_fixU}, we first offer an additional interpretation of the definition of $f\left( \userset, \creatorset\right)$. Recall that we defined the function with Equation~\eqref{eqn:submodular_fixU_func_def} above, as the maximum engagement from any recommendations that give a stable set $\left( \userset, \creatorset\right)$. Alternatively, $f\left( \userset, \creatorset\right)$ is also the objective value of the following integer linear program:
\begin{alignat*}{4}
    & \max_{x_{ij}} & \sum_{i=1}^U \sum_{j=1}^C x_{ij} u_i^T c_j &&&&& \quad \text{(total engagement)} \\
    & \text{subject to} \quad & x_{ij} \left( u_i^T c_j - \ebar \right) & \geq 0, \quad && \forall 1\leq i\leq U, 1\leq j\leq C, && \quad \text{(user constraints)} \\
    && \sum_{i=1}^U x_{ij} &\geq z_j \abar, \quad && \forall 1\leq j\leq C, && \quad \text{(creator constraints)}  \\
    && \sum_{j=1}^C x_{ij} &= K, && \forall 1\leq i\leq U, && \quad \text{(number of recommendations)}  \\
    && x_{ij} &\leq y_i, \quad && \forall 1\leq i\leq U, 1\leq j\leq C,  && \\
    && x_{ij} &\leq z_j, \quad && \forall 1\leq i\leq U, 1\leq j\leq C,  && \\
    && y_i &= \mathbf{1}_{\userset}\left(i\right), \quad && \forall 1\leq i\leq U,  && \\
    && z_j &= \mathbf{1}_{\creatorset}\left(j\right), \quad && \forall 1\leq j\leq C,  && \\
    && x_{ij} &\in \{0,1\}, \quad && \forall 1\leq i\leq U, 1\leq j\leq C,  &&
\end{alignat*}
where $\mathbf{1}_S(a)$ is an indicator function whose value is 1 if $a\in S$ and 0 otherwise. This is similar to the integer linear program for the general maximum stable set problem in Section~\ref{sec:stable_set}, but with $y_i$ and $z_j$ given according to $\userset$ and $\creatorset$, instead of being part of the optimization problem.

\subsubsection{High-Level Summary of the Proof of Proposition~\ref{prop:submodular_fixU}}  \label{sec:submodular_fixU_summary}
Without loss of generality, assume $\left| \userset \right| \geq \abar$. We will focus on the general case where $\left\lvert \creatorset \right\rvert \geq K$, but make a remark in Section~\ref{sec:submod_less_than_K} on cases where $\left\lvert \creatorset \right\rvert < K$.

Let $\rec, \rec^1$ and $\rec^{0,1}$ be the optimal recommendations that give maximum engagement for $f\left( \userset, \creatorset \right)$, $f\left( \userset, \creatorset \cup \left\{ c_1 \right\} \right)$ and $f\left( \userset, \creatorset \cup \left\{ c_0, c_1 \right\} \right)$ respectively. At a conceptual level, we compare the changes from $\rec^1$ to $\rec^{0,1}$ when creator $c_0$ is added, and show that they correspond to at least $\abar$ \textit{paths} of reassignments of users (which we call ``red paths''), and each path contributes one user to the audience of $c_0$. We then construct a recommendation $\rec^0$ on $\left( \userset, \creatorset \cup \left\{ c_0 \right\} \right)$ by attempting to reproduce the same changes on $\rec$. We show that this can be done by combining a subset of red edges (from $\rec^1$ to $\rec^{0,1}$) with a subset of edges that describe changes from $\rec$ to $\rec^1$ (which we call ``blue paths''), such that the resultant matching $\rec^0$ satisfies all participation constraints with at least as much difference in engagement. Refer to Figure~\ref{fig:submod_proof_outline} in the main text for an illustration of the high-level concepts.

Due to conflicts in notation, we use $\eng\left( \userset, \creatorset, \rec \right)$ to denote engagement of stable sets for the rest of this proof, and use $E$ to denote a set of edges in a graph. We will use either $c$ or $j$ to denote creators and their types, and either $u$ or $i$ to denote users and their types.

\subsubsection{Cases When \texorpdfstring{$\creatorset<K$}{C<K}}  \label{sec:submod_less_than_K}
We first prove Proposition~\ref{prop:submodular_fixU} holds when $\left\lvert \creatorset \right\rvert < K$. Recall that when this is true, $f\left( \userset, \creatorset \right)$ is defined as the maximum engagement of any stable set $\left( \userset, \creatorset \right)$ in a modified instance where each user receives $\left\lvert \creatorset \right\rvert$ recommendations instead of $K$. It is easy to see that the recommendation $\rec$ associated with any stable set must assign each user to all creators in $\creatorset$. Furthermore, this condition is sufficient for a stable set as it satisfies all participation constraints, due to $\ebar = 0$ and $\abar \leq \left|\userset \right|$. Therefore, the associated recommendation $\rec$ is uniquely defined in this case, and
\begin{align}
    f\left( \userset, \creatorset \right) &= \sum_{u \in \userset} \sum_{c \in \creatorset} u^T c.  \label{eqn:submodular_less_than_K}
\end{align}
Note that \eqref{eqn:submodular_less_than_K} is still true when $\left| \creatorset \right| = K$. Thus, \eqref{eqn:submodular_less_than_K} shows \eqref{eqn:submodular_fixU_appendix} when $\left| \creatorset \right| \leq K-2$.

It remains to show \eqref{eqn:submodular_fixU_appendix} holds when $\left| \creatorset \right| = K-1$. For any such $\creatorset$ and two other creators $c_0$, $c_1$, let $\rec$ be the recommendation associated with the stable set $\left( \userset, \creatorset \cup \left\{ c_0, c_1 \right\} \right)$ that gives maximum engagement. Then,
\begin{align*}
    f\left( \userset, \creatorset \cup \left\{ c_0, c_1 \right\} \right) - f\left( \userset, \creatorset \cup \left\{ c_1 \right\} \right) &= \engagement \left( \userset, \creatorset \cup \left\{ c_0, c_1 \right\}, \rec \right) - \sum_{u \in \userset} \sum_{c \in \creatorset \cup \left\{ c_1 \right\}} u^T c  \\
    &= \sum_{u \in \userset} \sum_{c \in \rec\left( u\right) } u^T c - \sum_{u \in \userset} \sum_{c \in \creatorset \cup \left\{ c_1 \right\}} u^T c  \\
    &\leq \sum_{u \in \userset} \sum_{c \in \creatorset \cup \left\{ c_0, c_1 \right\}} u^T c - \sum_{u \in \userset} \sum_{c \in \creatorset \cup \left\{ c_1 \right\}} u^T c  \\
    &= \sum_{u \in \userset} \sum_{c \in \creatorset \cup \left\{ c_0 \right\}} u^T c - \sum_{u \in \userset} \sum_{c \in \creatorset} u^T c  \\    
    &= f\left( \userset, \creatorset \cup \left\{ c_0 \right\} \right) - f\left( \userset, \creatorset \right),
\end{align*}
which shows \eqref{eqn:submodular_fixU_appendix}.

\subsubsection{Relocation Graphs that Represent Differences Between Recommendations}
We assume $\left\lvert \creatorset \right\rvert \geq K$ in the remainder of the proof. In this section, we first quantify the intuition of capturing changes between two recommendations, by constructing multi-graphs whose vertices are creators and whose edges are labeled with users.

Recall that any recommendation can be seen as a many-to-many \textit{matching} between users and creators. We consider a matching \textit{well-defined} if it satisfies all user constraints (each user is assigned to $K$ creators that she is happy with), but not necessarily creator constraints. (For the rest of the proof, we assume any matchings we mention are well-defined unless explicitly stated.) In addition, we say a (well-defined) matching is \textit{feasible} if it also satisfies all creator constraints (audience sizes are at least $\abar$), and \textit{optimal} if it maximizes total engagement out of all feasible matchings on the same sets of users and creators.

First, we introduce the notion of relocation triplets when recommendations change from $\rec$ to $\rec'$ (e.g., by adding a creator):
\begin{definition}[Relocation triplets]  \label{def:relocation_triplet}
    Let $\rec$ and $\rec'$ be two (well-defined) matchings on the same set of users $\userset$, and let $\phi = \left\{ \phi_u: u\in \userset \right\}$ be a collection of bijections such that for each $u$, $\phi_u$ is a bijection from $\rec\left(u\right) \backslash \rec'\left(u\right)$ to $\rec'\left(u\right) \backslash \rec\left(u\right)$ (note both sets are of the same size). Then, the set of \textit{relocation triplets} w.r.t. $\rec$, $\rec'$ and $\phi$ is $\left\{ \left( c, \phi_u(c), u \right) : u\in \userset, c\in \rec\left(u\right) \backslash \rec'\left(u\right) \right\}$.    
\end{definition}
Intuitively, each relocation triplet implies one of user $u$'s recommendations changed from $c$ to $c'$. For any two matchings $\rec$ and $\rec'$, the sets of all relocation triplets are generally not unique, as there are potentially many choices of $\phi$.

We now define a graph that captures all relocations from $\rec$ to $\rec'$ when they have the same users:
\begin{definition}[Relocation graphs]  \label{def:relocation_graph}
    Let $\rec$ and $\rec'$ be two matchings on users and creators $\left( \userset, \creatorset \right)$ and $\left( \userset, \creatorset' \right)$ respectively, such that they have the same set of users. A \textit{relocation graph} $\relocg \left( \rec, \rec' \right)$ w.r.t. bijections $\phi$ is a directed weighted multi-graph $G=\left( V, E, w \right)$ with edge labels, such that $V=\creatorset \cup \creatorset'$, and every relocation triplet $\left( c, c', u \right)$ given by $\phi$ becomes an edge $e$ from $c$ to $c'$ with label $u$. The weight of an edge is given by a function defined on its vertices and edge label: $w\left( e \right) = w\left( c, c', u \right) = u^T c' - u^T c$ (the change in engagement from the old matching to the new one).
\end{definition}
At a conceptual level, a relocation graph captures the ``flow of audiences'' between creators. Each edge from $c$ to $c'$ with label $u$ means that when recommendations change from $\rec$ to $\rec'$, creator $c$ gives away one of her audiences, user $u$, to $c'$. Any creator with a greater outdegree than indegree loses some audience, while a greater indegree than outdegree means she gains audience. Note that $\relocg \left( \rec, \rec' \right)$ is not unique given $\rec$ and $\rec'$ due to non-uniqueness of relocation triplets.

By definition, two matchings with the same set of users induce a set of relocation triplets and thus a relocation graph. Conversely, given a graph and the initial matching, the final matching is induced if and only if the graph satisfies certain properties, which we call \textit{consistency}:
\begin{lemma}[Consistency of relocation graphs]  \label{lemma:relocation_graph_consistency}
    Let $\rec$ be a matching between users $\userset$ and creators $\creatorset$, and $G=\left( V, E, w \right)$ be a directed weighted multi-graph such that $\creatorset \subseteq V$, all edge labels are in $\userset$, and edge weights $w$ are given by Definition~\ref{def:relocation_graph}. Then, $G=\relocg \left( \rec, \rec' \right)$ for some well-defined matching $\rec'$ if and only if for each edge $\left( c, c', u \right) \in E$, the following conditions hold:
    \begin{enumerate}[label=C\arabic*., ref=C\arabic*]
        \item $c\in \rec\left(u\right)$.  \label{enum:relocation_graph_consistency_recom_out}
        \item $c'\notin \rec\left(u\right)$.  \label{enum:relocation_graph_consistency_recom_in}
        \item $u^T c' \geq \ebar$.  \label{enum:relocation_graph_consistency_engagement}
        \item For any other creator $c''$, $\left( c, c'', u \right) \notin E$ and $\left( c'', c', u \right) \notin E$.  \label{enum:relocation_graph_consistency_disjoint}
    \end{enumerate}
    When all conditions hold, $\rec'$ satisfies $c' \in \rec'\left(u\right)$ and $c \notin \rec'\left(u\right)$, and we say $G$ is \textit{consistent} with the given matching $\rec$.

\end{lemma}
Conceptually, $G$ being consistent with $\rec$ means that we can start from the matching $\rec$, apply changes described by edges in $G$, and obtain another matching $\rec'$. Consistency ensures $\rec'$ is well-defined. The proof of Lemma~\ref{lemma:relocation_graph_consistency} is largely by construction, and is omitted. 

We observe three more properties of relocation graphs. Firstly, the sum of all edge weights in $\relocg \left( \rec, \rec' \right)$ is the difference between total engagements of the two matchings:
\begin{align}
    \sum_{e\in E} w\left(e\right) &= \eng\left( \userset, \creatorset', \rec' \right) - \eng\left( \userset, \creatorset, \rec \right).  \label{eqn:relocation_graph_sum_weights}
\end{align}
Additionally, the audience size of creator $c$ in a matching, which we denote as $a^\rec_c$, satisfies:
\begin{align}
    a^{\rec'}_c &= a^{\rec}_c + \deg^G_{in}\left(c\right) - \deg^G_{out}\left(c\right),  \label{eqn:relocation_graph_audience_degrees}
\end{align}
where $\deg^G_{in}\left(c\right)$ and $\deg^G_{out}\left(c\right)$ are the indegrees and outdegrees of $c$ in $G$ respectively. Lastly:
\begin{lemma}  \label{lemma:relocation_graphs_vertex_disjoint_labelu}
    Given a relocation graph $G = \relocg \left( \rec, \rec' \right)$ and a user $u$, the subset of edges in $G$ with label $u$ is vertex-disjoint.
\end{lemma}
\begin{proofCharles}
    If edges with label $u$ are not vertex disjoint, there must exist creators $c, c', c''$ such that at least one of the following pairs of edges co-exist in $G$: 
    \begin{itemize}
        \item $\left( c, c', u \right)$ and $\left( c, c'', u \right)$. However, this violates the consistency of $G$.
        \item $\left( c', c, u \right)$ and $\left( c'', c, u \right)$. However, this violates the consistency of $G$. 
        \item $\left( c, c', u \right)$ and $\left( c', c'', u \right)$. However, the first edge implies $c'\notin \rec\left(u\right)$ and the second implies $c'\in \rec\left(u\right)$, which is a contradiction.
    \end{itemize}
\end{proofCharles}

\subsubsection{Structure of Relocation Graphs Between Optimal Matchings with One Creator Added}
Recall that a matching $\rec$ defined on users $\userset$ and creators $\creatorset$ is \textit{optimal} if it is feasible and maximizes total engagement out of all feasible matchings. In other words, $\eng\left( \userset, \creatorset, \rec \right) = f\left( \userset, \creatorset \right)$.

Now consider two optimal matchings whose sets of players differ by a single creator. We obtain the following property:
\begin{proposition}[Relocation graphs are decomposable into paths]  \label{prop:relocation_graph_decompose}
    Fix users $\userset \subseteq \userset_0$, $c^* \in \creatorset_0$ and $\creatorset \subseteq \creatorset_0 \backslash \left\{ c^* \right\}$. For any optimal matching $\rec$ between players $\left( \userset, \creatorset \right)$ and optimal matching $\rec^*$ between $\left( \userset, \creatorset \cup \left\{ c^* \right\} \right)$, the relocation graph $\relocg\left( \rec, \rec^* \right)$ is a directed acyclic graph (DAG) whose edge set can be decomposed into edge-disjoint paths that all end at $c^*$, without loss of generality.
\end{proposition}
Intuitively, the ``flow of audiences'' between the two \textit{optimal} matchings is such that only the new creator $c^*$ is gaining users. Some other creators will lose users, and these lost audiences will eventually ``transfer'' to $c^*$ directly or indirectly via the decomposed paths. (For example, $u_1$ may be reassigned from $c_1$ to $c_2$, then $u_2$ reassigned from $c_2$ to $c_3$, and $u_3$ from $c_3$ to $c^*$.)

\begin{proofCharles}
    We first present a useful fact: given an initial matching $\rec$ and a relocation graph $G$ that is consistent with $\rec$, any subgraph of $G$ with a subset of edges is also consistent with $\rec$. This is because if all edges in $G$ satisfy conditions in Lemma~\ref{lemma:relocation_graph_consistency}, then a subset of them also satisfy the conditions. (In other words, only doing a subset of the relocations in $G$ still ensures each user receives $K$ recommendations, and does not assign a user to a creator that she is not happy with.)

    Choose an arbitrary optimal matching $\rec^*$ and corresponding graph $G=\relocg\left( \rec, \rec^* \right)$. It suffices to show the following statements:
    \begin{enumerate}
        \item $G$ contains no cycles.  \label{enum:decompose_no_cycles}
        \item $\deg^G_{out} \left(c^* \right) = 0$.  \label{enum:decompose_c*_sink}
        \item For all vertices $c\neq c^*$, $\deg^G_{in} \left(c \right) \leq \deg^G_{out} \left(c \right)$.  \label{enum:decompose_no_outflow}
    \end{enumerate}
    As long as all three statements are true, a path decomposition can be found by repeatedly starting a traversal from any source vertex with indegree 0, finding a path with arbitrary edges until a vertex with outdegree 0 is reached (which is necessarily $c^*$), and removing these paths from the graph until all edges have been removed.

    Statement~\ref{enum:decompose_c*_sink} is true because $c^* \notin \creatorset$, so no relocation triplets start with $c^*$. To show Statement~\ref{enum:decompose_no_cycles}, suppose there exists a simple cycle $c_1, c_2, \dots, c_k$. Note that $c^*$ is not in the cycle due to Statement~\ref{enum:decompose_c*_sink}. Consider the weight of the cycle $W:=\sum_{i=1}^K w\left( c_i \right)$:
    \begin{itemize}
        \item If $W=0$, consider another graph $G'$ that removes this cycle from $G$. The new graph $G'$ is consistent with $\rec$ as it's a subgraph of $G$, so let $G'=\relocg \left( \rec, \Tilde{\rec}^* \right)$. Observe that $\Tilde{\rec}^*$ is feasible due to \eqref{eqn:relocation_graph_audience_degrees}, since a cycle does not change the net degree (indegree minus outdegree) of any vertex and thus the audience size of any creator. The sum of edge weights in $G$ and $G'$ are equal, so by \eqref{eqn:relocation_graph_sum_weights}, $\rec^*$ and $\Tilde{\rec}^*$ have the same total engagement. This means $\Tilde{\rec}^*$ is another optimal matching on $\left( \userset, \creatorset \cup \left\{ c^* \right\} \right)$ with at least one fewer zero-weight cycle than $\rec^*$, so without loss of generality, we can repeat this until $\rec^*$ contains no zero-weight cycles. (Informally, any cyclic relocation with zero impact on engagement could have been ignored.)
        \item If $W<0$, consider $G'$ with this cycle removed, similar to the previous case. The new graph $G'$ has a greater sum of edge weights than $G$, so $\eng \left( \userset, \creatorset \cup \left\{ c^* \right\}, \Tilde{\rec}^* \right) = \eng \left( \userset, \creatorset \cup \left\{ c^* \right\}, \rec^* \right) - W$, which contradicts optimality of $\rec^*$. (Informally, any cyclic relocation that reduces engagement could have been ignored to obtain even higher engagement.)
        \item If $W>0$, consider a new graph $G''$ that only contains this cycle. The new graph $G''$ is consistent with $\rec$, so let $G''=\relocg \left( \rec, \Tilde{\rec} \right)$, and note that $\Tilde{\rec}$ is feasible. Thus, \eqref{eqn:relocation_graph_sum_weights} suggests $\eng \left( \userset, \creatorset, \Tilde{\rec} \right) = \eng \left( \userset, \creatorset, \rec \right) + W$, which contradicts optimality of $\rec$. (Informally, if adding creator $c^*$ results in a cyclic relocation that improves engagement, the same cycle could have been applied before adding $c^*$ to achieve even higher engagement on $\creatorset$.)
    \end{itemize}
    To show Statement~\ref{enum:decompose_no_outflow}, suppose there exists vertex $c$ such that $\deg^G_{in} \left(c \right) > \deg^G_{out} \left(c \right)$. Then there must exist a path $\left( c_1, \dots, c_k, c \right)$, where $\deg^G_{in} \left(c_1 \right) = 0$. Note that $c^*$ must not be on the path as it has no outgoing edges. Consider a graph $G'$ with this path removed, and another graph $G''$ that only contains this path. Observe that they are both consistent with $\rec$, so let $G'=\relocg \left( \rec, \Tilde{\rec}^* \right)$ and $G''=\relocg \left( \rec, \Tilde{\rec} \right)$. Meanwhile, feasibility of $\rec^*$ means $a^{\rec^*}_{c_1} \geq \abar$, so \eqref{eqn:relocation_graph_audience_degrees} gives $a^{\rec}_{c_1} > \abar$. This means $\Tilde{\rec}$ is feasible, because compared to $\rec$, only $c_1$ loses a user but her audience size still meets her constraints. Likewise, $a^{\rec}_{c} \geq \abar$ gives $a^{\rec^*}_{c} > \abar$, and thus $\Tilde{\rec}^*$ is also feasible. Using the same argument as above, either this path can be removed from $\rec^*$ without loss of generality, or optimality of either $\rec$ or $\rec^*$ is violated. (Informally, any path that does not end at $c^*$ can either be removed from the final matching if it has negative engagement, or be added to the original matching if it has positive engagement.)
\end{proofCharles}

{}

\subsubsection{Red and Blue Graphs}
We now apply Proposition~\ref{prop:relocation_graph_decompose} to specific pairs of creator sets that differ by one creator, namely, three terms in our submodularity objective \eqref{eqn:submodular_fixU_appendix}: $\creatorset$ to $\creatorset \cup \left\{ c_1 \right\}$ to $\creatorset \cup \left\{ c_0, c_1 \right\}$. We use blue paths to denote changes from $\creatorset$ to $\creatorset \cup \left\{ c_1 \right\}$, and red paths to denote changes from $\creatorset \cup \left\{ c_1 \right\}$ to $\creatorset \cup \left\{ c_0, c_1 \right\}$. We then derive additional properties.

Formally, fix arbitrary $\userset$, $c_0$, $c_1$ and $\creatorset$ as defined in Proposition~\ref{prop:submodular_fixU}. Define the following quantities:
\begin{itemize}
    \item $\rec$, $\rec^{1}$ and $\rec^{0,1}$ are optimal matchings on $\left( \userset, \creatorset \right)$, $\left( \userset, \creatorset \cup \left\{ c_1 \right\} \right)$ and $\left( \userset, \creatorset \cup \left\{ c_0, c_1 \right\} \right)$ respectively.
    \item $G^{1} := \relocg \left( \rec, \rec^{1} \right)$ and $G^{0,1} := \relocg \left( \rec^{1}, \rec^{0,1} \right) $ are corresponding relocation graphs. Denote their edge sets with $E^1$ and $E^{0,1}$ respectively.
    \item $G=\left( \creatorset \cup \left\{ c_0,c_1 \right\}, E^1 \cup E^{0,1} \right)$ is the union of the two relocation graphs $G^1$ and $G^{0,1}$. %
\end{itemize}
Notice that $G^{1}$ deals with addition of creator $c_1$, and $G^{0,1}$ deals with addition of creator $c_0$. Additionally, the left hand side of \eqref{eqn:submodular_fixU_appendix} is the sum of edge weights in $G^{0,1}$. Proposition~\ref{prop:relocation_graph_decompose} suggests there exist choices of $\rec^{1}$ and $\rec^{0,1}$ such that $G^{1}$ and $G^{0,1}$ can both be decomposed into paths that end at $c_1$ and $c_0$ respectively. We call the paths in $G^{1}$ \textit{blue} paths, and the paths in $G^{0,1}$ \textit{red} paths. (All edges retain their original color in $G$, and it can be shown that there cannot be any duplicate edges between $G^{1}$ and $G^{0,1}$ with the same user label, so each edge $\left( c, c', u \right) \in G$ is either red or blue.)

We show a technical result about the union graph $G$, which will be relevant in the next section.
\begin{lemma}  \label{lemma:submodular_fixU_length3}
    Without loss of generality, $G$ does not contain paths of length at least 3 without duplicate vertices, whose edges share the same user as their labels (regardless of their edge colors).
\end{lemma}
\begin{proofCharles}
    We present Algorithm~\ref{alg:submodular_fixU_length3_removal}, which takes in the blue graph $G^1$ and the red graph $G^{0,1}$, and augments them such that they are still relocation graphs that represent the same matchings, and their union does not contain paths of length 3 with the same user labels.
    
    \begin{algorithm}[h!]
    \caption{Augmentation of $G^1$ and $G^{0,1}$ to remove paths of length 3}
     \label{alg:submodular_fixU_length3_removal}
    \begin{algorithmic}[1]
        \State $G^1 = \left( \creatorset \cup \left\{ c_1 \right\}, E^1 \right) \gets \relocg \left( \rec, \rec^{1} \right)$  \Comment{Blue graph}
        \State $G^{0,1} = \left( \creatorset \cup \left\{ c_0, c_1 \right\}, E^{0,1} \right) \gets \relocg \left( \rec^{1}, \rec^{0,1} \right)$  \Comment{Red graph}
        \For{user $u \in \userset$}
            \State $E_u = \left\{ e\in E^1 \cup E^{0,1} : e \text{ has label } u \right\}$  \Comment{Subgraph of all blue and red edges with label $u$} \label{algline:length3_removal_Eu}
            \While{$E_u$ has paths of length 3}
                \State $\left( c_p, c_q, c_r, c_s \right) \gets $ directed path of length 3 in $E_u$  \label{algline:length3_removal_path}
                \If{$\left( c_p, c_q, u \right)$ is blue}
                    \State $E^1 \gets E^1 \backslash \left\{ \left( c_p, c_q, u \right), \left( c_r, c_s, u \right) \right\}$
                    \State $E^1 \gets E^1 \cup \left\{ \left( c_p, c_s, u \right), \left( c_r, c_q, u \right) \right\}$  \Comment{Swap end points, keep original colors}
                \Else
                    \State $E^{0,1} \gets E^{0,1} \backslash \left\{ \left( c_p, c_q, u \right), \left( c_r, c_s, u \right) \right\}$
                    \State $E^{0,1} \gets E^{0,1} \cup \left\{ \left( c_p, c_s, u \right), \left( c_r, c_q, u \right) \right\}$
                \EndIf
            \EndWhile
        \EndFor
        \State \Return{modified $G^1$ and $G^{0,1}$}
    \end{algorithmic}
    \end{algorithm}
    In summary, for every path $\left( c_p, c_q, c_r, c_s \right)$ of length 3 with user label $u$, Algorithm~\ref{alg:submodular_fixU_length3_removal} reconnects $c_p$ to $c_s$ and $c_r$ to $c_q$. This can be done because conceptually, the effects of these two edges are that part of user $u$'s recommendations changed from $\left\{ c_p, c_r \right\}$ to $\left\{ c_q, c_s \right\}$, but it does not matter whether the ``slot'' for $c_p$ is replaced by $c_q$ or $c_s$. This reassignment turns the path into a cycle $c_q \rightarrow c_r \rightarrow c_q$ and a single edge $c_p \rightarrow c_s$.

    To formally prove the correctness of Algorithm~\ref{alg:submodular_fixU_length3_removal}, let $\Tilde{G}^1$ and $\Tilde{G}^{0,1}$ denote the augmented blue and red graphs, with modified edges $\Tilde{E}^1$ and $\Tilde{E}^{0,1}$. We first show that in the augmented union graph $\Tilde{G}$ with edges $\Tilde{E} := \Tilde{E}^1 \cup \Tilde{E}^{0,1}$, no paths of length 3 with the same user labels exist. This is equivalent to showing the set $\Tilde{E}_u := \left\{ e\in \Tilde{E} : e \text{ has label } u \right\}$ contains no paths of length 3.

    Lemma~\ref{lemma:relocation_graphs_vertex_disjoint_labelu} implies that in the initial $E_u$ on Line~\ref{algline:length3_removal_Eu} (before any augmentation), all blue edges are vertex-disjoint, and so are all red edges. Thus, for any $\left( c_p, c_q, c_r, c_s \right)$ identified on Line~\ref{algline:length3_removal_path}, $\left( c_p, c_q \right)$ and $\left( c_r, c_s \right)$ must be of the same color while $\left( c_q, c_r \right)$ is of the opposite color. Without loss of generality, suppose $\left( c_p, c_q \right)$ is blue and $\left( c_r, c_s \right)$ is red.

    Consider the set of edges in $E_u$ that are incident on $c_q$. Before this iteration of the while loop, $\left( c_p, c_q \right)$ is the only such blue edge (due to Lemma~\ref{lemma:relocation_graphs_vertex_disjoint_labelu}), and $\left( c_q, c_r \right)$ is the only red edge. This means after the iteration, $\left( c_r, c_q \right)$ is $c_q$'s only incident blue edge in $E_u$ and $\left( c_q, c_r \right)$ remains its only red edge. The same applies to $c_r$; therefore, the pair of vertices $\left( c_q, c_r \right)$ forms a loop and becomes detached from the rest of $E_u$. This implies that after each iteration of the while loop, the number of vertices that are in paths of length 3 is reduced by at least 2, so the while loop eventually terminates and the final $\Tilde{E}_u$ has no paths of length 3.

    Now we show $\Tilde{G}^1$ and $\Tilde{G}^{0,1}$ are still relocation graphs for the same matchings as before: $\Tilde{G}^1 = \relocg \left( \rec, \rec^{1} \right)$ and $\Tilde{G}^{0,1} = \relocg \left( \rec^{1}, \rec^{0,1} \right)$. (Recall that relocation triplets, and thus relocation graphs, are not unique and depend on the bijections $\phi$ between differences in recommendations for each user $u$ (Definition~\ref{def:relocation_triplet}), so both $G^1$ and $\Tilde{G}^1$ can be relocation graphs $\relocg \left( \rec, \rec^{1} \right)$ for the same matchings.)

    Consider the blue graphs $G^1$ and $\Tilde{G}^1$. For a fixed $u$, let $\creatorset_u^+$ be all vertices with an incoming blue edge in the original $E_u$ (creators who gain user $u$), and $\creatorset_u^-$ be all vertices with an outgoing blue edge (creators who lose user $u$). The original bijection $\phi_u$, which gives relocation triplets that define the relocation graph $G^1$, maps $\creatorset_u^-$ to $\creatorset_u^+$. Notice that Algorithm~\ref{alg:submodular_fixU_length3_removal} does not change the ``blue'' indegree and outdegree of each vertex, so $\creatorset_u^+$ and $\creatorset_u^-$ remain the same under $\Tilde{E}_u$. In addition, all blue edges in $\Tilde{E}_u$ are still vertex-disjoint. This means we can define another bijection $\Tilde{\phi}_u$, such that $\Tilde{\phi}_u \left( c \right) = c'$ if and only if $\left( c, c', u \right) \in \Tilde{E}_u$. The collection of bijections $\Tilde{\phi} = \left\{ \Tilde{\phi}_u : u\in \userset \right\}$ still gives the same final matching $\rec^1$ when applied to the initial matching $\rec$, and all its relocation triplets correspond exactly to the new graph $\Tilde{G}^1$, so $\Tilde{G}^1 = \relocg \left( \rec, \rec^{1} \right)$. The statement on $\Tilde{G}^{0,1}$ can be shown identically.
\end{proofCharles}

\subsubsection{Junctions Between Blue and Red Paths}  %
Given the blue graph $G^1$, red graph $G^{0,1}$ and their union $G$, at a high level, we want to apply all red paths (each of which contributes one user to creator $c_0$) to the baseline instance $\left( \userset, \creatorset \right)$ and its optimal matching $\rec$. However, not all red edges can be applied directly to $\rec$, because the red graph $G^{0,1}$ is consistent with $\rec^1$ but not necessarily with $\rec$. In this section, we define the notion of a \textit{junction} to quantify situations in which a red edge cannot be applied directly to $\rec$.

\begin{definition}[Junctions]  \label{def:junction}
    Given a vertex $c \in \creatorset \cup \left\{ c_1 \right\}$ and a user $u$, we call $c$ a \textit{junction} w.r.t. $u$ if one of the following holds:
    \begin{enumerate}
        \item There exists an incoming blue edge $\left( c', c, u \right)$ and an outgoing red edge $\left( c, c'', u \right)$ both with label $u$. In this case, we call $c$ a \textit{blue-red junction}.
        \item There exists an incoming red edge $\left( c', c, u \right)$ and an outgoing blue edge $\left( c, c'', u \right)$ both with label $u$. In this case, we call $c$ a \textit{red-blue junction}.
    \end{enumerate}
\end{definition}
Conceptually, both blue-red and red-blue junctions describe \textit{prerequisites} for the red edge, as they indicate that if we want to construct a new set of relocations from $\rec$, the red edge cannot be added unless the blue edge is. Specifically:
\begin{itemize}
    \item When $c$ is a blue-red junction, the incoming blue edge $\left( c', c, u \right)$ implies $c\notin \rec\left(u\right)$, so the outgoing red edge $\left( c, c'', u \right)$ alone will not be consistent with $\rec$ without the incoming blue edge. The blue edge is needed to first ``supply'' creator $c$ to user $u$'s recommendations, so that the red edge can then reassign that recommendation to $c''$.
    \item When $c$ is a red-blue junction, the outgoing blue edge $\left( c, c'', u \right)$ implies $c\in \rec\left(u\right)$, so the incoming red edge $\left( c', c, u \right)$ alone will not be consistent with $\rec$ without the outgoing blue edge. The blue edge is needed to first divert creator $c$ away from user $u$'s recommendations, and ``make room'' in $\rec\left(u\right)$ such that $c$ can be assigned to $u$ again, which the red edge then does.    
\end{itemize}
We observe three more properties of junctions:
\begin{enumerate}[label=J\arabic*., ref=J\arabic*]
    \item If $c$ is a junction w.r.t. $u$, there can only be one unique blue edge in $G^1$ and one unique red edge in $G^{0,1}$ that satisfy Definition~\ref{def:junction} (due to Lemma~\ref{lemma:relocation_graphs_vertex_disjoint_labelu}). Thus, the blue and red paths that contain them are also uniquely given by $u$. We say these two edges and two paths are \textit{involved} in the junction $c$ w.r.t. $u$.  \label{enum:junction_unique_edges}
    \item For any edge $e$, a fixed vertex $c$ can be either a blue-red junction that involves $e$ or a red-blue junction that involves $e$, but not both. In addition, there can be at most one edge $e'$ of the opposite color such that $c$ is a junction that involves both $e$ and $e'$. This is due to Property~\ref{enum:junction_unique_edges} and the fact that such a junction at $c$ must be w.r.t. the same user that $e$ is labeled with.\footnote{Note that replacing edges with paths makes Property~\ref{enum:junction_one_per_kind_per_edge} no longer true: for any path $p$ instead of a single edge $e$, $c$ can be both a blue-red junction that involves $p$ and $p'$, and a red-blue junction that involves $p$ and $p''$. This means both the incoming and outgoing edge in $p$ are used to define the two different junctions. In this case, there is at most one such $p'$ and at most one such $p''$, and these two paths may be the same or different.}  \label{enum:junction_one_per_kind_per_edge}
    \item There cannot be two consecutive junctions with respect to the same user: that is, for any edge $\left( c, c', u \right)$ of any color, at most one of $c$ and $c'$ can be a junction w.r.t. $u$, without loss of generality. This is due to Lemma~\ref{lemma:submodular_fixU_length3}, because otherwise $G$ contains a path of length 3 with user label $u$, with the middle edge being $\left( c, c', u \right)$.  \label{enum:junction_no_two_consecutive}
\end{enumerate}

\subsubsection{Constructing a Matching that Satisfies Submodularity}
We present Algorithm~\ref{alg:submodular_fixU_backward_tracing}, the ``backward tracing'' algorithm, that constructs another graph $G^{0}$. We will show that it is a consistent relocation graph $\relocg \left( \rec, \rec^{0} \right)$ that adds creator $c_0$ to the baseline of $\left( \userset, \creatorset \right)$, with sum of weights at least as much as $G^{0,1}$. This induces matching $\rec^{0}$ between players $\left( \userset, \creatorset \cup \left\{ c_0 \right\} \right)$.

\begin{algorithm}[h!]
\caption{Backward tracing algorithm that constructs $G^{0}$}
 \label{alg:submodular_fixU_backward_tracing}
\begin{algorithmic}[1]
    \State $G^{1} = \left( \creatorset \cup \left\{ c_1 \right\}, E^1 \right) \gets \relocg \left( \rec, \rec^{1} \right)$  \Comment{Blue graph}
    \State $G^{0,1} = \left( \creatorset \cup \left\{ c_0, c_1 \right\}, E^{0,1} \right) \gets \relocg \left( \rec^{1}, \rec^{0,1} \right)$  \Comment{Red graph}
    \State $G \gets \left( \creatorset \cup \left\{ c_0,c_1 \right\}, E^1 \cup E^{0,1} \right)$  \Comment{Union graph}
    \State Add auxiliary vertex $c_X$ to $G$
    \For{every red and blue path $p$, whose first vertex is denoted $c_p$}
        \State Add auxiliary edge $\left( c_X, c_p, \emptyset \right)$ to $G$ as the new first edge in $p$, with the same color as $p$
    \EndFor  \Comment{Multiple auxiliary edges between $c_X$ and $c$ are allowed}

    \State $E^0 \gets \emptyset$  \Comment{Sets of chosen edges to be in the new graph $G^0$}

    \State
    \Function{Prev}{$e$}  \Comment{When traversing edge $e$, return a preceding edge to be traversed next}
        \State $c \gets$ origin of edge $e$
        \State $E^{in}_b, E^{in}_r \gets $ sets of all blue and red incoming edges into $c$ in $G$, respectively
        \If{$\left( E^{in}_b \cap E^0 \right) \neq E^{in}_b $}  \Comment{First $\left| E^{in}_b \right|$ visits to $c$: always go to a blue incoming edge}
            \If{$e$ is blue, and its preceding edge $e_{prev}$ in $p$ is not in $E^0$}
                \State \Return{$e_{prev}$}  \label{algline:backward_tracing_prev_blue_preceding}
            \ElsIf{$e$ is red, and $c$ is a blue-red junction that involves $e$}
                \If{$e_b$, the blue edge in this junction, is not in $E^0$}
                    \State \Return{$e_b$}  \label{algline:backward_tracing_prev_bluered_junction}
                \EndIf
            \EndIf
            \State \Return{any blue edge in $\left( E^{in}_b \backslash E^0 \right)$}
        \Else  \Comment{Subsequent $\left| E^{in}_r \right|$ visits to $c$: always go to a red incoming edge}
            \State \Return{a red edge $e_r$ in $\left( E^{in}_r \backslash E^0 \right)$, such that its succeeding edge in its path is in $E^0$, and if $c$ is a red-blue junction that involves $e_r$, the blue edge in this junction is also in $E^0$}  \label{algline:backward_tracing_prev_red}
        \EndIf
    \EndFunction
    \State
    
    \While{$c_0$ has an incoming red edge that is not in $E^0$}  \label{algline:backward_tracing_outer_while}
        \State $e = \left( c, c_0, u \right) \gets$ red edge in $\left( E^{0,1} \backslash E^0 \right)$ that ends at $c_0$
        \State $E^0 \gets E^0 \cup \left\{ e \right\}$
        \While{$e$ does not originate from $c_X$}  \Comment{Traversing $e$}    \label{algline:backward_tracing_traversal_while}
            \State $e \gets$ \Call{Prev}{$e$}  \label{algline:backward_tracing_traversal_prev_call}
            \State $E^0 \gets E^0 \cup \left\{ e \right\}$
        \EndWhile  \label{algline:backward_tracing_traversal_endwhile}
    \EndWhile  \label{algline:backward_tracing_outer_endwhile}
    \State Remove all auxiliary edges from $E^0$  \label{algline:backward_tracing_remove_auxiliary}
    \While{$E^0$ contains edges $\left( c', c, u \right)$ and $\left( c, c'', u \right)$ for some user $u$ and creators $c, c', c''$}  \label{algline:backward_tracing_compress_while}
        \State $E^0 \gets E^0 \backslash \left\{ \left( c', c, u \right), \left( c, c'', u \right) \right\} $  \Comment{Remove conflicting junction edges}  \label{algline:backward_tracing_compress_remove}
        \State $E^0 \gets E^0 \cup \left\{ \left( c', c'', u \right) \right\}$, \text{colored \textbf{purple}}  \Comment{Add new edge that bypasses the junction}  \label{algline:backward_tracing_compress_add}
    \EndWhile
    \State \Return{$G^0 = \left( \creatorset \cup \left\{ c_0 \right\}, E^0 \right)$}
\end{algorithmic}
\end{algorithm}

Conceptually, Algorithm~\ref{alg:submodular_fixU_backward_tracing} makes repeated backward traversals that starts at an red edge into $c_0$, the creator that we want to add. During each traversal (inner loop on Line~\ref{algline:backward_tracing_traversal_while}), it goes backwards and sequentially adds unvisited edges that precede one another, prioritizing blue edges over red edges whenever available. It also ensures that the chosen edges ``resolve'' all junctions: each red edge in a junction is only added if its corresponding blue edge is. Each traversal stops when the source of an original blue or red path is reached (indicated with an auxiliary edge), and the entire process stops when all incoming edges into $c_0$ have been visited. The algorithm terminates in finite time, as each edge in $G$ is traversed at most once.

We first show key properties about the choice of preceding edges in \textsc{Prev}$\left(e\right)$:
\begin{lemma}  \label{lemma:submodular_fixU_alg_prev_function}
    After each iteration of the outer while loop on Line~\ref{algline:backward_tracing_outer_while} in Algorithm~\ref{alg:submodular_fixU_backward_tracing}, the current edge set $E^0$ satisfies the following properties:
    \begin{enumerate}[label=(\alph*)]
        \item For any $c\in \creatorset \cup \left\{ c_1 \right\}$, its indegree and outdegree in $E^0$ are equal: $\deg^{E^0}_{in} \left(c\right) = \deg^{E^0}_{out} \left(c\right)$. In addition, $\deg^{E^0}_{in} \left(c_X\right) = \deg^{E^0}_{out} \left(c_0\right) = 0$, and $\deg^{E^0}_{out} \left(c_X\right) = \deg^{E^0}_{in} \left(c_0\right)$.  \textbf{(``Flow conservation'')} \label{enum:submodular_prev_flow_conservation}
        \item For every junction $c$ that involves blue edge $e_b$ and red edge $e_r$, $e_r\in E^0$ implies $e_b \in E^0$. \textbf{(``Junctions are resolved'')}  \label{enum:submodular_prev_junctions_resolved}
        \item If a red edge is in $E^0$ and it is not the last edge in its red path $r$, its succeeding edge in $r$ must also be in $E^0$. \textbf{(``Red tails'')}  \label{enum:submodular_prev_red_tails}
        \item If a blue edge is in $E^0$ and it is not the first edge in its blue path $b$, its preceding edge in $b$ must also be in $E^0$. \textbf{(``Blue heads'')}  \label{enum:submodular_prev_blue_heads}
    \end{enumerate}
    In particular, whenever Line~\ref{algline:backward_tracing_prev_red} is reached, at least one red edge in $\left( E^{in}_r \backslash E^0 \right)$ can be chosen that satisfies these properties.
\end{lemma}
\begin{proofCharles}
    Property~\ref{enum:submodular_prev_flow_conservation} is easy to see, as the inner loop on Line~\ref{algline:backward_tracing_traversal_while} repeatedly traverses edges that precede one another. This means a single iteration of the outer loop on Line~\ref{algline:backward_tracing_outer_while} adds to $E^0$ a ``mixed-color path'' that starts at $c_X$ and ends at $c_0$ (which may contain duplicate vertices), so flows on all intermediate vertices are preserved.

    We break down Properties~\ref{enum:submodular_prev_junctions_resolved}-\ref{enum:submodular_prev_blue_heads} by vertices. Fix a vertex $c\notin \left\{ c_X, c_0 \right\}$, and define $E^{in}_b = \left\{ e^{in}_{b,1}, \dots, e^{in}_{b,n} \right\}$, $E^{in}_r = \left\{ e^{in}_{r,1}, \dots, e^{in}_{r,m} \right\}$ to be the sets of all blue and red incoming edges into $c$ in $G$, and $E^{out}_b = \left\{ e^{out}_{b,1}, \dots, e^{out}_{b,n'} \right\}$, $E^{out}_r = \left\{ e^{out}_{r,1}, \dots, e^{out}_{r,m} \right\}$ to be the sets of all blue and red outgoing edges from $c$. Note that $\left| E^{in}_r \right| = m = \left| E^{out}_r \right|$, and $\left| E^{in}_b \right| = n \geq n' = \left| E^{out}_b \right|$, with $n'=0$ if $c=c_1$ and $n'=n$ otherwise. This is because after adding auxiliary edges, the source of every blue and red path is $c_X$, while all red paths have sink $c_0$ and all blue paths have sink $c_1$. We number the edges according to the paths they belong to: that is, for all $1\leq i\leq n'$, $e^{in}_{b,i}$ and $e^{out}_{b,i}$ belong in the same path, and red edges are numbered similarly.
    
    Then, Properties~\ref{enum:submodular_prev_junctions_resolved}-\ref{enum:submodular_prev_blue_heads} are satisfied if the following are true for all such $c$ and all $1\leq i\leq n$, $1\leq j\leq m$, after each time an incoming edge into $c$ has been added to $E^0$:
    \begin{enumerate}[label=(b\arabic*)]
        \item If $c$ is a blue-red junction that involves $e^{in}_{b,i}$ and $e^{out}_{r,j}$, $e^{out}_{r,j}\in E^0$ implies $e^{in}_{b,i} \in E^0$.  \label{enum:submodular_prev_junctions_resolved_c1}
        \item If $c$ is a red-blue junction that involves $e^{in}_{r,j}$ and $e^{out}_{b,j}$, $e^{in}_{r,j}\in E^0$ implies $e^{out}_{b,i} \in E^0$.  \label{enum:submodular_prev_junctions_resolved_c2}
    \end{enumerate}
    \begin{enumerate}[label=(\alph*),start=3]
        \item $e^{in}_{r,j}\in E^0$ implies $e^{out}_{r,j}\in E^0$.  \label{enum:submodular_prev_red_tails_c}
        \item $e^{out}_{b,i}\in E^0$ implies $e^{in}_{b,i}\in E^0$, if $i\leq n'$.  \label{enum:submodular_prev_blue_heads_c}
    \end{enumerate}

    We show that these properties are maintained after each visit to $c$ (i.e., after both an outgoing edge and an incoming edge have been added to $E^0$). During the $t^{\text{th}}$ visit to $c$, an outgoing edge $e^{out}$ has been added, while \textsc{Prev}$\left( e^{out} \right)$ is called to return an incoming edge to be added next. At the time of this call, $t-1$ incoming edges and $t$ outgoing edges are in $E^0$.

    First, suppose $t\leq n$. These must be when $E^{in}_b \cap E^0 \neq E^{in}_b $, so \textsc{Prev} always returns a blue incoming edge; thus, $E^{in}_r \cap E^0 = \emptyset$, so \ref{enum:submodular_prev_junctions_resolved_c2} and \ref{enum:submodular_prev_red_tails_c} are automatically satisfied. If $e^{out}$ is blue, \ref{enum:submodular_prev_blue_heads_c} must be satisfied after this iteration because of Line~\ref{algline:backward_tracing_prev_blue_preceding}; if $e^{out}$ is red, \ref{enum:submodular_prev_junctions_resolved_c1} must be satisfied after this iteration because of Line~\ref{algline:backward_tracing_prev_bluered_junction}.\footnote{There can only be at most one such blue edge $e_b$ on Line~\ref{algline:backward_tracing_prev_bluered_junction} that is in a blue-red junction with $e^{out}$ at $c$, due to Property~\ref{enum:junction_one_per_kind_per_edge}.} In both cases, the required edge is either already in $E^0$ or the return value of \textsc{Prev}$\left( e^{out} \right)$.

    Now suppose $t>n$, when \textsc{Prev} always returns a red incoming edge via Line~\ref{algline:backward_tracing_prev_red}. Since all of $E^{in}_b$ are in $E^0$, \ref{enum:submodular_prev_junctions_resolved_c1} and \ref{enum:submodular_prev_blue_heads} are automatically satisfied; in addition, an edge that meets the requirements on Line~\ref{algline:backward_tracing_prev_red} also satisfies \ref{enum:submodular_prev_junctions_resolved_c2} and \ref{enum:submodular_prev_red_tails_c}. We now show there is always a red incoming edge that satisfy the two requirements.

    Note that $t-n-1$ edges in $E^{in}_r$ are in $E^0$, so the other $m-\left(t-n\right)+1$ edges are still available. Denote $x:=m-\left(t-n\right)+1$, and the available edges as $E^{in}_r \backslash E^0 = \left\{ e^{in}_{r,\pi\left( 1 \right)}, \dots, e^{in}_{r,\pi\left( x \right)} \right\}$. Define disjoint prerequisite sets $S_k$ such that 
    $$ S_k = \begin{cases}
        \left\{ e^{out}_{r,\pi\left( k \right)}, e^{out}_{b,l} \right\}, & \text{if $c$ is a red-blue junction that involves $e^{in}_{r,\pi\left( k \right)}$ and $e^{out}_{b,l}$}  \\
        \left\{ e^{out}_{r,\pi\left( k \right)} \right\}, & \text{if $c$ is not a red-blue junction that involves $e^{in}_{r,\pi\left( k \right)}$}
    \end{cases}, \quad \text{for all } k=1,\dots,x. $$
    In other words, $S_k$ contains all outgoing edges that need to be in $E^0$ in order for $e^{in}_{r,\pi\left( k \right)}$ to meet the requirements. Suppose $y$ of these $S_k$'s have size 2 and the other $x-y$ of them have size 1.\footnote{There can only be at most one such blue edge $e^{out}_{b,l}$ that is in a red-blue junction with $e^{in}_{r,\pi\left( k \right)}$ at $c$, due to Property~\ref{enum:junction_one_per_kind_per_edge}. Thus, $\left|S_k\right| \leq 2$.} Denoting $E^{out} := \left( E^{out}_r \cup E^{out}_b \right)$ as the set of all outgoing edges, we get $\left| E^{out} \right| = m+n' \leq m+n$, $\left| E^{out} \cap E^0 \right| = t$, $\bigcup_{k=1}^{x} S_k \subseteq E^{out}$, and $\left| \bigcup_{k=1}^{x} S_k \right| = x+y$. Thus,
    \begin{align*}
        \left| \left( \bigcup_{k=1}^{x} S_k \right) \cap E^0 \right| &= \left| \left( \bigcup_{k=1}^{x} S_k \right) \cap \left( E^{out} \cap E^0 \right) \right|  \\
        &= \left| \bigcup_{k=1}^{x} S_k \right| + \left| E^{out} \cap E^0 \right| - \left| \left( \bigcup_{k=1}^{x} S_k \right) \cup \left( E^{out} \cap E^0 \right) \right|  \\
        &\geq \left| \bigcup_{k=1}^{x} S_k \right| + \left| E^{out} \cap E^0 \right| - \left| E^{out} \right|  \\
        &\geq \left( x+y \right) + t - \left( m+n \right)  \\
        &= y+1.
    \end{align*}
    This means at least $y+1$ edges among all $S_k$'s have been added to $E^0$. Since there are only $y$ sets of size 2, at least one $S_{k^*}$ must be fully added to $E^0$. Thus, $e^{in}_{r,\pi\left( k^* \right)}$ can be chosen as \textsc{Prev}$\left( e^{out} \right)$ to satisfy \ref{enum:submodular_prev_junctions_resolved_c2} and \ref{enum:submodular_prev_red_tails_c}, which completes the proof.
\end{proofCharles}

We now show properties of the output graph $G^0$:
\begin{proposition}  \label{prop:submodular_fixU_alg_output}
    Suppose the blue graph $G^1$ and the red graph $G^{0,1}$ can both be decomposed into paths, such that their union does not contain paths of length 3 with the same user labels (holds without loss of generality due to Lemma~\ref{lemma:submodular_fixU_length3}). The output graph $G^0$ of Algorithm~\ref{alg:submodular_fixU_backward_tracing} satisfies:
    \begin{enumerate}
        \item $G^0$ is consistent with $\rec$, and thus, is a relocation graph $G^0 = \relocg \left( \rec, \rec^0 \right)$ for a matching $\rec^0$.  \label{enum:submodular_alg_consistent}
        \item $\rec^0$ is a feasible matching between users $\userset$ and creators $\creatorset \cup \left\{ c_0 \right\}$.  \label{enum:submodular_alg_feasible}
        \item The sum of edge weights in $G^0$ is at least that of $G^{0,1}$.  \label{enum:submodular_alg_weights}
    \end{enumerate}
\end{proposition}
Below is a proof sketch of Proposition~\ref{prop:submodular_fixU_alg_output}:
\begin{itemize}
    \item Statement~\ref{enum:submodular_alg_consistent} (consistency): Any possible inconsistencies must involve edges in junctions, but Lemma~\ref{lemma:submodular_fixU_alg_prev_function}\ref{enum:submodular_prev_junctions_resolved} ensures that all prerequisites identified by junctions have been met.
    \item Statement~\ref{enum:submodular_alg_feasible} (feasibility): The only creators who lose users must be incident on auxiliary edges (due to Lemma~\ref{lemma:submodular_fixU_alg_prev_function}\ref{enum:submodular_prev_flow_conservation}), but that implies they are sources of blue and red paths who lose at least as many users in corresponding relocation graphs, so they must have excess audience to spare in $\rec$.
    \item Statement~\ref{enum:submodular_alg_weights} (improvement in edge weights): Lemma~\ref{lemma:submodular_fixU_alg_prev_function}\ref{enum:submodular_prev_red_tails} and \ref{lemma:submodular_fixU_alg_prev_function}\ref{enum:submodular_prev_blue_heads} allow us to pair each red path $r$ with a blue path $b$ that shares a common vertex $c$, such that the set of chosen edges in $r$ is its ``tail'' after $c$ and the set of chosen edges in $b$ is its ``head'' before $c$. We then show that the head of $b$ (chosen in $G^0$) must have weight at least that of the head of $r$ (in $G^{0,1}$). If not, optimality of either $\rec$ or $\rec^1$ must be violated, as we can augment the blue graph $G^1$ by replacing part of the head of $b$ with part of the head of $r$, maintaining consistency without reducing the sum of edge weights from $G^1$.
\end{itemize}

\begin{proofCharles}
    In this proof, we use $\Tilde{E}^0$ to denote the intermediate value of the edge set $E^0$ on Line~\ref{algline:backward_tracing_outer_endwhile} after the outer while loop ends, and use $E^0$ to denote its final value at the end of the algorithm. Compared to $E^0$, $\Tilde{E}^0$ contains auxiliary edges and may contain both edges that are involved in junctions. Note that Lemma~\ref{lemma:submodular_fixU_alg_prev_function} applies to $\Tilde{E}^0$ but not necessarily $E^0$.
    \\~\\
    \textbf{Statement~\ref{enum:submodular_alg_consistent} (consistency):} We show all consistency conditions in Lemma~\ref{lemma:relocation_graph_consistency}. That is, any edge $\left( c, c', u \right) \in E^0$ satisfies: $c\in \rec\left(u\right)$ (Condition~\ref{enum:relocation_graph_consistency_recom_out}), $c'\notin \rec\left(u\right)$ (Condition~\ref{enum:relocation_graph_consistency_recom_in}), $u_u^T c_{c'} \geq \ebar$ (Condition~\ref{enum:relocation_graph_consistency_engagement}), and for any other $c''$, $\left( c, c'', u \right) \notin E^0$ and $\left( c'', c', u \right) \notin E^0$ (Condition~\ref{enum:relocation_graph_consistency_disjoint}).

    Suppose $e=\left( c, c', u \right) \in E^0$. We first show Conditions~\ref{enum:relocation_graph_consistency_recom_out} and \ref{enum:relocation_graph_consistency_recom_in}. If $e$ is blue, both conditions are given by its consistency in $G^1$. If $e$ is red, its consistency in $G^{0,1}$ gives $c\in \rec^1\left(u\right)$ and $c'\notin \rec^1\left(u\right)$. To show that they imply Conditions~\ref{enum:relocation_graph_consistency_recom_out} and \ref{enum:relocation_graph_consistency_recom_in}:
    \begin{itemize}
        \item (Condition~\ref{enum:relocation_graph_consistency_recom_out}) Suppose for contradiction that $c\notin \rec\left(u\right)$. Since $c\in \rec^1\left(u\right)$, there must be a blue edge $e_b=\left( c'', c, u \right)$ for some $c''$.\footnote{To see this, since $c\notin \rec\left(u\right)$ and $c\in \rec^1\left(u\right)$, $c$ must appear in the range of any bijection $\phi_u: \rec\left(u\right) \backslash \rec^1\left(u\right) \mapsto \rec^1\left(u\right) \backslash \rec\left(u\right)$ that describes relocations from $\rec$ to $\rec^1$ (Definition~\ref{def:relocation_triplet}). Thus, in any set of relocation triplets and thus any (blue) relocation graphs from $\rec$ to $\rec^1$, there must be exactly one vertex $c'' = \phi^{-1}_u \left(c\right)$, so $\left( c'', c, u \right)$ is a relocation triplet and thus a blue edge.} This means $c$ is a blue-red junction w.r.t. $u$, with blue incoming edge $e_b$ and red outgoing edge $e$. Since $e\in E^0$, $e\in \Tilde{E}^0$, so Lemma~\ref{lemma:submodular_fixU_alg_prev_function}\ref{enum:submodular_prev_junctions_resolved} implies $e_b\in \Tilde{E}^0$. However, this means $e_b$ and $e$ would have been combined in the final loop on Line~\ref{algline:backward_tracing_compress_while} to form a purple edge,\footnote{Note that $e$ cannot be an auxiliary edge, because it originates from a junction $c$, while all auxiliary edges originate from $c_X$ which is not a junction; and $e_b$ cannot be an auxiliary edge, because it shares the same user label as $e$, while auxiliary edges have no user labels.} which contradicts $e\in E^0$. Thus, $c\in \rec\left(u\right)$.
        \item (Condition~\ref{enum:relocation_graph_consistency_recom_in}) Suppose for contradiction that $c'\in \rec\left(u\right)$. Since $c'\notin \rec^1\left(u\right)$, there must be a blue edge $e_b=\left( c', c'', u \right)$ for some $c''$. This means $c'$ is a red-blue junction w.r.t. $u$, with red incoming edge $e$ and blue outgoing edge $e_b$. Since $e\in E^0$, $e\in \Tilde{E}^0$, so Lemma~\ref{lemma:submodular_fixU_alg_prev_function}\ref{enum:submodular_prev_junctions_resolved} implies $e_b\in \Tilde{E}^0$. However, this means $e_b$ and $e$ would have been combined in the final loop to form a purple edge, which contradicts $e\in E^0$. Thus, $c'\in \rec\left(u\right)$.
    \end{itemize}
    If $e$ is purple, it must have been combined from blue or red\footnote{Neither of these two edges can be purple, as that would imply at least two consecutive junctions w.r.t. $u$, which violates Property~\ref{enum:junction_no_two_consecutive}.} edges $\left( c, c_j, u \right)$ and $\left( c_j, c', u \right)$ in $\Tilde{E}^0$. This means $c_j$ is a junction w.r.t. $u$. Moreover, neither $c$ nor $c'$ can be junctions w.r.t. $u$ (Property~\ref{enum:junction_no_two_consecutive}).
    Consider the type of junction that $c_j$ is w.r.t. $u$:
    \begin{itemize}
        \item If $c_j$ is a blue-red junction, consistency of the blue incoming edge $\left( c, c_j, u \right)$ gives $c\in \rec\left(u\right)$. Consistency of the red outgoing edge $\left( c_j, c', u \right)$ gives $c' \notin \rec^{1}\left(u\right)$. This must imply $c' \notin \rec\left(u\right)$, because otherwise there must be a blue edge $\left( c', c'', u \right)$ for some $c''$, which contradicts that $c'$ is not a junction w.r.t. $u$. 
        \item If $c_j$ is a red-blue junction, consistency of the blue outgoing edge $\left( c_j, c', u \right)$ gives $c' \notin \rec\left(u\right)$. Consistency of the red incoming edge $\left( c, c_j, u \right)$ gives $c\in \rec^1\left(u\right)$. This must imply $c\in \rec\left(u\right)$, because otherwise there must be a blue edge $\left( c'', c, u \right)$ for some $c''$, which contradicts that $c$ is not a junction w.r.t. $u$. 
    \end{itemize}
    Condition~\ref{enum:relocation_graph_consistency_engagement} is easy to see for edge $e$ of any color. If $e$ is blue or red, it is given by its consistency in the original relocation graph. If $e$ is purple and combined from edges $\left( c, c_j, u \right)$ and $\left( c_j, c', u \right)$, it is given by consistency of the outgoing edge $\left( c_j, c', u \right)$.

    Finally, we show Condition~\ref{enum:relocation_graph_consistency_disjoint}. Suppose for contradiction that $e=\left( c, c', u \right) \in E^0$ and $e'=\left( c, c'', u \right) \in E^0$ share the same origin $c$ and the same user label $u$. If any of them is purple, there must be another red or blue edge $\left( c, c_j, u \right) \in \Tilde{E}^0$ that was combined to make the purple edge. Thus, in all cases without loss of generality, there must exist red or blue edges $e=\left( c, c', u \right)$ and $e'=\left( c, c'', u \right)$ in the union graph $G$.

    Using consistency Condition~\ref{enum:relocation_graph_consistency_disjoint} on relocation graphs $G^1$ and $G^{0,1}$ means $e$ and $e'$ must be of different color. Without loss of generality, suppose $e$ is blue and $e'$ is red. This means $c\in \rec\left(u\right)$ (given by $e$) and $c\in \rec^1\left(u\right)$ (given by $e'$). However, this means that creator $c$ cannot appear in any bijection $\phi_u: \rec\left(u\right) \backslash \rec^1\left(u\right) \mapsto \rec^1\left(u\right) \backslash \rec\left(u\right)$ (Definition~\ref{def:relocation_triplet}), and thus cannot appear in any relocation triplets and relocation graphs from $\rec$ to $\rec^1$. This violates the existence of $e$ in the blue relocation graph $G^1$.

    We can similarly show that there cannot be two edges $e=\left( c', c, u \right)$ and $e'=\left( c'', c, u \right)$ in $E^0$ that share the same destination and user, by showing that it implies $c\in \rec^1\left(u\right)$ and $c\in \rec^{0,1}\left(u\right)$ which contradicts the existence of a red edge. Therefore, Condition~\ref{enum:relocation_graph_consistency_disjoint} holds for all edges in $E^0$.
    \\~\\
    \textbf{Statement~\ref{enum:submodular_alg_feasible} (feasibility):}
    We first note that $c_1$ is not involved in the new matching $\rec^0$ as it does not have any incident edges in $E^0$. The key observation is that $c_1$ cannot have any blue outgoing edges, and the presence of any red outgoing edge implies it must be a blue-red junction, whose two edges with the same user label must have both been added to $\Tilde{E}^0$ and then combined into a purple edge that bypasses $c_1$. The full proof is left as an exercise.

    {}
    
    Now we show that $\rec^0$ satisfies all creator constraints. For any creator $c\notin \left\{ c_X, c_0 \right\}$, her loss of audience when going from $\rec$ to $\rec^0$ is
    \begin{align}
        a^{\rec}_c - a^{\rec^0}_c &= \deg^{G^{0}}_{out} \left(c\right) - \deg^{G^{0}}_{in} \left(c\right)  \label{eqn:relocation_feasibility_audience_degrees_G0} \\ 
        &= \deg^{\text{non-aux edges in }\Tilde{E}^0}_{out} \left(c\right) - \deg^{\text{non-aux edges in }\Tilde{E}^0}_{in} \left(c\right)  \label{eqn:relocation_feasibility_audience_degrees_nonaux} \\
        &= \text{number of auxiliary edges into $c$ in $\Tilde{E}^0$},  \label{eqn:relocation_feasibility_audience_num_aux_edges_E0}
    \end{align}
    where \eqref{eqn:relocation_feasibility_audience_degrees_G0} results from \eqref{eqn:relocation_graph_audience_degrees}; \eqref{eqn:relocation_feasibility_audience_degrees_nonaux} uses the fact that the final while loop, particularly Lines~\ref{algline:backward_tracing_compress_remove} and \ref{algline:backward_tracing_compress_add}, have no effects on net degrees of any vertex; and \eqref{eqn:relocation_feasibility_audience_num_aux_edges_E0} is because Lemma~\ref{lemma:submodular_fixU_alg_prev_function}\ref{enum:submodular_prev_flow_conservation} gives that $\deg^{\Tilde{E}^0}_{in} \left(c\right) = \deg^{\Tilde{E}^0}_{out} \left(c\right)$, where both terms count both auxiliary and non-auxiliary edges, and all $c\neq c_X$ have no outgoing auxiliary edges. Note that \eqref{eqn:relocation_feasibility_audience_num_aux_edges_E0} counts multiple auxiliary edges from $c_X$ to $c$ separately (which may happen if $c$ is the source of multiple red and blue paths). Additionally,

    \begin{align}
        a^{\rec}_c - a^{\rec^0}_c %
        &= \text{number of auxiliary edges into $c$ in $\Tilde{E}^0$}  \nonumber \\ %
        &\leq \text{number of auxiliary edges into $c$ in $G$}  \nonumber \\ %
        &= \left| \left\{ b \in \mathcal{B} : b\text{ starts at }c \right\} \right| + \left| \left\{ r \in \mathcal{R} : r\text{ starts at }c \right\} \right|  \label{eqn:relocation_feasibility_audience_num_path_starts} \\
        &= \left( \deg^{G^{1}}_{out} \left(c\right) - \deg^{G^{1}}_{in} \left(c\right) \right) + \left( \deg^{G^{0,1}}_{out} \left(c\right) - \deg^{G^{0,1}}_{in} \left(c\right) \right)  \label{eqn:relocation_feasibility_audience_degrees_G1_G01} \\
        &= \left( a^{\rec}_c - a^{\rec^1}_c \right) + \left( a^{\rec^1}_c - a^{\rec^{0,1}}_c \right)  \label{eqn:relocation_feasibility_audience_diff_audiences} \\
        &= a^{\rec}_c - a^{\rec^{0,1}}_c,  \nonumber
    \end{align}
    where \eqref{eqn:relocation_feasibility_audience_num_path_starts} is because one auxiliary edge is added to $G$ per red or blue path, and \eqref{eqn:relocation_feasibility_audience_diff_audiences} results from \eqref{eqn:relocation_graph_audience_degrees}. 
    Note that \eqref{eqn:relocation_feasibility_audience_degrees_G1_G01} counts degrees in the original blue and red graphs $G^1$ and $G^{0,1}$ without auxiliary edges. Since $\rec^{0,1}$ is feasible, $a^{\rec^{0,1}}_c \geq \abar$, which gives $a^{\rec^0}_c \geq \abar$ and that $c$'s creator constraint is satisfied. 
    
    Finally, we verify that $c_0$'s creator constraint is satisfied. Note that the total number of incoming edges into $c_0$ in $\Tilde{E}^0$ equals to its indegree in $G^{0,1}$ due to Line~\ref{algline:backward_tracing_outer_while}, and each such edge is either preserved in $E^0$ or combined into a purple edge. Thus, $a^{\rec^0}_{c_0} = \deg^{G^0}_{in} = \deg^{G^{0,1}}_{in} = a^{\rec^{0,1}}_{c_0} \geq \abar$.
    \\~\\
    \textbf{Statement~\ref{enum:submodular_alg_weights} (sum of edge weights):} Let $\mathcal{B}$ be the set of all blue paths and $\mathcal{R}$ be all red paths in $G$ (with auxiliary edges). Let $\mathcal{R}'$ be the set of red paths that contain some edges that are not in $\Tilde{E}^0$, and $\mathcal{B}'$ be the set of blue paths that contain at least one edge in $\Tilde{E}^0$. 

    Lemma~\ref{lemma:submodular_fixU_alg_prev_function}\ref{enum:submodular_prev_red_tails} implies that for every red path $r\in \mathcal{R}$, $r \cap \Tilde{E}^0$ equals to all edges in $r$ after some vertex $c$, with $c \neq c_X$ if and only if $r\in \mathcal{R}'$. We define function $start\left(r\right):=c$. Similarly, Lemma~\ref{lemma:submodular_fixU_alg_prev_function}\ref{enum:submodular_prev_blue_heads} implies that for every blue path $b\in \mathcal{B}$, $b \cap \Tilde{E}^0$ equals to all edges in $b$ before some vertex $c$, with $c \neq c_X$ if and only if $b\in \mathcal{B}'$. We define $end\left(b\right):=c$. %
    
    We first construct a function $p: \mathcal{R}' \mapsto \mathcal{B}'$ with the following properties for any red path $r\in \mathcal{R}'$:
    \begin{enumerate}[label=(\roman*)]
        \item $start\left( r \right) = end\left( p\left(r\right) \right) = c$ for some $c \notin \left\{ c_X, c_0 \right\}.$  \label{enum:submodular_weights_pathmap_headtail} %
        \item $p$ is a bijection from $\mathcal{R}'$ to $\mathcal{B}'$.  \label{enum:submodular_weights_pathmap_bijection}
        \item The sum of edge weights of the part of (blue) $p\left(r\right)$ before $c$ is at least that of the part of (red) $r$ before $c$. (We define auxiliary edges to have weight zero.)  \label{enum:submodular_weights_pathmap_headweight} %
    \end{enumerate}
    \underline{Definition of $p$ and proof of \ref{enum:submodular_weights_pathmap_headtail} and \ref{enum:submodular_weights_pathmap_bijection}:} For any vertex $c \notin \left\{ c_X, c_0 \right\}$, let $\mathcal{B}_c$ be the set of blue paths containing $c$ whose incoming edge into $c$ is in $\Tilde{E}^0$ but outgoing edge from $c$ is not in $\Tilde{E}^0$. Conversely, let $\mathcal{R}_c$ be the set of red paths containing $c$ whose incoming edge into $c$ is not in $\Tilde{E}^0$ but outgoing edge from $c$ is in $\Tilde{E}^0$. Lemma~\ref{lemma:submodular_fixU_alg_prev_function}\ref{enum:submodular_prev_flow_conservation} implies $\left| \mathcal{B}_c \right| = \left| \mathcal{R}_c \right|$, and by definition, $start\left(r\right) = end\left(b\right) = c$ for all $r\in \mathcal{R}_c$ and $b\in \mathcal{B}_c$. Thus, there exists a bijection $\pi_c: \mathcal{R}_c \mapsto \mathcal{B}_c$. Let $p\left(r\right) = \pi_c\left(r\right)$ for all $r\in \mathcal{R}_c$, and this definition satisfies \ref{enum:submodular_weights_pathmap_headtail} for red paths that are in some $\mathcal{R}_c$.

    To show \ref{enum:submodular_weights_pathmap_headtail} for all $r\in \mathcal{R}'$ and to show \ref{enum:submodular_weights_pathmap_bijection}, note that for any $r\in \mathcal{R}'$, $start\left(r\right) \notin \left\{ c_X, c_0 \right\}$. Thus, $\mathcal{R}' = \bigcup_{c \notin \left\{ c_X, c_0 \right\}} \left\{ r\in \mathcal{R}' : start\left(r\right) = c \right\} = \bigcup_{c \notin \left\{ c_X, c_0 \right\}} \mathcal{R}_c$, which completes \ref{enum:submodular_weights_pathmap_headtail}. Similarly, for any $b\in \mathcal{B}'$, $end\left(b\right) \notin \left\{ c_X, c_0 \right\}$, so $\mathcal{B}' = \bigcup_{c \notin \left\{ c_X, c_0 \right\}} \mathcal{B}_c$. This shows $p\left(\mathcal{R}'\right)$ is surjective onto $\mathcal{B}'$. It can also be shown that $p\left(r\right) \neq p\left(r'\right)$ for any $r\neq r'$ (proof omitted), which proves \ref{enum:submodular_weights_pathmap_bijection}.
    \\~\\
    \underline{Proof of \ref{enum:submodular_weights_pathmap_headweight}:} For any red path $r$, let $b = p\left( r \right)$, $c = start\left(r\right) = end\left(b\right)$, and denote $r_{head}$ and $b_{head}$ to be the parts of $r$ and $b$ before $c$ respectively. We want to show $\sum_{e \in b_{head}} w\left(e\right) \geq \sum_{e \in r_{head}} w\left(e\right)$. We will prove this by dividing both $b_{head}$ and $r_{head}$ into equal numbers of segments, and showing that the sum of weights of a single segment in $b_{head}$ is at least that of the corresponding segment in $r_{head}$, because otherwise in the blue graph $G^1$, we can replace that blue segment with the red segment to obtain a feasible matching with higher total engagement than the optimal matching $\rec^1$.

    Construct a list of vertices $L = \left\{ c_{(0)}, \dots, c_{(k)} \right\}$ as follows: Let the first vertex be $c_{(0)} = c_X$. For each $i\geq 1$, $c_{(i)}$ is the first vertex $v$ in $r_{head}$ after $c_{(i-1)}$ that satisfies: $v$ is a junction that involves an edge in $r_{head}$ and an edge in $b_{head}$, and $v$ also comes after $c_{(i-1)}$ in $b_{head}$. If no such $v$ exists, set $c_{(i)}=c$ as the end of $L$.
    Clearly, $L$ divides both $r_{head}$ and $b_{head}$ into $k$ segments that follow one another. For each $1\leq i\leq k$, let $r_i$ and $b_i$ be the segments of $r_{head}$ and $b_{head}$ from $c_{(i-1)}$ to $c_{(i)}$, respectively, and let $W^r_i$ an $W^b_i$ be the sums of their edge weights (auxiliary edges have weight zero). A key property is that there are no junctions whose red edge is in $r_i$ and blue edge is in $b_i$.\footnote{Edges in $r_i$ may form junctions with $b_{head}$, but by construction, such junctions can only involve the part of $b_{head}$ before $c_{(i)}$.}

    Suppose for contradiction that $W^r_i > W^b_i$ for some $1\leq i\leq k$. Consider a graph $\Tilde{G}^1$ modified from the blue relocation graph $G^1$ that replaces $b_i$ with $r_i$ while keeping all other blue edges the same, with the slight exception that if any of $c_{(i-1)}$ and $c_{(i)}$ is a junction with edges $\left( c', c_{(j)}, u \right)$ and $\left( c_{(j)}, c'', u \right)$, these two edges are combined into a single edge $\left( c', c'', u \right)$.
    We can show that $\Tilde{G}^1$ is consistent by showing that each red edge in $r_i$ satisfies consistency conditions. The key observation is that the red edge can only possibly be inconsistent if it is involved with a junction at either end, but all intermediate vertices in $r_i$ can only form junctions with blue edges that are not in $b_i$, which are still in $\Tilde{G}^1$;\footnote{If any intermediate vertex $v$ in $r_i$ is a junction that involves both a red edge in $r_i$ and a blue edge in $b_i$, $v$ must come after $c_{(i-1)}$ in both $r_{head}$ and $b_{head}$, but this means $v$ would have been chosen as $c_{(i)}$ instead.} and at both $c_{(i-1)}$ and $c_{(i)}$, any possible junctions have been combined.\footnote{The first red edge in $r_i$, which is the red outgoing edge from $c_{(i-1)}$, can only be involved in a blue-red junction at $c_{(i-1)}$ with a blue incoming edge into that vertex, which must still be in $\Tilde{G}^1$. Likewise, the last red edge in $r_i$ can only be involved in a red-blue junction at $c_{(i)}$ with a blue outgoing edge from that vertex, which must also be in $\Tilde{G}^1$ (this also applies to when $i=k$ and thus $c_{(i)}=c$).} Thus, $\Tilde{G}^1$ induces a matching $\Tilde{\rec}^1$ that can be shown to be feasible, and since $W^r_i > W^b_i$, $\Tilde{G}^1$ has higher total weight than $G^1$, so $\Tilde{\rec}^1$ has higher total engagement than $\rec^1$. However, this contradicts the optimality of $\rec^1$.
    Therefore, $W^b_i \geq W^r_i$ for all $1\leq i\leq k$, so $\sum_{e \in b_{head}} w\left(e\right) = \sum_{i=1}^k W^b_i \geq \sum_{i=1}^k W^r_i = \sum_{e \in r_{head}} w\left(e\right)$.
    \\~\\    
    \underline{Proof of Statement~\ref{enum:submodular_alg_weights} using \ref{enum:submodular_weights_pathmap_headtail}-\ref{enum:submodular_weights_pathmap_headweight}:} Since the final while loop on Line~\ref{algline:backward_tracing_compress_while} does not change the total edge weights, and we assumed that auxiliary edges (removed on Line~\ref{algline:backward_tracing_remove_auxiliary}) have weight 0, it suffices to show that the sum of weights of chosen edges in $\Tilde{E}^0$ is at least that of all red edges (including auxiliary edges). Denote $W\left(S\right) := \sum_{e\in S} w\left(e\right)$ as the total weight of any set of edges $S$, $r_{head}$ and $r_{tail}$ as the parts of any red edge $r$ before and after $start\left(r\right)$ respectively, and $b_{head}$ as the part of any blue edge $b$ before $end\left(b\right)$. Then,
    \begin{align}
        W\left( \Tilde{E}^0 \right) &= \sum_{r \in \mathcal{R}} W\left( r_{tail} \right) + \sum_{b \in \mathcal{B}'} W\left( b_{head} \right)  \nonumber \\
        &= \sum_{r \in \mathcal{R} \backslash \mathcal{R}'} W\left( r \right) + \sum_{r \in \mathcal{R}'} \left( W\left( r \right) - W\left( r_{head} \right) \right) + \sum_{b \in \mathcal{B}'} W\left( b_{head} \right)  \nonumber \\
        &= \sum_{r \in \mathcal{R}} W\left( r \right) - \sum_{r \in \mathcal{R}'} W\left( r_{head} \right) + \sum_{r \in \mathcal{R}'} W\left( p\left(r\right)_{head} \right)  \label{eqn:submodular_alg_weights_bijection} \\
        &= \sum_{r \in \mathcal{R}} W\left( r \right) + \sum_{r \in \mathcal{R}'} \left( W\left( p\left(r\right)_{head} \right) - W\left( r_{head} \right)\right)  \nonumber \\
        &\geq \sum_{r \in \mathcal{R}} W\left( r \right) = W\left( E^{0,1} \right), \quad \text{total weight of all red edges},  \label{eqn:submodular_alg_weights_head_nonneg}
    \end{align}
    where \eqref{eqn:submodular_alg_weights_bijection} uses \ref{enum:submodular_weights_pathmap_bijection} and \eqref{eqn:submodular_alg_weights_head_nonneg} uses \ref{enum:submodular_weights_pathmap_headweight}.
\end{proofCharles}

Finally, Proposition~\ref{prop:submodular_fixU_alg_output} leads to a proof of Proposition~\ref{prop:submodular_fixU}:
\begin{proofCharles}[Proof of Proposition~\ref{prop:submodular_fixU}]
    \begin{align}
        &\phantom{{}={}} f\left( \userset, \creatorset \cup \left\{ c_0, c_1 \right\} \right) - f\left( \userset, \creatorset \cup \left\{ c_1 \right\} \right)  \nonumber \\
        &= \eng\left( \userset, \creatorset \cup \left\{ c_0, c_1 \right\}, \rec^{0,1} \right) - \eng\left( \userset, \creatorset \cup \left\{ c_1 \right\}, \rec^{1} \right)  \label{eqn:submodular_fixU_thm_R01} \\
        &= \sum_{e\in G^{0,1}} w\left(e \right)  \label{eqn:submodular_fixU_thm_G01} \\
        &\leq \sum_{e\in G^{0}} w\left(e \right)  \label{eqn:submodular_fixU_thm_G0} \\
        &= \eng\left( \userset, \creatorset \cup \left\{ c_0 \right\}, \rec^{0} \right) - \eng\left( \userset, \creatorset, \rec \right)  \label{eqn:submodular_fixU_thm_R0} \\
        &\leq f\left( \userset, \creatorset \cup \left\{ c_0 \right\} \right) - f\left( \userset, \creatorset \right),  \label{eqn:submodular_fixU_thm_f0}
    \end{align}
where \eqref{eqn:submodular_fixU_thm_R01} uses optimality of $\rec^{0,1}$ and $\rec^{1}$, \eqref{eqn:submodular_fixU_thm_G01} uses \eqref{eqn:relocation_graph_sum_weights}, \eqref{eqn:submodular_fixU_thm_G0} uses Statement \ref{enum:submodular_alg_weights} in Proposition~\ref{prop:submodular_fixU_alg_output}, \eqref{eqn:submodular_fixU_thm_R0} again uses \eqref{eqn:relocation_graph_sum_weights}, and \eqref{eqn:submodular_fixU_thm_f0} uses optimality of $\rec$ and the fact that $\rec^0$ is a feasible matching between $\userset$ and $\creatorset \cup \left\{ c_0 \right\}$ that is not necessarily optimal.
\end{proofCharles}

\subsubsection{Comparison to Proof in \texorpdfstring{\cite{mladenov2020optimizing}}{Mladenov et al. (2020)}}
We briefly compare our proof of Proposition~\ref{prop:submodular_fixU} to the analogous Theorem~\ref{thm:mladenov_submodular} in \citeauthor{mladenov2020optimizing}. %

Our high-level proof outline in Section~\ref{sec:submodular_fixU_summary} and Figure~\ref{fig:submod_proof_outline} is inspired by \citeauthor{mladenov2020optimizing} and near identical. Definitions of key concepts (relocation triplets, relocation graphs and their consistency, junctions) are also very similar, but with slight modifications.\footnote{Some examples include: \citeauthor{mladenov2020optimizing} allow relocation graphs to have paths like $c \xrightarrow{u} c' \xrightarrow{u} c''$ but we do not; our consistency Condition~\ref{enum:relocation_graph_consistency_engagement} accounts for user constraints $\ebar$.} The biggest differences stem from their assumption that $K=1$: This means their instances do not contain any red-blue junctions, because each red-blue junction with an incoming red edge $\left( c', c, u \right)$ and an outgoing blue edge $\left( c, c'', u \right)$ implies $\left\{ c', c'' \right\} \subseteq \rec^1\left(u\right)$. Allowing $K>1$ also requires us to explicitly state the bijection $\phi_u$ in Definition~\ref{def:relocation_triplet} and all four consistency conditions~\ref{enum:relocation_graph_consistency_recom_out}-\ref{enum:relocation_graph_consistency_disjoint} in Lemma~\ref{lemma:relocation_graph_consistency}, whereas the bijection is implicitly uniquely defined when $K=1$ as its domain and range both have size 1.

However, there are major differences between our Algorithm~\ref{alg:submodular_fixU_backward_tracing} that constructs $G^0$ and their analogous algorithm. \citeauthor{mladenov2020optimizing} instead uses an algorithm that repeatedly scans for junctions (entirely blue-red) between red path $r$ and blue path $b$ such that it is the last junction in $b$, then adds the head of $b$ and the tail of $r$ to the chosen set of edges $E^0$, and removes both paths from future consideration.

The different design of our Algorithm~\ref{alg:submodular_fixU_backward_tracing} is both to account for red-blue junctions,\footnote{For example, existence of both blue-red and red-blue junctions means that a vertex $c$ can be a blue-red junction that involves red path $r$ and blue path $b$, a red-blue junction that involves $r$ and $b'$, and a red-blue junction that involves $r'$ and $b$ concurrently. Resolving any one of the junctions alone is insufficient in ensuring all prerequisites given by all junctions are met.} and to fix an error with their proof that results from complete removal of paths. As a counter-example, suppose a red path $r$ has blue-red junctions $c_1, c_2, c_3$ in order (from the source of $r$ to its sink), that involve blue paths $b_1, b_2, b_3$ each of which does not have any other junctions. If $c_1$ is examined first, it would have removed $r$ and $b_1$ from consideration while adding parts of them to $E^0$; however, this means $E^0$ only contains the red edge in junction $c_3$ but not the blue edge, so it is inconsistent. This indicates that the problem of pairing red paths with blue paths (which we solve when proving Statement~\ref{enum:submodular_alg_weights} of Proposition~\ref{prop:submodular_fixU_alg_output}) has greater complexity than that captured by a simple constraint of ``each junction examined is the last on $b$''.

We resolve this issue with the \textsc{Prev} function and Lemma~\ref{lemma:submodular_fixU_alg_prev_function}, which systematically chooses an incoming edge at each vertex (which may or may not be a junction) that explicitly satisfies all junction constraints, while still maintaining the property that only the tails of red paths and heads of blue paths can be added, thus allowing them to be paired. Note that each backward traversal (each iteration of the outer while loop on Line~\ref{algline:backward_tracing_outer_while}) adds a path that may combine parts of several blue paths and several red paths, unlike in \citeauthor{mladenov2020optimizing} where each blue or red path is only considered and added once. Nevertheless, our Algorithm~\ref{alg:submodular_fixU_backward_tracing} draws motivation from the high-level intuition of \citeauthor{mladenov2020optimizing}'s algorithm: whenever a red path and a blue path intersect, we can switch from the tail of the red path to the head of the blue path, and achieve consistency while obtaining at least as much total engagement. We do so in the \textsc{Prev} function by always prioritizing blue incoming edges over red ones whenever available.

\subsection{Inapplicability of Approximation Results to the General Case}  \label{sec:Mladenov_inapproximability}
As a final note on this special case of $\ebar=0$, we now show that our approximation result in Theorem~\ref{thm:approx_bigK} does not apply to the general model that allows $\ebar>0$, and neither does Theorem~\ref{thm:mladenov_constant_approx} from \citeauthor{mladenov2020optimizing}. This is because the two key ingredients for the approximation, namely the objective functions ($f\left( \userset, \creatorset \right)$ in our result and $g\left(\creatorset\right)$ in \citeauthor{mladenov2020optimizing}) being submodular in $\creatorset$ and their feasibility being characterized by a single knapsack constraint, are not necessarily true when $K>1$ or $\ebar>0$.
\\~\\
\textbf{Insufficiency of knapsack constraints.} Proposition~\ref{prop:mladenov_knapsack} is violated when users have non-trivial participation constraints, i.e., $\ebar > 0$. In this case, the knapsack constraint $\left| \creatorset \right| \abar \leq UK$ is no longer sufficient for either $f\left( \userset, \creatorset \right)$ or $g\left(\creatorset\right)$ to be well-defined, even if $K=1$. An example is in the proof of Proposition~\ref{prop:hard_regular}, in which we reduced the \MIS\ problem on regular graphs to \MSS\ instances with $K=1$. On such instances, we showed that $g\left(\creatorset\right)$ is well-defined if and only if the set of vertices that correspond to creators in $\creatorset$ forms an independent set. This is strictly stronger than the knapsack constraint, which, on these instances, only requires that no more than half of the vertices are chosen. Additionally, $f\left( \userset, \creatorset \right)$ is well-defined if and only if $\creatorset$ satisfies the aforementioned property and $\userset$ corresponds to all edges that are incident on those vertices, which is again much stronger than the knapsack constraint.
\\~\\
\textbf{Lack of submodularity in creators when users are not exogenously fixed.} We can also show that Theorem~\ref{thm:mladenov_submodular} from \citeauthor{mladenov2020optimizing} no longer holds when $K>1$, or equivalently, our Proposition~\ref{prop:submodular_fixU} no longer holds when the set of users $\userset$ is not exogenously fixed, necessitating the use of $g\left(\creatorset \right)$ instead of $f\left( \userset, \creatorset \right)$. In other words,  if the platform needs to choose an optimal subset of users as part of the optimization problem, the objective $g\left(\creatorset \right)$ is no longer submodular in the choice of creators.
We show two possible scenarios that violate the submodularity condition \eqref{eqn:Mladenov_submodular} defined on $g\left(\creatorset \right)$ when $K>1$. %
\begin{enumerate}[label=(\alph*)]
    \item $g\left( \creatorset \right)$ and $g\left( \creatorset \cup \left\{ c_0, c_1 \right\} \right)$ are both feasbile, but $g\left( \creatorset \cup \left\{ c_0 \right\} \right)$ and $g\left( \creatorset \cup \left\{ c_1 \right\} \right)$ may not be.  \label{enum:submodularity_case_infeasible}
    \begin{enumerate}[label=(\roman*)]
        \item Consider an instance where the only two creators on the platform are $\creatorset_0 = \left\{ c_0, c_1 \right\}$, $K=2$, and a non-empty stable set exists. It must hold that $\creatorset = \emptyset$ as it is the only valid choice. Then, $g\left( \creatorset \right)$ is well-defined as it corresponds to the empty stable set. However, any non-empty stable set must include both creators (due to $K=2$), so $g\left( \creatorset \cup \left\{ c_0, c_1 \right\} \right)$ is well-defined, but $g\left( \creatorset \cup \left\{ c_0 \right\} \right)$ and $g\left( \creatorset \cup \left\{ c_1 \right\} \right)$ are not.  \label{enum:submodularity_case1}

        \item As another example, consider the example shown in Figure~\ref{fig:example_submodularity_case2}, where $\creatorset = \left\{ c_2, c_3, c_4, c_5 \right\}$. It is possible to keep either only creators in $\creatorset$ or all creators $\creatorset \cup \left\{ c_0, c_1 \right\}$, as illustrated in the figure. However, retaining exactly $\creatorset \cup \left\{ c_0 \right\}$ or $\creatorset \cup \left\{ c_1 \right\}$ is impossible: for creator $c_0$ to stay, all the middle $\left( \abar-2 \right)$ users also need to be in the stable set (because her potential audience has size $\abar$ and includes all these $\left( \abar-2 \right)$ users), but these users will not have any other creator that they are happy with if $c_1$ is not also in the stable set.  \label{enum:submodularity_case2}

        \begin{figure}[h]
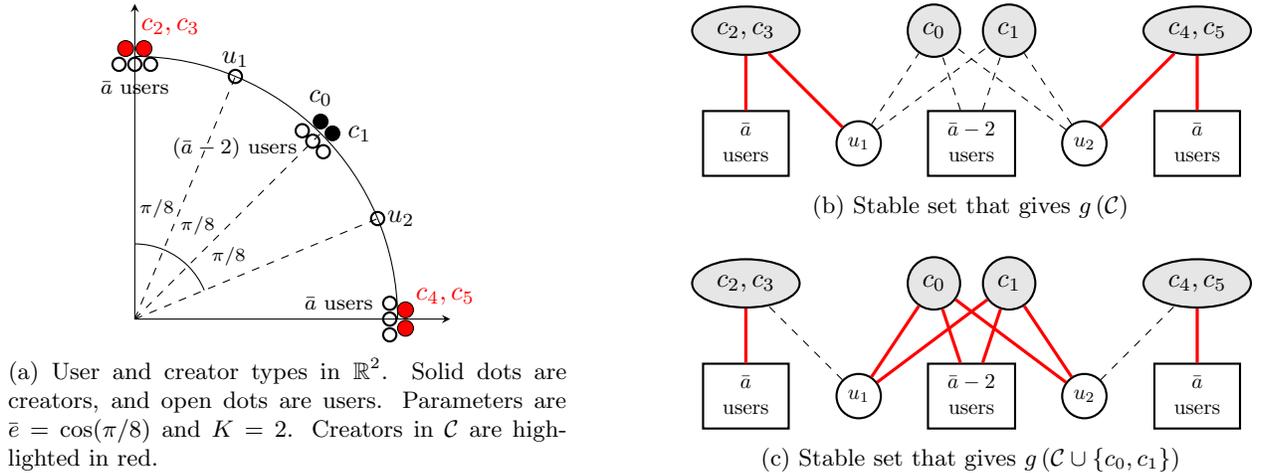

            \centering
            \begin{subfigure}[c]{0.45\textwidth}
                \centering
                \includegraphics[width=0.7\textwidth]{tikz/submod2_vectors.tikz}
                \caption{User and creator types in $\mathbb{R}^2$. Solid dots are creators, and open dots are users. Parameters are $\ebar = \cos (\pi/8)$ and $K=2$. Creators in $\creatorset$ are highlighted in red.}
                \label{fig:example_submod2_vectors}
            \end{subfigure}
            \hfill%
            \begin{minipage}[c]{0.45\textwidth}
                \begin{subfigure}{\textwidth}
                    \centering
                    \includegraphics{tikz/submod2_C.tikz}
                    \caption{Stable set that gives $g\left( \creatorset \right)$}
                    \label{fig:example_submod2_C}
                \end{subfigure}
                \par\bigskip
                \begin{subfigure}{\textwidth}
                    \centering
                    \includegraphics{tikz/submod2_Cc0c1.tikz}
                    \caption{Stable set that gives $g\left( \creatorset \cup \left\{ c_0, c_1 \right\} \right)$}
                    \label{fig:example_submod2_Cc0c1}
                \end{subfigure}
            \end{minipage}
            \caption{Illustration of example in case~\ref{enum:submodularity_case_infeasible}\ref{enum:submodularity_case2}}
            \label{fig:example_submodularity_case2}
        \end{figure}
    \end{enumerate}
    
    \item All of $g\left( \creatorset \right), g\left( \creatorset \cup \left\{ c_0 \right\} \right), g\left( \creatorset \cup \left\{ c_1 \right\} \right)$ and $g\left( \creatorset \cup \left\{ c_0, c_1 \right\} \right)$ are feasible, but they do not satisfy the submodularity condition in \eqref{eqn:Mladenov_submodular}. Consider the example shown in Figure~\ref{fig:example_submodularity_case3}, where $\creatorset = \left\{ c_2, c_3, c_4, c_5 \right\}$. Any stable set that does not include both $c_0$ and $c_1$ concurrently can only include at most $3\abar$ users and achieve total engagement at most $6\abar \cos \left( \pi/8 \right)$, as the middle $x$ users cannot be happy with $K=2$ creators unless both $c_0$ and $c_1$ are on the platform. In contrast, $g\left( \creatorset \cup \left\{ c_0, c_1 \right\} \right)$ keeps $\left( 3\abar + x \right)$ users and achieves engagement $6\abar \cos \left( \pi/8 \right) + 2x$. Thus, submodularity is violated.  \label{enum:submodularity_case3}

    \begin{figure}[h]
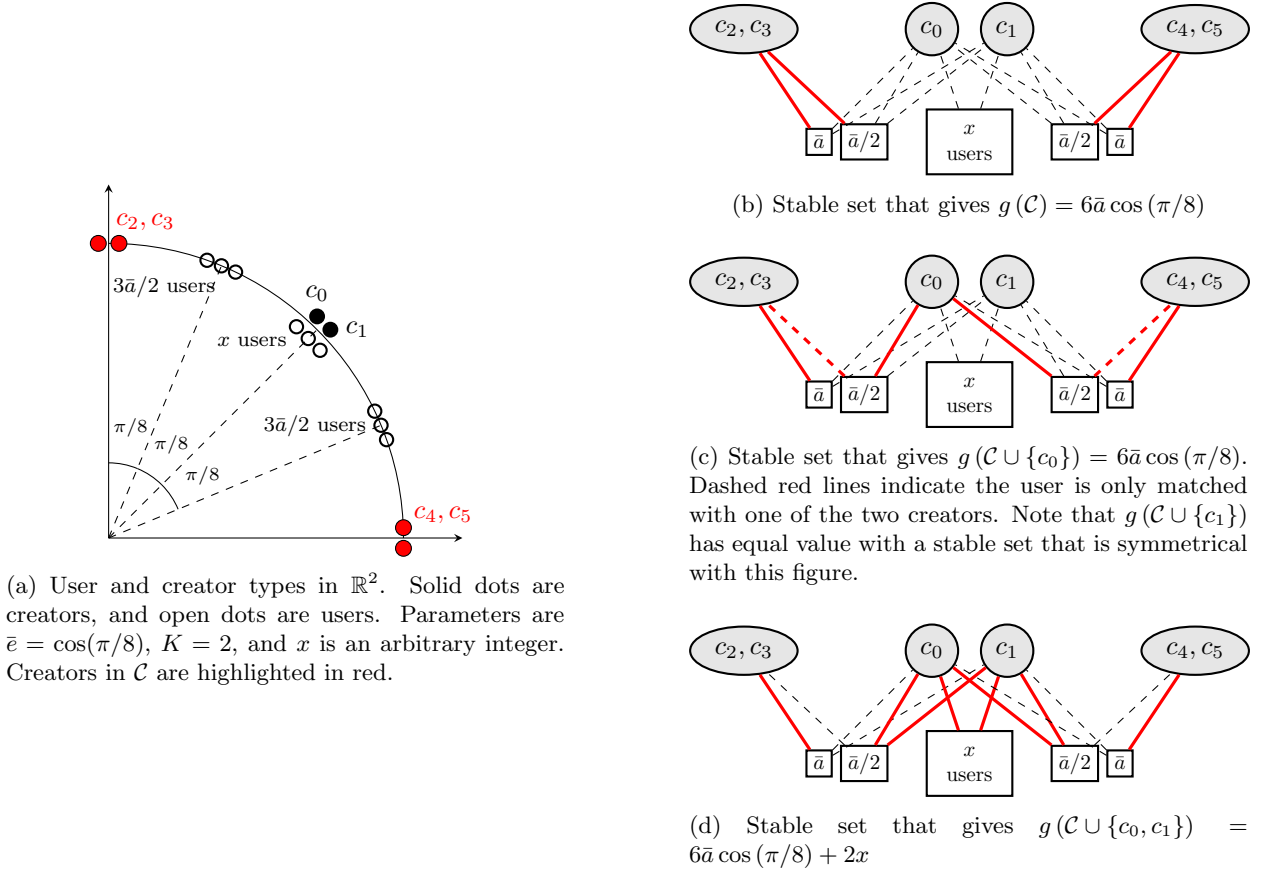

        \centering
        \begin{subfigure}[c]{0.45\textwidth}
            \centering
            \includegraphics[width=0.7\textwidth]{tikz/submod3_vectors.tikz}
            \caption{User and creator types in $\mathbb{R}^2$. Solid dots are creators, and open dots are users. Parameters are $\ebar = \cos (\pi/8)$, $K=2$, and $x$ is an arbitrary integer. Creators in $\creatorset$ are highlighted in red.}
            \label{fig:example_submod3_vectors}
        \end{subfigure}
        \hfill%
        \begin{minipage}[c]{0.45\textwidth}
            \begin{subfigure}{\textwidth}
                \centering
                \includegraphics{tikz/submod3_C.tikz}
                \caption{Stable set that gives $g\left( \creatorset \right) = 6\abar \cos \left( \pi/8 \right)$}
                \label{fig:example_submod3_C}
            \end{subfigure}
            \par\bigskip
            \begin{subfigure}{\textwidth}
                \centering
                \includegraphics{tikz/submod3_Cc0.tikz}
                \caption{Stable set that gives $g\left( \creatorset \cup \left\{ c_0 \right\} \right) = 6\abar \cos \left( \pi/8 \right)$. Dashed red lines indicate the user is only matched with one of the two creators. Note that $g\left( \creatorset \cup \left\{ c_1 \right\} \right)$ has equal value with a stable set that is symmetrical with this figure.}
                \label{fig:example_submod3_Cc0}
            \end{subfigure}
            \par\bigskip
            \begin{subfigure}{\textwidth}
                \centering
                \includegraphics{tikz/submod3_Cc0c1.tikz}
                \caption{Stable set that gives $g\left( \creatorset \cup \left\{ c_0, c_1 \right\} \right) = 6\abar \cos \left( \pi/8 \right) + 2x$}
                \label{fig:example_submod3_Cc0c1}
            \end{subfigure}
        \end{minipage}
        \caption{Illustration of example in case~\ref{enum:submodularity_case3}}
        \label{fig:example_submodularity_case3}
    \end{figure}
\end{enumerate}

\section{Proofs on \IE's Expected Performance on Random Instances}  \label{sec:proof_random_instances}
In this section, we prove Proposition~\ref{prop:increase_creators} and Theorem~\ref{thm:increase users}, two statements that describe scenarios in which \IE's expected performance worsens relative to the first-best \FL\ on random instances.

\subsection{Proof of Proposition~\ref{prop:increase_creators}: Increase \texorpdfstring{$C$}{C}}  \label{sec:proof_increase_creators}
Recall Proposition~\ref{prop:increase_creators} from Section~\ref{sec:increase_creators_condensed}:
\begin{customTheorem}{Proposition~\ref{prop:increase_creators}}
    Fix $U$, $K$, $\ebar$, $\abar$ such that $U\geq \abar$, $\ebar<1$, and assume $u_i, c_j \sim \distribution$ i.i.d. for all $1\leq i\leq U$ and $1\leq j\leq C$. When $C\rightarrow \infty$, %
    \begin{align*}
        \lim_{C\rightarrow \infty} \mathbb{E}\left[ \frac{\longtermengagement\left(\UCseq\right)}{\longtermengagement\left(\FLseq\right)} \, \middle| \, \longtermengagement\left(\FLseq\right)>0 \right] &= 0.
    \end{align*}
\end{customTheorem}

\begin{proofCharles}[Proof of Proposition~\ref{prop:increase_creators}]
    It suffices to show that $\lim_{C\rightarrow \infty} \mathbb{E}\left[ \longtermengagement\left(\UCseq\right) \right] = 0$ (which implies $\longtermengagement\left(\UCseq\right) = 0$ with probability 1 as $C\rightarrow \infty$) and $\lim_{C\rightarrow \infty} \mathbb{E}\left[ \longtermengagement\left(\FLseq\right) \right] > 0$. We first show $\lim_{C\rightarrow \infty} \mathbb{E}\left[ \longtermengagement\left(\UCseq\right) \right] = 0$; in particular, we show that it holds even when fixing a particular realization of user types $u_1, \dots, u_U$. 
    
    Let $c\in \mathbb{R}^D$ be an arbitrary non-negative unit vector. Let $a_{c}$ be the audience size of a creator with type $c$ under \IE\ at $t=0$. Then, its expectation assuming the $C-1$ other creators $c_1, \dots, c_{C-1}$ are drawn i.i.d. from $\distribution$ is:
    \begin{align}
        \mathbb{E}_{c_1, \dots, c_{C-1}} \left[ a_{c} \right] &= \sum_{i=1}^{U} \Pr \left( \text{user $i$ is assigned to creator at $c$ under $UC_0$} \right)  \nonumber \\
        &= \sum_{i=1}^{U} \Pr \left( \text{$c$ is one of the $K$ nearest creators from $u_i$} \right)  \nonumber \\
        &= \sum_{i=1}^{U} \Pr \left( \text{at most $K-1$ of $c_1, \dots, c_{C-1}$ are closer to $u_i$ than $c$ is} \right).  \label{eqn:at_most_K-1_nearer}
    \end{align}
    Define $S:=\left\{ x\in \R^D_{\geq 0}: \left\lVert x \right\rVert = 1 \right\}$ to be the region that contains all non-negative unit vectors in $\mathbb{R}^D$, and for each $1\leq i\leq U$, define $A_i^c:=\left\{ x\in S: \left\lVert x-u_i \right\rVert < \left\lVert c-u_i \right\rVert \right\}$ to be the region containing vectors that are closer to $u_i$ than $c$ is. Let $X_i^c$ be the number of vectors among $c_1, \dots, c_{C-1}$ that fall in $A_i^c$, and let $p:=\left|A_i^c\right| / \left|S\right|$, the ratio of the volume of $A_i^c$ over that of $S$. Note that $0\leq p<1$ w.p. 1, and $p$ is also the probability that a vector drawn uniformly at random from $S$ falls within $A_i^c$. Thus, $X_i^c$ follows a binomial distribution $X_i^c \sim B\left( C-1, p \right)$, and
    \begin{align*}
        \eqref{eqn:at_most_K-1_nearer} &= \sum_{i=1}^{U} \Pr \left( \text{at most $K-1$ of $c_1, \dots, c_{C-1}$ are in $A_i^c$} \right)  \\
        &= \sum_{i=1}^{U} \Pr \left( X_i^c\leq K-1 \right),
    \end{align*}
    As sample size $C-1 \rightarrow \infty$, $\Pr \left( X_i^c\leq K-1 \right) \rightarrow 0$. (A more formal argument can be shown with Central Limit Theorem approximating $X_i^c$ as a normal distribution.)
        Therefore, $\mathbb{E} \left[ a_{c} \right] \rightarrow 0$ as $C\rightarrow \infty$. This holds for any potential type vector of a creator, which means that any creator's expected audience size under $UC_0$ will converge to 0, and no creators will stay. This completes the proof of $\lim_{C\rightarrow \infty} \mathbb{E}\left[ \longtermengagement\left(\UCseq\right) \right] = 0$.

    We still need to show $\lim_{C\rightarrow \infty} \mathbb{E}\left[ \longtermengagement\left(\FLseq\right) \right] > 0$, i.e., the \FL\ algorithm receives positive long-term engagement with non-zero probability even as $C\rightarrow \infty$. To see this, let $B\subseteq S$ be an arbitrary neighborhood such that $x^T y \leq \ebar$ for all $x,y\in B$, i.e., all users in $B$ are happy with all creators in $B$.\footnote{Recall we defined the \textit{happy distance} in Section~\ref{sec:clustering_alg_def} denoted $d$. Thus, $\left\lVert x-y\right\rVert \leq d$ for all $x,y\in B$. This also means $B\neq \emptyset$.} There is a positive probability that at least $\abar$ users fall in $B$.\footnote{More precisely, this probability is $\sum_{i=\abar}^{U} {\binom{U}{i}} \left( |B|/|S| \right)^i \left( 1-|B|/|S| \right)^{U-i}$.} Also, when $C\geq K$, there is a positive probability that at least $K$ creators fall in $B$. Both events happen in conjunction with positive probability due to their independence; when they do, there exists a stable set consisting of all users and creators in $B$ with a complete matching between them, and the stable set has positive total engagement w.p. 1. In such instances, $\longtermengagement\left(\FLseq\right) > 0$. Therefore, $\lim_{C\rightarrow \infty} \mathbb{E}\left[ \longtermengagement\left(\FLseq\right) \right] > 0$, which completes the proof.
\end{proofCharles}

\subsection{Proof of Theorem~\ref{thm:increase users}: Increase \texorpdfstring{$U$ and $\abar$}{U and a} Proportionally}  \label{sec:proof_increase_users}
Recall Theorem~\ref{thm:increase users} from Section~\ref{sec:increase_users_condensed}:
\begin{customTheorem}{Theorem~\ref{thm:increase users}}
    Fix constant integers $r$, $C$ and $K$ such that $K \in \left( C/2, C \right)$, and let $U=rC$, $\abar=rK$, $D=2$, and $\ebar \in [0,1]$ be arbitrary. Assume $u_i, c_j \sim \userdistribution$ i.i.d. for all $1\leq i\leq U, 1\leq j\leq C$, 
    and define $\epsilon := \Pr \left( \longtermengagement\left(\FLseq\right)=0 \right)$ given by the parameters $U, C, K, \ebar$ and $\abar$. 
    Then, when $r\rightarrow \infty$, 
    \begin{align}
        &\phantom{{}={}} \lim_{r\rightarrow \infty} \mathbb{E}\left[ \frac{\longtermengagement\left(\UCseq\right)}{\longtermengagement\left(\FLseq\right)} \, \middle| \, \longtermengagement\left(\FLseq\right)>0 \right]  \nonumber \\
        &\leq  \frac{1}{1-\epsilon} \left[ \sum_{i=1}^{C-K+1} \Pr \left( \frac{X_i + X_{K+i}}{2} \geq \frac{K}{C} \text{ and } \frac{X_{i-1} + X_{K+i-1}}{2} \leq 1-\frac{K}{C} \right) \right],  \label{eqn:thm_increase_users_appendix}
    \end{align}
    where $X_1, \dots, X_C$ are sorted from $C$ i.i.d. random variables drawn from $\text{Uniform}(0,1)$, such that $X_1 \leq X_2 \leq \dots \leq X_C$, and we define $X_0 := -\infty$ and $X_{C+1} := +\infty$ deterministically. In addition, the right hand side of \eqref{eqn:thm_increase_users_appendix} evaluates to
    
    \begin{align}
        \eqref{eqn:thm_increase_users_appendix} &= \frac{1}{1-\epsilon} \left[ \sum_{i=1}^{C-K+1} \frac{C!}{\left(i-2\right)! \left(K-2\right)! \left(C-K-i+1\right)!} \int_{0}^{\min\left\{ 1-K/C, 3-4K/C \right\}} \int_{\max\left\{ a, 2K/C-1 \right\}}^{2-2K/C-a} \int_b^{2-2K/C-a} \right.  \nonumber \\
        &\phantom{{}={}} \left.  a^{i-2} \left(c-b\right)^{K-2} \left(1-\max\left\{ c, 2\frac{K}{C}-b \right\} \right)^{C-K-i+1} \ dc \ db \ da \right].  \label{eqn:thm_increase_users_bound_appendix}
    \end{align}
\end{customTheorem}
We break down Theorem \ref{thm:increase users} into a series of lemmas. Note that some of these statements also apply to more generic priors, where user types are drawn from $\userdistribution$ and creator types are drawn from $\creatordistribution$ that may not be equal and may not be the uniform prior $\distribution$.

\subsubsection{Upper bounding expected approximation ratio by a probability}
\begin{lemma}  \label{lemma:approx_ratio_prob_K_stay}
    Under any priors for users $\userdistribution$ and creators $\creatordistribution$, and for any values of $U,C,K,D, \ebar$ and $\abar$,
    \begin{align*}
        \mathbb{E} \left[ \frac{\longtermengagement\left(\UCseq\right)}{\longtermengagement\left(\FLseq\right)} \, \middle| \, \longtermengagement\left(\FLseq\right)>0 \right] &\leq \Pr\left( \text{at least }K \text{ creators stay at }t=0 \text{ under }UC_0 \, \middle| \, \longtermengagement\left(\FLseq\right)>0 \right).
    \end{align*}
\end{lemma}
\begin{proofCharles}
    Conditioned on instances that satisfy $\longtermengagement\left(\FLseq\right)>0$, under the \IE\ algorithm, two possible scenarios may occur at $t=0$:
    \begin{itemize}
        \item Fewer than $K$ creators stay on the platform. In the next time step, all users will leave because they do not receive enough recommendations. Thus, $\longtermengagement\left(\UCseq\right)=0$.
        \item At least $K$ creators stay on the platform. In this case, $\longtermengagement\left(\UCseq\right)$ may or may not be nonzero, but $\longtermengagement\left(\UCseq\right)\leq \longtermengagement\left(\FLseq\right)$ by maximality of the \FL\ algorithm.
    \end{itemize}
    Therefore,
    \begin{align*}
        &\phantom{{}={}} \mathbb{E} \left[ \frac{\longtermengagement\left(\UCseq\right)}{\longtermengagement\left(\FLseq\right)} \, \middle| \, \longtermengagement(FL)>0 \right] \\
        &= \Pr\left( \text{at least }K \text{ creators stay under } UC_0 \, \middle| \, \longtermengagement\left(\FLseq\right)>0 \right)  \\
        &\phantom{{}={}} \cdot \mathbb{E} \left[ \frac{\longtermengagement\left(\UCseq\right)}{\longtermengagement\left(\FLseq\right)} \, \middle| \, \longtermengagement\left(\FLseq\right)>0, \text{at least }K \text{ creators stay under } UC_0 \right]  \\
        &\phantom{{}={}} + \Pr\left( \text{fewer than }K \text{ creators stay under } UC_0 \, \middle| \, \longtermengagement\left(\FLseq\right)>0 \right)  \\
        &\phantom{{}={}}  \cdot \mathbb{E} \left[ \frac{\longtermengagement\left(\UCseq\right)}{\longtermengagement\left(\FLseq\right)} \, \middle| \, \longtermengagement\left(\FLseq\right)>0, \text{fewer than }K \text{ creators stay under } UC_0 \right]  \\
        &\leq \Pr\left( \text{at least }K \text{ creators stay under } UC_0 \, \middle| \, \longtermengagement\left(\FLseq\right)>0 \right) \cdot 1  \\
        &\phantom{{}={}}  + \Pr\left( \text{fewer than }K \text{ creators stay under } UC_0 \, \middle| \, \longtermengagement\left(\FLseq\right)>0 \right) \cdot 0  \\
        &= \Pr\left( \text{at least }K \text{ creators stay at }t=0 \text{ under }UC_0 \, \middle| \, \longtermengagement\left(\FLseq\right)>0 \right).
    \end{align*}
\end{proofCharles}
A natural corollary comes from Lemma~\ref{lemma:approx_ratio_prob_K_stay}:
\begin{corollary}  \label{cor:approx_ratio_prob_K_stay_fl>0}
    Let $\epsilon := \Pr \left( \longtermengagement\left(\FLseq\right)=0 \right)$. Under any priors for users $\userdistribution$ and creators $\creatordistribution$, and for any values of $U,C,K,D$ and $\abar$ that satisfy $C\geq K$, $U\geq \abar$ and $\epsilon < 1$,
    \begin{align}
        \mathbb{E} \left[ \frac{\longtermengagement\left(\UCseq\right)}{\longtermengagement\left(\FLseq\right)} \, \middle| \, \longtermengagement\left(\FLseq\right)>0 \right] &\leq \frac{1}{1-\epsilon} \Pr\left( \text{at least }K \text{ creators stay at }t=0 \text{ under }UC_0 \right).  \label{eqn:cor_approx_ratio_prob_K_stay_fl>0}
    \end{align}

\end{corollary}
\begin{proofCharles}
    \begin{align*}
        &\phantom{{}={}} \Pr\left( \text{at least }K \text{ creators stay at }t=0 \text{ under }UC_0 \right)  \\
        &= \Pr\left( \longtermengagement\left(\FLseq\right)>0 \right) \cdot \Pr\left( \text{at least }K \text{ creators stay at }t=0 \text{ under }UC_0 \, \middle| \, \longtermengagement\left(\FLseq\right)>0 \right) \\
        &\phantom{{}={}} + \Pr\left( \longtermengagement\left(\FLseq\right)=0 \right) \cdot \Pr\left( \text{at least }K \text{ creators stay at }t=0 \text{ under }UC_0 \, \middle| \, \longtermengagement\left(\FLseq\right)=0 \right)  \\
        &\geq \left( 1-\epsilon \right) \Pr\left( \text{at least }K \text{ creators stay at }t=0 \text{ under }UC_0 \, \middle| \, \longtermengagement\left(\FLseq\right)>0 \right) + \epsilon \cdot 0,
    \end{align*}
    which gives \eqref{eqn:cor_approx_ratio_prob_K_stay_fl>0} upon rearranging and using Lemma~\ref{lemma:approx_ratio_prob_K_stay}.
\end{proofCharles}

The rest of the proof upper bounds the limit of $\Pr\left( \text{at least }K \text{ creators stay at }t=0 \text{ under }UC_0 \right)$ as $r\rightarrow \infty$, by analyzing the number of creators that \UC\ is able to retain at $t=0$ and comparing it to $K$. 

\subsubsection{Characterizing All Possible Recommendations into Regions}
Suppose we fix a realization of creator types. We first derive a series of regions that cover the entire support of the prior, such that the \IE\ algorithm gives different recommendations for users in different regions. We then express the number of users in these regions as a multinomial distribution with probabilities as their sizes.
\begin{lemma}  \label{lemma:regions}
    Assume $D=2$, $K<C$, and $\userdistribution = \creatordistribution = \distribution$, the uniform distribution over the set of non-negative unit vectors in $\R^2$. Sort all creators in increasing order of their angular representation: $\creatorset_0 = \left\{ c_1, c_2, \dots, c_C \right\}$, where $c_i = \left( \cos X_i \pi/2, \sin X_i \pi/2 \right)$, and $X_1, \dots, X_C$ are random variables that are sorted in increasing order from i.i.d. uniform random variables on $\left[0,1\right]$. Define regions $R_1, \dots, R_{C-K+1} \subseteq [0,1]$ as follows:
    \begin{align*}
        R_1 &= \left[ 0, \frac{X_1+X_{K+1}}{2} \right],  \\
        R_i &= \left( \frac{X_{i-1} + X_{i+K-1}}{2}, \frac{X_{i} + X_{i+K}}{2} \right], \quad \forall \, 2\leq i\leq C-K,  \\
        R_{C-K+1} &= \left( \frac{X_{C-K} + X_{C}}{2}, 1 \right].
    \end{align*}
    Then, at $t=0$, \IE\ assigns a user with type $u = \left( \cos y \pi/2, \sin y \pi/2 \right)$ where $y\in [0,1]$ to creators $\left\{ i, i+1, \dots, i+K-1 \right\}$ if and only if $y\in R_i$.
\end{lemma}
Note that $\bigcup_{i=1}^{C-K+1} R_i = [0,1]$ and $R_i \cap R_j = \emptyset$ for all $i\neq j$. Therefore, Lemma~\ref{lemma:regions} completely characterizes all possible recommendations that \IE\ can give at $t=0$.

\begin{proofCharles}
     First, because all user and creator types are non-negative unit vectors in $\mathbb{R}^2$, user engagement $u^T c_j$ is the cosine of the angular distance between the two vectors. Therefore, we can equivalently represent all user and creator types as scalars on $[0,1]$, and \IE\ assigns each user to the $K$ nearest creators (which give maximum user engagement).
     
     When $y=0$, the user's recommendations are creators $1,2,\dots,K$. Consider what happens as $y$ increases from $0$ to $1$. Once $y$ increases beyond $X_1$, its distance from creator 1 starts increasing, and is also the greatest among all creators whose distances from the user are increasing. On the other hand, the user's distance from creator $K+1$ is decreasing, and is the smallest among the remaining creators that are not part of the recommendations. Therefore, the first change in recommendations must be replacing creator 1 with creator $K+1$, which happens at $y=\frac{X_1 + X_{K+1}}{2}$, beyond which $y$ is closer to $X_{K+1}$ than $X_1$. This gives endpoints of $R_1$ and the start of $R_2$.

     Similarly, at $\frac{X_2 + X_{K+2}}{2}$, creator $2$ is replaced with creator $K+2$ in the recommendations, which gives the end of $R_2$. Repeating this gives definitions of all regions $R_1, \dots, R_{C-K+1}$.
\end{proofCharles}

\begin{definition}  \label{def:probs_users}
    For all $1\leq i\leq C-K+1$, let $P_i$ be the length of $R_i$, and $U_i$ be the number of users in $R_i$ in a random instance. Specifically, $P_i$'s are defined such that
    \begin{align*}
        P_1 &= \frac{X_1+X_{K+1}}{2},  \\
        P_i &= \frac{X_{i} + X_{i+K}}{2} - \frac{X_{i-1} + X_{i+K-1}}{2}, \quad \forall \, 2\leq i\leq C-K,  \\
        P_{C-K+1} &= 1 - \frac{X_{C-K} + X_{C}}{2}.
    \end{align*}
\end{definition}
Note that with uniform priors, $P_i$ can also be interpreted as the probability that a random user falls within region $R_i$. This means the random variables $U_1, \dots, U_{C-K+1}$ follow a multinomial distribution:
\begin{align}
    \left( U_1, \dots, U_{C-K+1} \right) \sim \text{Mult} \left( U, \left( P_1, \dots, P_{C-K+1} \right) \right).  \label{eqn:users_multinomial}
\end{align}

\subsubsection{Properties of Staying Creators}
We now analyze the set of creators that can stay on the platform after $t=0$. This will allow us to establish \eqref{eqn:thm_increase_users_appendix}, which upper bounds the expected approximation ratio of \IE\ by the probability of an unlikely event that is defined on a sequence of sorted uniform random variables, whose definitions are not in the context of our problem.

\begin{lemma}  \label{lemma:creators_adjacent}
    Assume $K>C/2$ and $U=rC$ for some integer $r$, in addition to assumptions in Lemma~\ref{lemma:regions}. Then, the set of creators that stay at $t=0$ under $UC_0$ must be $\left\{ i, i+1, \dots, j-1, j \right\}$ for some $i,j\in [1, C]$. Furthermore, $j-i+1 \leq K$, i.e., at most $K$ creators can stay.
\end{lemma}

\begin{proofCharles}
    Using Lemma~\ref{lemma:regions}, creator $i$'s audience size $a_{i,0}$ is:
    \begin{align*}
        a_{i,0}&= \begin{cases}
            U_1 + \dots + U_i, & 1 \leq i \leq C-K,  \\
            U, & C-K+1\leq i \leq K,  \\
            U_{i-K+1} + \dots + U_{C-K+1}, & K+1 \leq i \leq C.
        \end{cases}  %
    \end{align*}
    
    Recall that each creator needs $\bar{a}=rK$ users, and note that creators $C-K+1,\dots,K$ will always stay. Additionally, observe the following facts:
    \begin{enumerate}
        \item For any $1\leq i < j \leq C-K$, if creator $i$ stays, then creator $j$ stays. This is because $a_{j,0} = \sum_{k=1}^j U_k > \sum_{k=1}^i U_k = a_{i,0}$.  \label{enum:random_instances_staying_creators_heads}
        \item For any $K+1\leq i < j \leq C$, if creator $j$ stays, then creator $i$ stays. This is because $a_{j,0} = \sum_{k=j-K+1}^{C-K+1} U_k < \sum_{k=i-K+1}^{C-K+1} U_k = a_{i,0}$.  \label{enum:random_instances_staying_creators_tails}
        \item For any $i\in [1,C-K]$, creators $i$ and $i+K$ cannot stay at the same time. This is because $a_{i,0} + a_{i+K,0} = \sum_{k=1}^i U_k + \sum_{k=i+1}^{C-K+1} U_k = \sum_{k=1}^{C-K+1} U_k = U$, so if creator $i$ stays, then $a_{i,0} \geq \bar{a} = rK > rC/2 = U/2$, so $a_{i+K} < U/2 < \bar{a}$ which means creator $i+K$ leaves; and vice versa.  \label{enum:random_instances_staying_creators_lengthK}
    \end{enumerate}
    
    Let $i$ and $j$ be the first and last staying creators respectively. By Facts~\ref{enum:random_instances_staying_creators_heads} and \ref{enum:random_instances_staying_creators_tails}, any creators in $[i+1, C-K]$ and $[K+1, j-1]$ must also stay; also, creators $[C-K+1, K]$ always stay. This means the set of staying creators must be a continuous range $\left\{ i, i+1, \dots, j-1, j \right\}$. By Fact~\ref{enum:random_instances_staying_creators_lengthK}, $i+K$ cannot be within this range, which gives $j-i+1\leq K$.
\end{proofCharles}

\begin{lemma}  \label{lemma:bound_U}
    Under all assumptions in Lemma~\ref{lemma:creators_adjacent}, 
    \begin{align*}
        &\phantom{{}={}} \Pr\left( \text{at least }K \text{ creators stay at }t=0 \text{ under }UC_0 \right) \\
        &= \sum_{i=1}^{C-K+1} \Pr \left( \frac{\sum_{j=1}^i U_j}{rC} \geq \frac{K}{C} \text{ and } \frac{\sum_{j=i}^C U_j}{rC} \geq \frac{K}{C} \right).
    \end{align*}
\end{lemma}
\begin{proofCharles}
    By Lemma~\ref{lemma:creators_adjacent}, at least $K$ creators stay if and only if exactly $K$ adjacent creators stay. Thus,
    \begin{align}
         &\phantom{{}={}} \Pr\left( \text{at least }K \text{ creators stay at }t=0 \text{ under }UC_0 \right) \nonumber \\
         &= \sum_{i=1}^{C-K+1} \Pr \left( \text{creators }\left\{ i, i+1, \dots, i+K-1 \right\} \text{ stay} \right).  \label{eqn:creators_stay_list}
    \end{align}
    Fact 1 in Lemma \ref{lemma:creators_adjacent} suggests that all creators in $[i, C-K]$ stay if and only if creator $i$ stays, and all creators in $[K+1, i+K-1]$ stay if and only if creator $i+K-1$ stays. This means
    \begin{align}
        \eqref{eqn:creators_stay_list} &= \sum_{i=1}^{C-K+1} \Pr \left( \text{creators } i \text{ and } i+K-1 \text{ stay} \right)  \nonumber \\
        &= \sum_{i=1}^{C-K+1} \Pr \left( \sum_{j=1}^i U_j \geq rK \text{ and } \sum_{j=i}^C U_j \geq rK \right)  \label{eqn:tail_creators_stay_U} \\
        &= \sum_{i=1}^{C-K+1} \Pr \left( \frac{\sum_{j=1}^i U_j}{rC} \geq \frac{K}{C} \text{ and } \frac{\sum_{j=i}^C U_j}{rC} \geq \frac{K}{C} \right),  \label{eqn:tail_creators_stay_ratio}
    \end{align}
    where \eqref{eqn:tail_creators_stay_U} uses $\bar{a}=rK$, and \eqref{eqn:tail_creators_stay_ratio} divides both inequalities by $U=rC$ so that both sides are finite even when $r\rightarrow \infty$. Intuitively, \eqref{eqn:tail_creators_stay_ratio} is defined on the \textit{proportion} of users that creators $i$ and $i+K-1$ receive, and asserts that they must both be at least $K/C$.
\end{proofCharles}

\begin{lemma} \label{lemma:bound_X}
    Under all assumptions in Lemma~\ref{lemma:creators_adjacent}, when $r\rightarrow \infty$ (thus $U \rightarrow \infty$ and $\bar{a}$ increases correspondingly), 
    \begin{align*}
        & \Pr\left( \text{at least }K \text{ creators stay at }t=0 \text{ under }UC_0 \right) \\
        \xrightarrow{r\rightarrow \infty}& \sum_{i=2}^{C-K} \Pr \left( \frac{X_i + X_{K+i}}{2} \geq \frac{K}{C} \text{ and } \frac{X_{i-1} + X_{K+i-1}}{2} \leq 1-\frac{K}{C} \right)  \\
    &+ \Pr \left( \frac{X_1 + X_{K+1}}{2} \geq \frac{K}{C} \right) + \Pr \left( \frac{X_{C-K} + X_C}{2} \leq 1-\frac{K}{C} \right).
    \end{align*}
\end{lemma}

\begin{proofCharles}
Recall from~\eqref{eqn:users_multinomial} that $\left( U_1, \dots, U_{C-K+1} \right) \sim \text{Mult} \left( U, \left( P_1, \dots, P_{C-K+1} \right) \right)$ and $U=rC$. Define 
$$ Y_i := \frac{U_i - rCP_i}{\sqrt{rCP_i}}, \quad \forall 1\leq i\leq C-K+1. $$
By Theorem 11.5 from \cite{georgii2012stochastics}, which gives the approximation of multinomial distributions by normal distributions,
\begin{align*}
    \begin{bmatrix}
        Y_1 \\
        \dots \\
        Y_{C-K} \\
        Y_{C-K+1}
    \end{bmatrix} &\xrightarrow[r\rightarrow \infty]{d} Q \begin{bmatrix}
        Z_1 \\
        \dots \\
        Z_{C-K} \\
        0
    \end{bmatrix},
\end{align*}
where $Q$ is an orthogonal matrix whose last column is $\left[ \sqrt{P_1}, \dots, \sqrt{P_{C-K+1}} \right]^T$, and $Z_1, \dots, Z_{C-K}$ are i.i.d. standard normal random variables. Additionally,
\begin{align*}
    \begin{bmatrix}
        1/\sqrt{rCP_1} \\
        \dots \\
        1/\sqrt{rCP_{C-K}} \\
        1/\sqrt{rCP_{C-K+1}}
    \end{bmatrix} &\xrightarrow[r\rightarrow \infty]{i.p.} \begin{bmatrix}
        0 \\
        \dots \\
        0 \\
        0
    \end{bmatrix},
\end{align*}
since the vector on the left hand side is deterministic and converges to $\mathbf{0}$ as $r\rightarrow \infty$. Thus, Slutsky's theorem gives convergence of their element-wise products in distribution:
\begin{align*}
    \begin{bmatrix}
        Y_1/\sqrt{rCP_1} \\
        \dots \\
        Y_{C-K}/\sqrt{rCP_{C-K}} \\
        Y_{C-K+1}/\sqrt{rCP_{C-K+1}}
    \end{bmatrix} &\xrightarrow[r\rightarrow \infty]{d} \begin{bmatrix}
        0 \\
        \dots \\
        0 \\
        0
    \end{bmatrix}.
\end{align*}
This means
\begin{align*}
    \begin{bmatrix}
        \frac{U_1}{rC} \\
        \dots \\
        \frac{U_{C-K+1}}{rC}
    \end{bmatrix} &= \begin{bmatrix}
        P_1 \left( 1 + \frac{U_1 - rCP_1}{rCP_1} \right) \\
        \dots \\
        P_{C-K+1} \left( 1 + \frac{U_{C-K+1} - rCP_{C-K+1}}{rCP_{C-K+1}} \right)
    \end{bmatrix}  \\
    &= \begin{bmatrix}
        P_1 \\
        \dots \\
        P_{C-K+1}
    \end{bmatrix} + \begin{bmatrix}
        P_1 \frac{Y_1}{\sqrt{rCP_1}} \\
        \dots \\
        P_{C-K+1} \frac{Y_{C-K+1}}{\sqrt{rCP_{C-K+1}}}
    \end{bmatrix}  \\
    &\xrightarrow[r\rightarrow \infty]{d} \begin{bmatrix}
        P_1 \\
        \dots \\
        P_{C-K+1}
    \end{bmatrix}.
\end{align*}
Thus, from Lemma~\ref{lemma:bound_U},
\begin{align}
    &\Pr\left( \text{at least }K \text{ creators stay at }t=0 \text{ under }UC_0 \right)  \nonumber \\
    =& \sum_{i=1}^{C-K+1} \Pr \left( \frac{\sum_{j=1}^i U_j}{rC} \geq \frac{K}{C} \text{ and } \frac{\sum_{j=i}^C U_j}{rC} \geq \frac{K}{C} \right)  \nonumber \\
    \xrightarrow{r\rightarrow \infty}& \sum_{i=1}^{C-K+1} \Pr \left( \sum_{j=1}^i P_j \geq \frac{K}{C} \text{ and } \sum_{j=i}^C P_j \geq \frac{K}{C} \right)  \nonumber \\
    =& \sum_{i=2}^{C-K} \Pr \left( \frac{X_i + X_{K+i}}{2} \geq \frac{K}{C} \text{ and } 1-\frac{X_{i-1} + X_{K+i-1}}{2} \geq \frac{K}{C} \right) \nonumber  \\
    &+ \Pr \left( \frac{X_1 + X_{K+1}}{2} \geq \frac{K}{C} \right) + \Pr \left( 1-\frac{X_{C-K} + X_C}{2} \geq \frac{K}{C} \right)  \label{eqn:creators_stay_X_bounds_working} \\
    =& \sum_{i=1}^{C-K+1} \Pr \left( \frac{X_i + X_{K+i}}{2} \geq \frac{K}{C} \text{ and } \frac{X_{i-1} + X_{K+i-1}}{2} \leq 1-\frac{K}{C} \right), \label{eqn:creators_stay_X_bounds} %
\end{align}
where \eqref{eqn:creators_stay_X_bounds_working} uses definitions of $P_i$'s, and \eqref{eqn:creators_stay_X_bounds} is obtained by defining $X_0 := -\infty$ and $X_{C+1} := +\infty$.
\end{proofCharles}

Lemma~\ref{lemma:bound_X} gives an explicit expression of $\Pr\left( \text{at least }K \text{ creators stay at }t=0 \text{ under }UC_0 \right)$ as a probability that does not depend on the problem setup, since $X_i$'s are simply sorted i.i.d. random variables drawn from $\text{Uniform}[0,1]$. This completes the first and major part of the proof of Theorem~\ref{thm:increase users}, i.e., Equation~\eqref{eqn:thm_increase_users_appendix}.

Before we present the rest of the proof, we first give an intuitive justification for why \eqref{eqn:creators_stay_X_bounds} is a low probability event when $K>C/2$. Note that each term in \eqref{eqn:creators_stay_X_bounds} requires the event $\left(X_i + X_{K+i}\right)/2 \geq K/C$ and $\left(X_{i-1} + X_{K+i-1}\right)/2 \leq 1-K/C$. When $K>C/2$, especially when $K$ is large, there is a significant gap between $1-K/C$ and $K/C$. On the other hand, the expected difference between $\left(X_{i-1} + X_{K+i-1}\right)/2$ and $\left(X_i + X_{K+i}\right)/2$ is $1/(C+1)$, which is much smaller. This means that such an event requires a much larger than average gap between $X_{i-1}$ and $X_i$, and between $X_{K+i-1}$ and $X_{K+i}$. Both pairs are adjacent, so there must not have been any other random variables taking on values in $(X_{i-1}, X_i)$ and $(X_{K+i-1}, X_{K+i})$, which becomes unlikely when the gap is large.

\subsubsection{Evaluating the Probability}

\begin{proofCharles}[Proof of Theorem~\ref{thm:increase users}]
    Lemma~\ref{lemma:bound_X} combined with Corollary~\ref{cor:approx_ratio_prob_K_stay_fl>0} already completes the first half of the proof, noting that $\eqref{eqn:thm_increase_users_appendix}=\frac{1}{1-\epsilon}\eqref{eqn:creators_stay_X_bounds}$. The only remaining task is to obtain \eqref{eqn:thm_increase_users_bound_appendix}, by evaluate \eqref{eqn:creators_stay_X_bounds} as an integral using joint PDF of the four random variables:
    \begin{align}
        & \Pr \left( \frac{X_i + X_{K+i}}{2} \geq \frac{K}{C} \text{ and } \frac{X_{i-1} + X_{K+i-1}}{2} \leq 1-\frac{K}{C} \right)  \nonumber \\
        =& \int_{0}^{\min\left\{ 1-K/C, 3-4K/C \right\}} \int_{\max\left\{ a, 2K/C-1 \right\}}^{2-2K/C-a} \int_b^{2-2K/C-a} \int_{\max\left\{ c, 2K/C-b \right\}}^1 f_{X_{i-1}, X_i, X_{K+i-1}, X_{K+i}}(a,b,c,d)  \nonumber \\
        & \ dd \ dc \ db \ da  \nonumber \\
        =& \int_{0}^{\min\left\{ 1-K/C, 3-4K/C \right\}} \int_{\max\left\{ a, 2K/C-1 \right\}}^{2-2K/C-a} \int_b^{2-2K/C-a} \int_{\max\left\{ c, 2K/C-b \right\}}^1 \left( \frac{C!}{\left(i-2\right)! \left(K-2\right)! \left(C-K-i\right)!} \right. \nonumber \\
        & \left. a^{i-2} \left(c-b\right)^{K-2} \left(1-d\right)^{C-K-i} \right) \ dd \ dc \ db \ da  \nonumber \\
        =& \frac{C!}{\left(i-2\right)! \left(K-2\right)! \left(C-K-i+1\right)!} \int_{0}^{\min\left\{ 1-K/C, 3-4K/C \right\}} \int_{\max\left\{ a, 2K/C-1 \right\}}^{2-2K/C-a} \int_b^{2-2K/C-a}  \nonumber \\
        &   a^{i-2} \left(c-b\right)^{K-2} \left(1-\max\left\{ c, 2\frac{K}{C}-b \right\} \right)^{C-K-i+1} \ dc \ db \ da.  \label{eqn:bound_exact_integrals}
    \end{align}
    This completes the proof as $\eqref{eqn:thm_increase_users_bound_appendix}=\frac{1}{1-\epsilon} \sum_{i=1}^{C-K+1} \eqref{eqn:bound_exact_integrals}$.
\end{proofCharles}

\section{Details and Proofs of the \CA\ Algorithm} 
 \label{sec:proof_clustering}
In this appendix, we present further discussions of the \CA\ algorithm. First, we formally define the concepts of neighborhood ball and kissing number, offering additional insights from the brief definition in Section~\ref{sec:clustering_alg_def}. We then prove Theorem~\ref{thm:neighbor_apx_ratio_3D}, which establishes the approximation guarantee of the algorithm. Finally, we present a counter-example where the \CA\ algorithm performs significantly worse than the first-best \FL\ algorithm; more generally, any neighborhood-based local matching algorithm is also suboptimal on this example, even if is uses more sophisticated techniques than the \CA\ algorithm.

\subsection{Clarifications of Concepts}
\subsubsection{Neighborhood Ball}
We first offer a more formal definition of a \textit{neighborhood ball} than we did in Section~\ref{sec:clustering_alg_def} with additional discussions:

\begin{definition}[Neighborhood ball]\label{def:neighbor_ball}
    A \textit{neighborhood ball centered at user} $i\in \userset_0$, denoted $B(u_i)$, is a ball centered at $u_i$ such that the intersection of its boundary and the unit-vector type space $\left\{ x\in \R^D_{\geq 0}: \left\lVert x \right\rVert = 1 \right\}$ is a sphere with diameter $d$ (the happy distance). Formally,  $$ B(u_i) = \left\{ x\in \R^D : \left\lVert x - u_i \right\rVert \leq r \right\} $$
    for some constant $r$ (the radius of neighborhood balls) such that 
    $$ \max_{\substack{x,y\in \mathbb{R}^D_{\geq 0} \\ \left\lVert x \right\rVert = \left\lVert y \right\rVert = 1 \\ \left\lVert x-u_i \right\rVert = \left\lVert y-u_i \right\rVert = r}} \left\lVert x - y \right\rVert = d.  $$
    In particular, we call $r$ the \textit{neighborhood radius}.
\end{definition}
A consequence of Definition~\ref{def:neighbor_ball} is that for any $i,k\in \left[1,U\right]$ and $j\in \left[1,C\right]$, if $u_k$ and $c_j$ are both in the ball $B\left( u_i \right)$, then $\left\lVert u_k - c_j \right \rVert \leq d$. This means user $k$ is happy with creator $j$, even if $i\neq k$. In other words, all users in $B\left( u_i \right)$ are happy with all creators in the ball. For this reason, we called $d$ the ``effective diameter'' of neighborhood balls in Section~\ref{sec:clustering_alg_def}. 

\begin{figure}[h]
 \centering
 \begin{subfigure}[b]{0.55\textwidth}
     \centering
     \includegraphics[width=\textwidth]{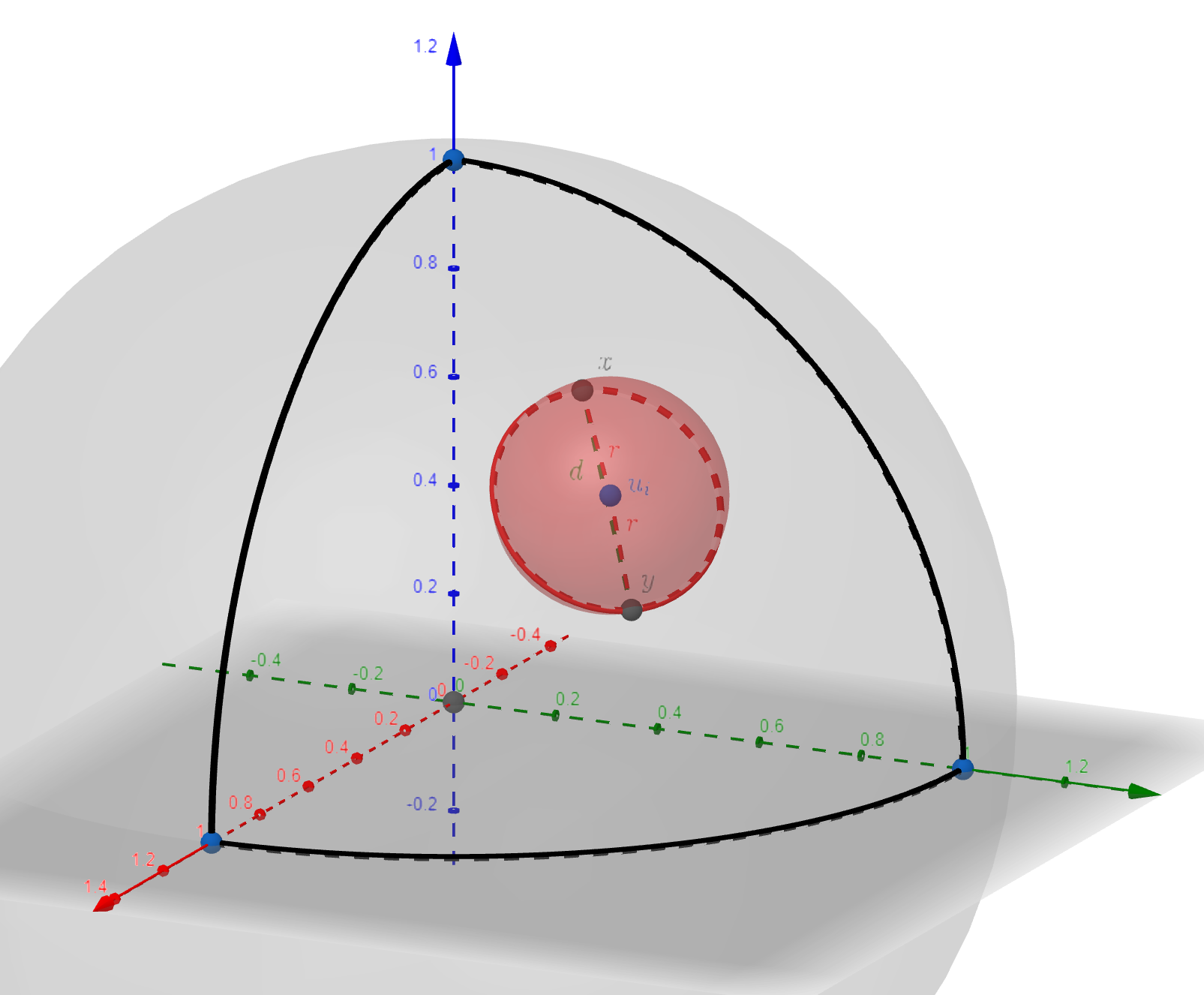}
     \caption{A neighborhood ball in $\mathbb{R}^3$. The red ball is $B(u_i)$, the grey ball is the set of all unit vectors, and the convex triangular region is the type space (non-negative unit vectors). Note the extreme points $x$ and $y$ satisfy $\left\lVert x-y \right\rVert = d$. }
     \label{fig:neighbor_APX_neighborhood_ball_3dfig}
 \end{subfigure}
 \hfill
 \begin{subfigure}[b]{0.4\textwidth}
     \centering
     \includegraphics[width=\textwidth]{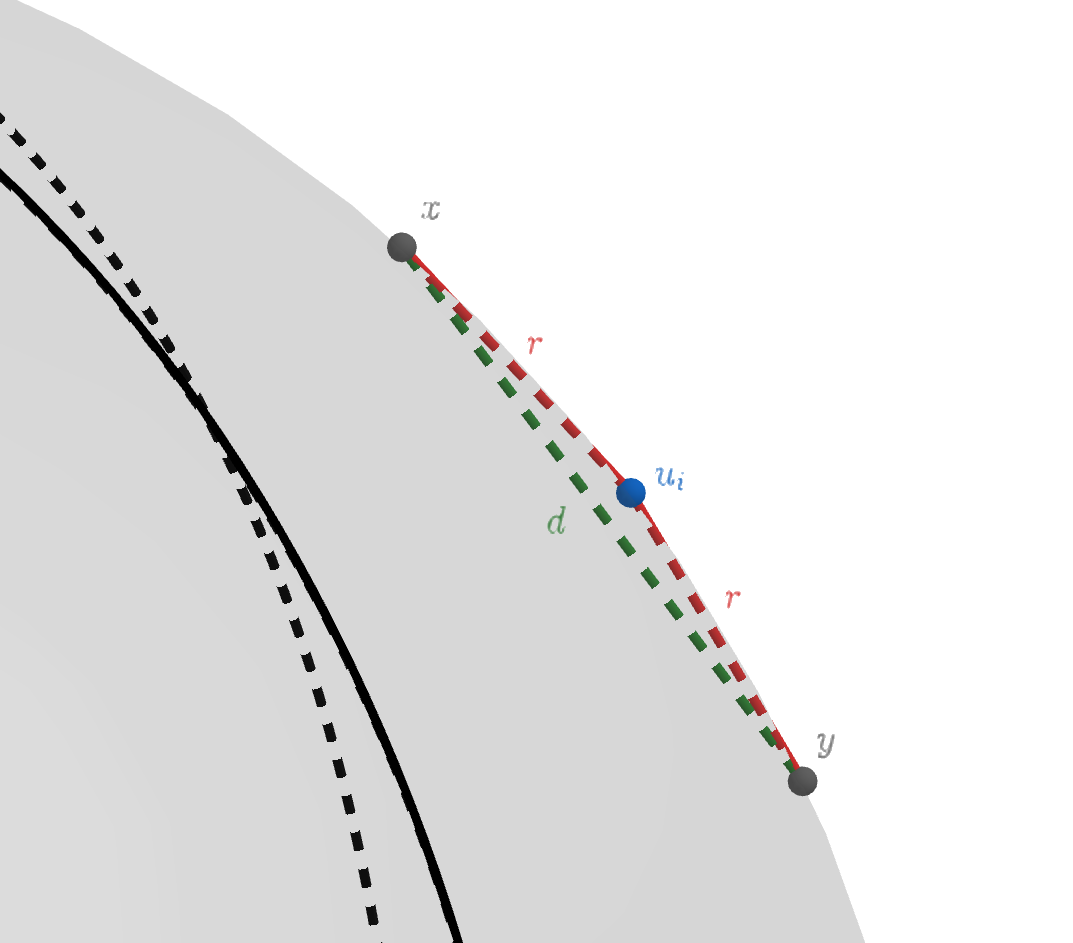}
     \caption{Cross section of the intersection of $B(u_i)$ and the type space. Notice $d<2r$ due to convexity of the type space $\left\{ x\in \mathbb{R}^D_{\geq 0}: \left\lVert x \right\rVert = 1 \right\}$. }
     \label{fig:neighbor_APX_neighborhood_ball_xsection}
 \end{subfigure}
    \caption{Neighborhood ball in $\mathbb{R}^3$}
    \label{fig:neighbor_APX_neighborhood_ball}
\end{figure}

However, the radius $r$ of all neighborhood balls (note they all have the same size) satisfies $r>d/2$, which means the ``effective diameter'' $d$ is actually smaller than the true diameter of neighborhood balls. This is because the unit-vector type space $\left\{ x\in \mathbb{R}^D_{\geq 0}: \left\lVert x \right\rVert = 1 \right\}$, which contains both $u_i$ and the endpoints $x,y$ in Definition~\ref{def:neighbor_ball}, is convex. See Figure \ref{fig:neighbor_APX_neighborhood_ball} for an illustration.

\subsubsection{Kissing Number}
We also formally define the concept of \textit{kissing number} from geometry, and discuss a consequence that is relevant to the \CA\ algorithm.

\begin{definition}[Kissing number]  \label{def:kissing}
    The \textit{kissing number} in $\mathbb{R}^D$, which we denote as $\kappa_D$, is the maximum number of non-overlapping unit spheres in $\mathbb{R}^D$ that can exist such that they all touch a common unit sphere.
\end{definition}
An alternative definition is as follows:
\begin{corollary}  \label{cor:neighbor_apx_kissing}
    Let $C=\left\{x\in \mathbb{R}^D: \left\lVert x \right\rVert =1 \right\}$ be a unit hypersphere, and let $n$ be the maximum number such that there exists points $\left\{ x_1, \dots, x_n \right\} \subset C$ that satisfy $\left\lVert x_i - x_j \right\rVert \geq 1$ for all $1\leq i,j\leq n$. Then, $n=\kappa_D$. 
\end{corollary}
Conceptually, Corollary~\ref{cor:neighbor_apx_kissing} plays a role in our proof of Theorem~\ref{thm:neighbor_apx_ratio_3D}, as it allows us to bound the number of points within a ball whose pairwise distances are sufficiently large, which can be seen as the ``density'' of points in the ball.

The kissing number is conjectured to grow exponentially with $D$. Lower and upper bounds when $D\leq 24$ have been shown, but no bounds are known for $D>24$. The exact values of $\kappa_D$ remain unknown aside from $D=1,2,3,4,8$ and $24$.

\subsection{Proof of Theorem~\ref{thm:neighbor_apx_ratio_3D}: Approximation Guarantee of the \CA\ Algorithm}
Recall Theorem~\ref{thm:neighbor_apx_ratio_3D} from Section~\ref{sec:clustering_guarantee_density}:
\begin{customTheorem}{Theorem~\ref{thm:neighbor_apx_ratio_3D}}[Approximation of \CA]
    There exists function $r\left( \ebar, D \right)$ which only depends on user constraint value $\ebar$ and number of dimensions $D$, such that on all deterministic instances with dimensionality $D$ that satisfy the density Assumption \ref{assumption:density}, 
    $$ \frac{\longtermengagement\left( \CAseq \right)}{\longtermengagement\left( \FLseq \right)} \geq r\left( \ebar, D \right).$$
    In addition, when $D=3$, a lower bound for the approximation ratio is given by $r\left( \ebar, 3 \right) \geq \frac{\ebar}{37}$.
\end{customTheorem}
We also offered a conjeture in Section~\ref{sec:clustering_guarantee_density} for how this result may generate to higher dimensions, which we formally state below:
\begin{conjecture}[Approximation of \CA\ when $D>3$]  \label{conj:neighbor_apx_ratio_highD}
    For any fixed $D$ such that $D>3$, 
    $ r\left( \ebar, D \right) \geq \frac{\ebar}{1+\kappa_D (\kappa_D + 1)}.$
\end{conjecture}
We show a number of smaller results that collectively prove Theorem~\ref{thm:neighbor_apx_ratio_3D}. Most of them give concrete results for arbitrary values of $D$. Lemma~\ref{lemma:neighbor_apx_ring_cover_3D} is the only exception, which shows existence for any $D$ but only provides a quantitative result for $D=3$. We offer an analogous Conjecture~\ref{conj:neighbor_apx_ring_cover_highD} that is our best guess for how Lemma~\ref{lemma:neighbor_apx_ring_cover_3D} may apply to $D>3$. This conjecture, combined with other results, will lead to Conjecture~\ref{conj:neighbor_apx_ratio_highD}.

\subsubsection{Proof Outline}
It is easy to see that Algorithm~\ref{alg:neighbor_apx} returns a stable set: each user $i$ with $\rec(i)\neq \emptyset$ gets $K$ recommendations, each creator gets either 0 or at least $\abar$ users (some may get multiples of $\abar$ if they are in multiple balls), and each user is happy with all assigned creators as they are all in the same neighborhood ball.

To prove Theorem~\ref{thm:neighbor_apx_ratio_3D}, we show the following statements:
\begin{enumerate}
    \item Each user $i$ that is not in the final matching $\rec$ can be associated with a user $k$ whose neighborhood ball $B(u_k)$ was chosen in the matching. Furthermore, $\left\lVert u_i - u_k \right\rVert \leq d$. (Lemma \ref{lemma:neighbor_apx_excluded_because})
    \item Define a \textit{large ball} around each user whose neighborhood ball is chosen, and a \textit{ring} to be the region in the large ball that is not in the corresponding neighborhood ball. Then, every user must be in at least one large ball. (Proposition \ref{prop:neighbor_apx_large_balls})
    \item An upper bound exists on the number of unassigned users in each ring (Proposition \ref{prop:neighbor_apx_ring_density}). To show this, we first show each ring can be covered by a constant number of balls of radius $r$ (not necessarily centered at users) (Lemma \ref{lemma:neighbor_apx_ring_cover_3D}), then show each such coverage ball has an upper bound on the number of unassigned users (Lemma \ref{lemma:neighbor_apx_ball_density}).
    \item The upper bound on the number of users in each ring implies the algorithm keeps at least a constant fraction of all users (Proposition \ref{prop:neighbor_apx_user_ratio}), thus achieves a constant approximation ratio on engagement (Theorem \ref{thm:neighbor_apx_ratio_3D}).
\end{enumerate}

\subsubsection{Relating Unassigned Users to Assigned Balls}
Let $\userset^*$ be the set of users whose neighborhood balls were chosen during Algorithm~\ref{alg:neighbor_apx}. This means the final $\rec$ consists entirely of matchings within each $B(u_k)$ for each $k\in \userset^*$. We call these users \textit{representative users}. Additionally, we call user $i$ an \textit{unassigned user} if $\rec(i)=\emptyset$ in the final matching, i.e., $i$ is not in the stable set produced by the algorithm.

\begin{lemma}  \label{lemma:neighbor_apx_excluded_because}
    For every unassigned user $i$, there exists representative user $k\in \userset^*$ such that $B(u_i)$ and $B(u_k)$ overlap, and their intersection contains at least one user. (We call this ``$i$ was excluded because of $k$''.) Additionally, $\left\lVert u_i - u_k \right\rVert \leq d$.
\end{lemma}
\begin{proofCharles}
    Recall that in Algorithm~\ref{alg:neighbor_apx}, we use $\userset_i$ to denote the set of unassigned users in $B(u_i)$ at the time when user $i$ was examined, and $\rec(i)$ is the recommendations given to user $i$. Since $\rec(i)=\emptyset$ in the final matching, $\rec(i)=\emptyset$ at the time when user $i$ was examined during the algorithm. This happens only when $\left| \userset_i \right| < \abar$, i.e., $B(u_i)$ contains fewer than $\abar$ unassigned users at that time. By the density assumption, $B(u_i)$ contains at least $\abar$ users, so at least one user $j$ with $u_j \in B(u_i)$ must have already been assigned, i.e., $\rec(j)\neq \emptyset$. This means $j$ must have been assigned earlier in the algorithm as part of another neighborhood ball $B(u_k)$, so $u_j \in B(u_i) \cap B(u_k)$.

    Furthermore, $u_j \in B(u_i)$ means $\left\lVert u_i - u_j \right\rVert \leq r$, and similarly $\left\lVert u_k - u_j \right\rVert \leq r$. This means $u_i, u_k \in B(u_j)$, so $\left\lVert u_i - u_k \right\rVert \leq d$.
\end{proofCharles}

Note that an unassigned user $i$ may be excluded because of multiple representative users.

\begin{lemma}  \label{lemma:neighbor_apx_unassigned_ball_density}
    For every unassigned user $i$, $B(u_i)$ contains at most $\abar-1$ unassigned users in the final matching.
\end{lemma}
\begin{proofCharles}
    As shown in the proof of Lemma~\ref{lemma:neighbor_apx_excluded_because}, $\rec(i)=\emptyset$ in the final matching only if $\left| \userset_i \right| < \abar$ when user $i$ was examined. In later steps of the algorithm, some users in $\userset_i$ may become assigned, but no users in $B(u_i)$ can change from assigned to unassigned. Thus, the final set of unassigned users in $B(u_i)$ is a subset of $\userset_i$, and therefore has size at most $\abar-1$.
\end{proofCharles}
Lemma~\ref{lemma:neighbor_apx_unassigned_ball_density} provides a way to bound the density of unassigned users in each neighborhood ball.

\subsubsection{Placing All Users into Large Balls}
\begin{figure}
     \centering
     \begin{subfigure}[b]{0.48\textwidth}
         \centering
         \includegraphics[width=\textwidth]{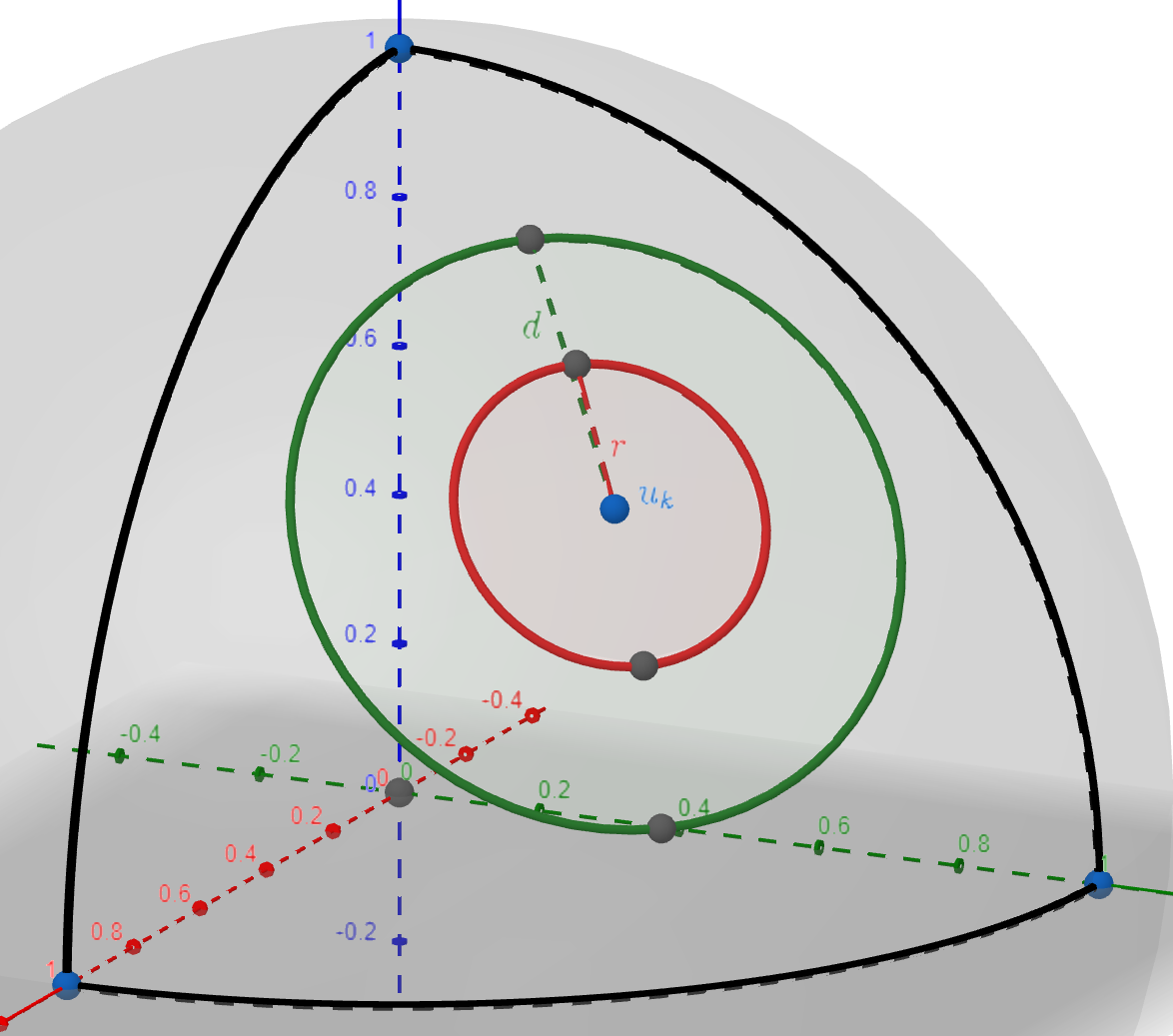}
         \caption{A neighborhood ball $B(u_k)$ (red) and a large ball $B'(u_k)$ (green) centered at $u_k$ in $\mathbb{R}^3$. Only the intersections of their boundaries with the type space are shown (the two circles). The ring $C(u_k)$ is the region between the boundaries.}
         \label{fig:neighbor_APX_large_ball_3dfig}
     \end{subfigure}
     \hfill
     \begin{subfigure}[b]{0.48\textwidth}
         \centering
         \includegraphics[width=\textwidth]{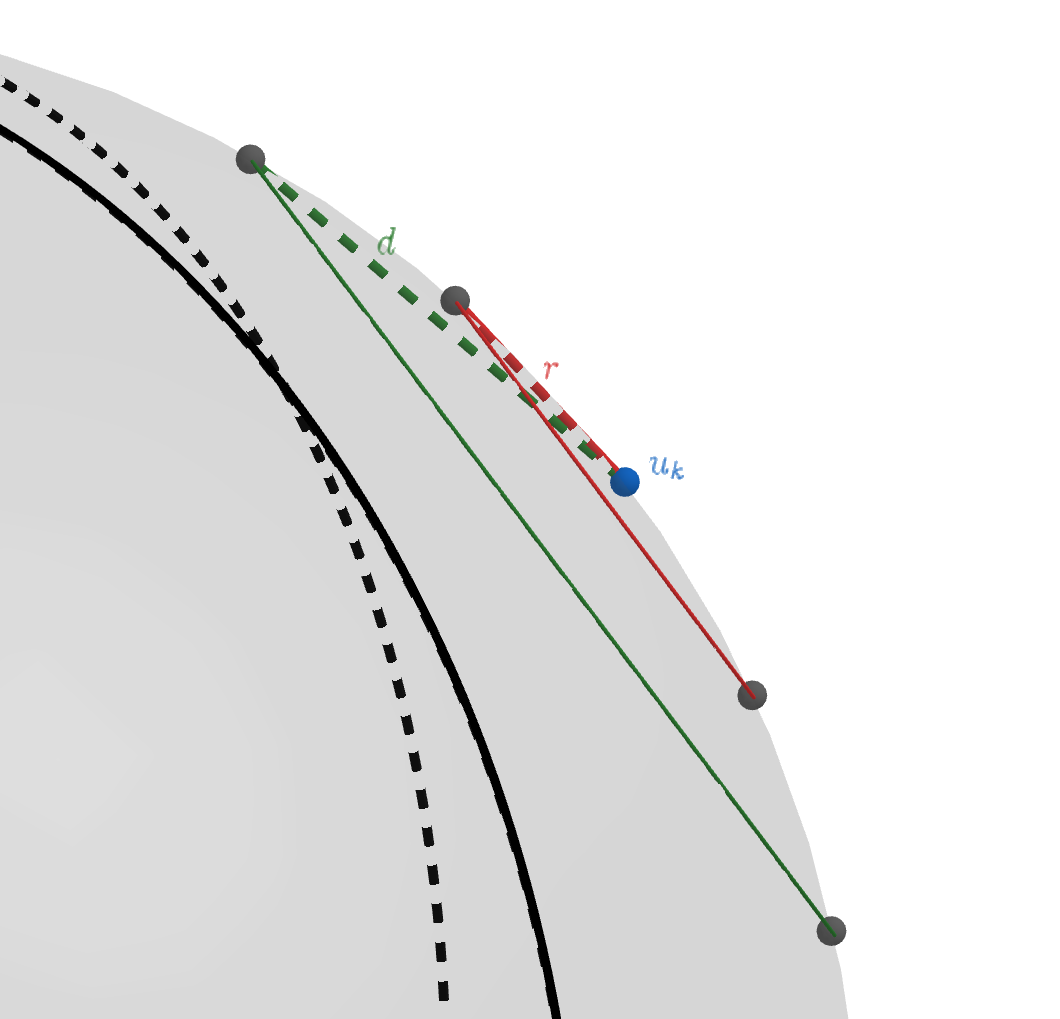}
         \caption{Cross section of the intersections of $B(u_k)$ and $B'(u_k)$ with the type space.}
         \label{fig:neighbor_APX_large_ball_xsection}
     \end{subfigure}
        \caption{Neighborhood ball, large ball and ring in $\mathbb{R}^3$}
        \label{fig:neighbor_APX_large_ball}
\end{figure}

\begin{definition}[Large balls and rings]
    For each representative user $k\in \userset^*$, define a \textit{large ball} centered at $k$, $B'(u_k)$, as a ball centered at $u_k$ with \textit{radius} $d$. Note that $B(u_k) \subseteq B'(u_k)$. Define a \textit{ring} centered at $k$ to be their difference, or the region in the large ball $B'(u_k)$ that is not in the neighborhood ball $B(u_k)$: $C(u_k) := B'(u_k) \backslash B(u_k)$.
\end{definition}
See Figure~\ref{fig:neighbor_APX_large_ball} for an illustration of large balls and rings.

\begin{proposition}  \label{prop:neighbor_apx_large_balls}
    Every user is in at least one large ball.
\end{proposition}
\begin{proofCharles}
    Consider any user $i$. If $i$ is unassigned, by Lemma~\ref{lemma:neighbor_apx_excluded_because}, there must exist representative user $k$ such that $\left\lVert u_i - u_k \right\rVert \leq d$, so $u_i \in B'(u_k)$. If $i$ is assigned, it must be within a neighborhood ball centered at a representative user $k$, so $u_i \in B(u_k) \subseteq B'(u_k)$.
\end{proofCharles}

Note that a user can be in multiple large balls, regardless of whether the user is assigned. Since Proposition~\ref{prop:neighbor_apx_large_balls} associates each user with at least one large ball, it offers us a way to bound the density of unassigned users in the entire instance by bounding the number of unassigned users in each large ball. Additionally, the neighborhood ball associated with each large ball must be fully assigned (by definition of large balls and representative users), so the bound only needs to be on each ring, which is our next step.

\subsubsection{Bounding the Number of Unassigned Users in Rings}

In this step, we provide an upper bound on the number of \textit{unassigned} users in each ring $C(u_k)$ when dimensionality is fixed. We do so by first showing each ring can be covered by a fixed number of ``coverage balls'' of radius $r$, even though they may not be centered at users. We then bound the number of unassigned users in each coverage ball.

\begin{lemma}  \label{lemma:neighbor_apx_ring_cover_3D}
    For any representative user $k$, the set of all unit vectors in the ring $C(u_k)$ can be covered by a finite number of balls of radius $r$, which we denote as \textit{coverage balls}, and the number of such coverage balls only depends on dimension $D$. Formally, there exists a function $f:\mathbb{N} \mapsto \mathbb{N}$ where $f\left(D\right)$ gives the number of coverage balls required in $D$ dimensions, and a list of coverage ball centers $x_1, \dots, x_{f\left(D\right)} \in \left\{ x\in \mathbb{R}^D_{\geq 0}: \left\lVert x \right\rVert = 1 \right\}$, such that 
    $$ \bigcup_{i=1}^{f\left(D\right)} B(x_i) \supseteq C(u_k) \cap \left\{ x\in \mathbb{R}^D_{\geq 0}: \left\lVert x \right\rVert = 1 \right\}, $$
    where $x_1, \dots, x_{f\left(D\right)}$ are not necessarily user types. In addition, when $D=3$, $f\left(D\right) = 6$.
\end{lemma}
    \begin{figure}[h!]
        \centering
        \includegraphics[width=0.5\textwidth]{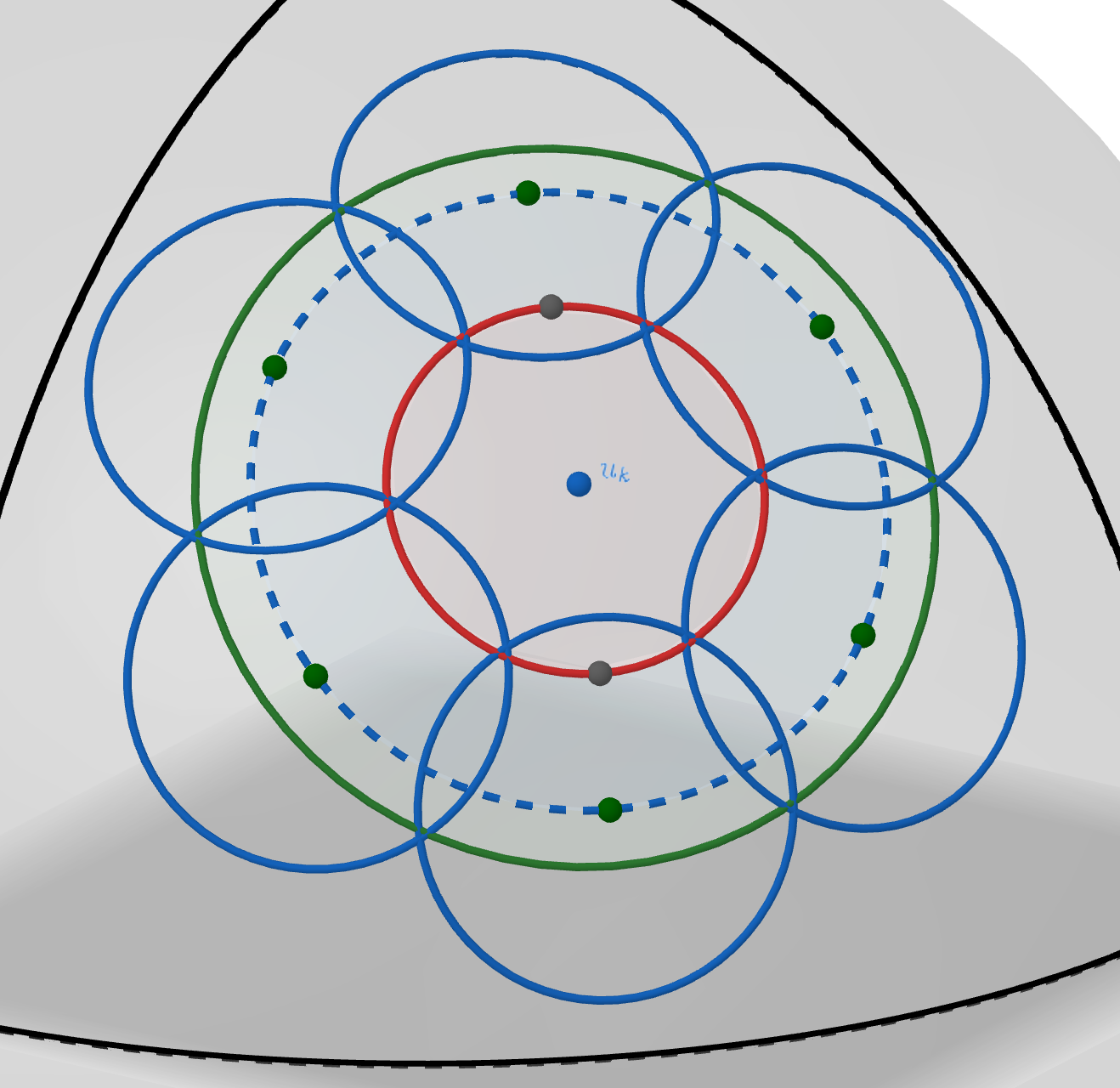}
        \caption[A covering of the ring $C(u_k)$ by 6 balls of radius $r$]{A covering of the ring $C(u_k)$ (region between blue and red circles) by 6 balls of radius $r$. The dotted blue circle represents the ``middle ball'' with radius $r'$, and the six points on it are $x_1,\dots,x_6$. Notice that on the type space, the intersection of all 6 balls covers the entire ring.}
        \label{fig:neighbor_apx_ring_cover}
    \end{figure}
\begin{proofCharles}
    It is easy to see the existence of a finite $f\left(D\right)$, as the ring $C(u_k)$ is a compact set in $\mathbb{R}^D$. When $D=3$, Figure~\ref{fig:neighbor_apx_ring_cover} illustrates the proof by construction. The sequence of coverage ball centers $\{x_1, \dots, x_6\}$ can be found by first defining a ``middle ball'' with radius $r'$ that is between $r$ (that of $B(u_k)$) and $d$ (that of $B'(u_k)$), then picking the 6 points on the intersection of its boundary and $\left\{ x\in \mathbb{R}^D_{\geq 0}: \left\lVert x \right\rVert = 1 \right\}$ that divides the circle evenly. %
\end{proofCharles}
\begin{conjecture}  \label{conj:neighbor_apx_ring_cover_highD}
    When $D>3$, a variant of Lemma~\ref{lemma:neighbor_apx_ring_cover_3D} is true by replacing $6$ with a constant that is at most $\kappa_D + 1$.
\end{conjecture}

\begin{figure}[h!]
        \centering
        \includegraphics[width=0.7\textwidth]{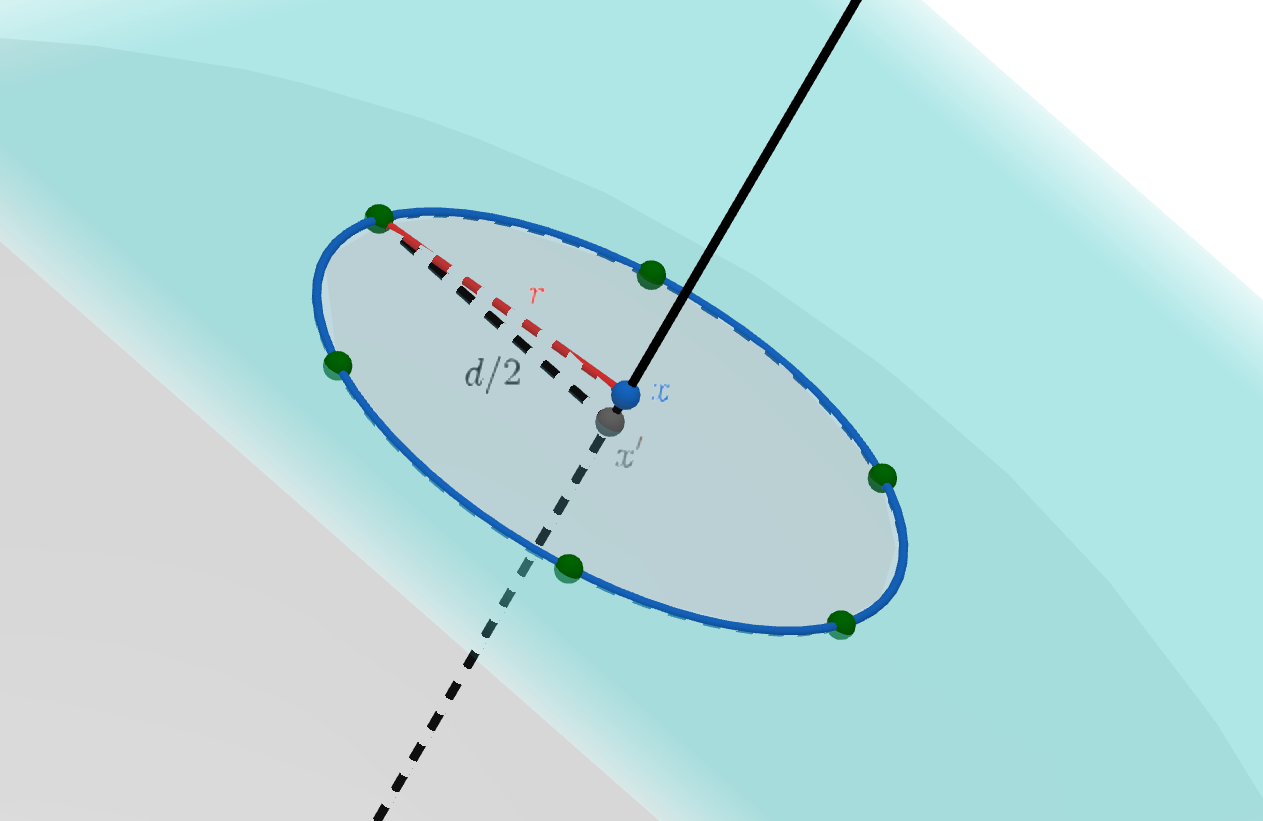}
        \caption[A ball of radius $r$ centered at $x$ and its projection onto the plane $P$]{A ball of radius $r$ centered at $x$ (blue) and its projection onto the plane $P$ (cyan). The point $x$ on the convex type space is projected to $x'\in P$, such that a radius of the ball (red, length $r$) between two unit vectors becomes a radius of the projected circle in $P$ (black, length $d/2$). In this case, both equalities in Conjecture \ref{conj:neighbor_apx_ball_proj} hold. Also note the 6 green points correspond to the ``kissing number arrangement'' used in the last part of proof of Lemma \ref{lemma:neighbor_apx_ball_density}, even though their distance before projection is also $d/2$ which is less than $r$.}
        \label{fig:neighbor_apx_projection}
    \end{figure}

We now bound the number of unassigned users in each coverage ball $B(x_i)$. Since $x_i$ does not necessarily correspond to a user type, Lemma~\ref{lemma:neighbor_apx_excluded_because} cannot be applied directly. However, we can derive a density bound on any ball of radius $r$ regardless of its center, including coverage balls. We first show a technical result:

\begin{lemma}  \label{conj:neighbor_apx_ball_proj}
    Let $x$ be any vector in $\mathbb{R}^D_{\geq 0}$ such that $\left\lVert x \right\rVert = 1$, $B(x)=\left\{ y\in \mathbb{R}^D: \left\lVert x-y \right\rVert \leq r \right\}$ be a ball of radius $r$ centered at $x$, and $P$ be the hyperplane containing the intersection of $B(x)$ and the set of unit vectors, namely the circle $\left\{ y\in \mathbb{R}^D: \left\lVert x-y \right\rVert = r \land \left\lVert y \right\rVert = 1 \right\}$. Let $a,b$ be arbitrary points on the convex surface $B(x) \cap \left\{ y\in \mathbb{R}^D: \left\lVert y \right\rVert = 1 \right\}$, and $a',b'$ be their projections onto $P$ in the direction of the vector $x$. Then,
    \begin{align}
        \left\lVert a-b \right\rVert \geq r \implies \left\lVert a'-b' \right\rVert \geq \frac{d}{2}.  \label{eqn:conj_proj_radius_d/2}
    \end{align}
    Note $d/2$ is the radius of the circle $\left\{ y\in \mathbb{R}^D: \left\lVert x-y \right\rVert = r \land \left\lVert y \right\rVert = 1 \right\}$.
\end{lemma}
Intuitively, Lemma~\ref{conj:neighbor_apx_ball_proj} establishes a relationship between the convex surface in a ball $B(x)$ of radius $r$ and its projection onto the planar surface containing its boundary. This distinction is necessary because the region of all unit vectors in $B(x)$ is convex, so there's a reduction in distance when projected onto the plane. The conjecture can then allow us to utilize the properties of the kissing number (Corollary~\ref{cor:neighbor_apx_kissing}) on the planar surface. Figure~\ref{fig:neighbor_apx_projection} provides a clearer illustration of the projection.

\begin{proofCharles}
    Note that both equalities in \eqref{eqn:conj_proj_radius_d/2} hold when $a=x$ and $b$ is on the boundary of $B(x)$. This choice of $a$ and $b$ makes the greatest angle with the hyperplane $P$ out of all choices that ensure $\left\lVert a-b \right\rVert \geq r$, and thus, the ratio $\left\lVert a'-b' \right\rVert / \left\lVert a-b \right\rVert$ is the smallest (i.e., the distance is reduced by the greatest factor during the projection). Thus, for any other $a,b$, the ratio $\left\lVert a'-b' \right\rVert / \left\lVert a-b \right\rVert $ is greater, resulting in a greater $\left\lVert a'-b' \right\rVert$.
\end{proofCharles}

We can now bound the density of unassigned users in any ball of radius $r$, which can be applied to coverage balls:

\begin{lemma}  \label{lemma:neighbor_apx_ball_density}
    Let $x$ be an arbitrary vector in $\mathbb{R}^D_{\geq 0}$ such that $\left\lVert x \right\rVert = 1$. Then, $B(x)$ contains at most $\kappa_D (\abar-1)$ unassigned users.
\end{lemma}
\begin{proofCharles}
    Let $\userset_x = \left\{ i\in \userset_0: u_i \in B(x) \land R(i) = \emptyset \right\}$ be all unassigned users in $B(x)$. Let $\userset'_x = \left\{ \phi(1), \dots, \phi(M) \right\}$ be an ordered sequence produced via the following procedure:
    \begin{itemize}
        \item Initialize $\userset:=\userset_x$. %
        \item Choose a user $i$ from $\userset$. Add $i$ to $\userset'_x$, and remove all users in $\userset$ with distance at most $r$ from $u_i$.
        \item Repeat until $\userset=\emptyset$. Denote the length of the final sequence with $M$.
    \end{itemize}
    Note that when user $\phi(i)$ is added to $\userset'_x$, the set of users removed from $\userset$ in the same step is a subset of users in the ball $B(u_{\phi(i)})$. This means for all unassigned users $j\in \userset_x$, there exists at least one $i\in [1,M]$ such that $u_j \in B(u_{\phi(i)})$, and $j$ was removed from $\userset$ when the earliest $\phi(i)$ was examined. Thus,  %
    \begin{align*}
        \userset_x &= \userset_x \cap \left( \bigcup_{i=1}^M B\left( u_{\phi(i)} \right) \right).  %
    \end{align*}
    Intuitively, the procedure divides all unassigned users in $\userset_x$ into $M$ neighborhood balls completely.
    Thus, the total number of unassigned users in $B(x)$ is at most
    \begin{align*}
        \left| \userset_x \right| &\leq \sum_{i=1}^M \left| \userset_x \cap B\left( u_{\phi(i)} \right) \right| \leq \sum_{i=1}^M \left( \text{number of unassigned users in } B\left( u_{\phi(i)} \right) \right) \leq M(\abar-1).
    \end{align*}
    The last inequality uses Lemma \ref{lemma:neighbor_apx_unassigned_ball_density}, which implies that each ball $B\left( u_{\phi(i)} \right)$ defined on an unassigned user $u_{\phi(i)}$ contains at most $\abar-1$ unassigned users.
    
    It remains to show $M\leq \kappa_D$. Note that $\left\lVert u_{\phi(i)} - u_{\phi(j)} \right\rVert > r$ whenever $i\neq j$. Let $u'_{\phi(i)}$ be the projection of $u_{\phi(i)}$ onto the plane $P$ containing the circle $\left\{ y\in \mathbb{R}^D: \left\lVert x-y \right\rVert = r \land \left\lVert y \right\rVert = 1 \right\}$ in the direction of $x$. By Conjecture \ref{conj:neighbor_apx_ball_proj}, $\left\lVert u'_{\phi(i)} - u'_{\phi(j)} \right\rVert \geq d/2$ for all $i\neq j$, and all $u'_{\phi(i)}$ lie within the circle with radius $d/2$. By Corollary \ref{cor:neighbor_apx_kissing}, there can be at most $\kappa_D$ such points, so $M\leq \kappa_D$. 
\end{proofCharles}

Combining Lemma \ref{lemma:neighbor_apx_ring_cover_3D} and Lemma \ref{lemma:neighbor_apx_ball_density} gives a bound on the number of unassigned users in $C(u_k)$:

\begin{proposition}  \label{prop:neighbor_apx_ring_density}
    For any representative user $k$, the ring $C(u_k)$ contains at most $\kappa_D f\left(D\right) \left(\abar-1\right)$ unassigned users. In particular, when $D=3$, this quantity is $36 \left(\abar-1\right)$.
\end{proposition}

\subsubsection{Deriving an Approximation Ratio from Ring Density}
We first consider the fraction of users retained by the algorithm:
\begin{proposition}  \label{prop:neighbor_apx_user_ratio}
    Let $\userset_{LC}$ be the set of all users included in the matching produced by the \CA\ algorithm. Then, $\frac{\left|\userset_{LC} \right|}{\left|\userset_0 \right|} \geq \frac{1}{1+\kappa_D f\left(D\right)} $. In particular, when $D=3$, $\left|\userset_{LC} \right|/\left|\userset_0 \right| \geq 1/37 $.   %
\end{proposition}
\begin{proofCharles}
    The number of all users, $\left| \userset_{0}\right|$, satisfies
    \begin{align*}
        \left| \userset_{0}\right| &\leq \sum_{k\in \userset^*} \left| \userset_{0} \cap B'(u_k) \right| \quad \text{(Proposition \ref{prop:neighbor_apx_large_balls})}  \\
        &= \sum_{k\in \userset^*} \left( \left| \userset_{0} \cap B(u_k) \right| + \left| \userset_{0} \cap C(u_k) \right| \right)  \\
        &\leq \sum_{k\in \userset^*} \left( \left| \userset_{0} \cap B(u_k) \right| + \kappa_D f\left(D\right) \left(\abar-1\right) \right) \quad \text{(Proposition \ref{prop:neighbor_apx_ring_density})}  \\
        &= \left| \userset_{LC}\right| + \sum_{k\in \userset^*} \kappa_D f\left(D\right) \left(\abar-1\right) \quad \text{(All users in $B(u_k)$ are assigned and they have no overlaps)},
    \end{align*}
    so
    \begin{align*}
        \frac{\left| \userset_{LC}\right|}{\left| \userset_{0}\right|} &\geq \frac{\left| \userset_{LC}\right|}{\left| \userset_{LC}\right| + \sum_{k\in \userset^*} \kappa_D f\left(D\right) \left(\abar-1\right)}  \\
        &\geq \frac{\sum_{k\in \userset^*} \abar}{\sum_{k\in \userset^*} \abar + \sum_{k\in \userset^*} \kappa_D f\left(D\right) \left(\abar-1\right)}  \quad \text{(Each neighborhood ball has at least $\abar$ assigned users)}  \\
        &= \frac{\abar}{\abar + \kappa_D f\left(D\right) \left(\abar-1\right)}  \\
        &\geq \frac{1}{1+\kappa_D f\left(D\right)}.
    \end{align*}
\end{proofCharles}

Finally, we prove Theorem~\ref{thm:neighbor_apx_ratio_3D} by deriving an approximation ratio on total engagement from number of users:
\begin{proofCharles}[Proof of Theorem~\ref{thm:neighbor_apx_ratio_3D}]
    In any stable set, each user gets engagement at least $K\cdot \ebar$ and at most $K$. Thus, 
    \begin{align*}
        \frac{\longtermengagement\left( \CAseq \right)}{\longtermengagement\left( \FLseq \right)} &\geq \frac{\left| \userset_{LC}\right| \cdot K \cdot \ebar}{\left| \userset_{FL}\right| \cdot K} \geq \frac{\left| \userset_{LC}\right| \cdot K \cdot \ebar}{\left| \userset_{0}\right| \cdot K} \geq \frac{\ebar}{1+\kappa_D f\left(D\right)}.
    \end{align*}
    where $\userset_{FL}$ is the set of users in the maximum stable set, i.e., the \FL\ algorithm's recommendations. Thus, $r\left( \ebar, D\right) \geq \frac{\ebar}{1+\kappa_D f\left(D\right)}$. In particular, when $D=3$, this lower bound evaluates to $\ebar/37$ since $f\left(D\right) = 6$.
\end{proofCharles}

The corresponding conjecture when $D>3$, Conjecture~\ref{conj:neighbor_apx_ratio_highD}, is derived by substituting $f\left(D\right)$ with a hypothesized value in Conjecture~\ref{conj:neighbor_apx_ring_cover_highD}, which is the only step in the entire proof that does not give a concrete numerical result on higher values of $D$.

\begin{remark}
    This proof shows a lower bound on the approximation ratio of the \CA\ algorithm. The lower bound is very loose, due to the approach used in bounding the density of rings (Lemmas~\ref{lemma:neighbor_apx_ring_cover_3D} and \ref{lemma:neighbor_apx_ball_density}) that give a maximum density of $\kappa_D f\left(D\right) \left(\abar-1\right)$ unassigned users, or $36 \left(\abar-1\right)$ unassigned users when $D=3$. We conjecture that a tighter bound for the $D=3$ case may be $18(\abar-1)$, which would give a $\ebar/19$ approximation if proven true.
\end{remark}

\subsection{Example where Neighborhood-Based Algorithms Fail}
The \CA\ algorithm is among a class of algorithms that we call ``neighborhood-based algorithms``, which examine local clusters of users and creators, and assign complete bipartite matchings within each cluster when applicable. This means when the density Assumption~\ref{assumption:density} holds, the matching that such algorithms output must be the union of several complete bipartite matchings between subsets of users and creators.

We present an example where such algorithms, including the \CA\ algorithm, only achieves a fraction of long-term total engagement relative to the \FL\ algorithm. This suggests that in these cases, any recommendation algorithm that aims to do well relative to \FL\ needs to capture the ``global structure'' of the instance when assigning matchings, instead of focusing on local clusters.

\begin{example}  \label{example:model_flower}
    Choose $\abar\geq 7$ to be an arbitrary integer, and $d\in [0,1]$ to be an arbitrary real value. Consider the instance shown in Figure~\ref{fig:example_model_flower}, with $9$ creators, $5\abar-20$ users, $K=5$, and $\ebar$ is chosen such that the happy distance is $d$. User and creator types lie on the unit sphere in $\mathbb{R}^3$, and their projection onto the $\mathbb{R}^2$ space is shown in Figure~\ref{fig:example_model_flower_vectors}: a non-negative unit vector is first chosen to be a ``center node'', then 5 other ``corner nodes'' are evenly distributed on a sphere of radius $d$ from the center node. This means each of the corners is of distance $d$ from the center, but the pairwise distances between any two corners are all greater than $d$. Place 4 creators and 5 users at the center, and 1 creator and $\left(\bar{a}-5\right)$ users at each corner. This means each corner user is only happy with the 4 center creators and the creator in its own corner, while each center user is happy with all creators.
    
    \begin{figure}[h!]
        \centering
        \begin{subfigure}[t]{0.45\textwidth}
            \centering
            \includegraphics{tikz/ex_flower_vectors.tikz}
            \caption{User and creator types projected onto $\mathbb{R}^2$. Solid dots are creators, and open dots are users. Each user receives $K=5$ recommendations, and $d$ is the happy distance.}
            \label{fig:example_model_flower_vectors}
        \end{subfigure}
        \hfill%
        \begin{subfigure}[t]{0.45\textwidth}
            \centering
            \includegraphics{tikz/ex_flower_fl.tikz}
            \caption{The \FL\ recommendation, which is the only matching that retains all users and creators. Red edges are recommendations for each of the $\left(\bar{a}-5\right)$ users at each corner, which are the nearest creator and all 4 center creators. Blue edges are recommendations for each of the 5 center users, which include all 5 corner creators.}
            \label{fig:example_model_flower_fl}
        \end{subfigure}
        \par\bigskip
        \begin{subfigure}{\textwidth}
            \centering
            \begin{subfigure}[t]{0.3\textwidth}
                \centering
                \includegraphics[width=\textwidth]{tikz/ex_flower_ie0.tikz}
                \caption*{$t=0$}
            \end{subfigure}
            \hspace{0.1\textwidth}
            \begin{subfigure}[t]{0.3\textwidth}
                \centering
                \includegraphics[width=\textwidth]{tikz/ex_flower_ie1.tikz}
                \caption*{$t\geq 2$}
            \end{subfigure}
            \caption{Matching given by \UC, \CA, and any neighborhood-based algorithm. It only obtains around $1/5$ of \FL's long-term engagement.}
            \label{fig:example_model_flower_ie}
        \end{subfigure}
        \caption{Illustration of Example~\ref{example:model_flower}}
        \label{fig:example_model_flower}
    \end{figure}

    The only matching that satisfies all user and creator constraints is the one given by the \FL\ algorithm, as shown in Figure~\ref{fig:example_model_flower_fl}. Each corner user is assigned to the nearest corner creator and 4 center creators (who are the only 5 creators that the user is happy with), and each center user is assigned to all 5 corner creators. It is easy to see that no other matching can satisfy all user and creator constraints: recommendations for corner users (red edges) are effectively fixed, so the only degree of freedom is in recommendations for the 5 center users, but each corner creator requires all 5 center users for her audience constraint to be satisfied.

    On the other hand, \UC, \CA\ and any neighborhood-based algorithm will give the matching in Figure~\ref{fig:example_model_flower_ie}, which only achieves approximately $1/5$ of the total engagement of \FL.

\end{example}

\begin{remark}
    Note that Example~\ref{example:model_flower} satisfies the density Assumption~\ref{assumption:density}. While this suggests an upper bound of around $1/5$ for the approximation ratio $r\left( \ebar, 3 \right)$ in 3 dimensions, this is not a tight upper bound for the \CA\ algorithm. There are other examples where \CA\ only achieves long-term engagement that is approximately $1/19$ of \FL, but other neighborhood-based algorithms that examine neighborhood balls in non-arbitrary order can do better.

    However, neighborhood-based algorithms may still not do well even when Assumption~\ref{assumption:density} holds. The choice of 5 corners is not a coincidence: 5 is one less than $\kappa_2=6$, so it is the maximum number of corner points that can be placed on the circumference while keeping all pairwise distances between them above $d$. This means the example can generalize to higher dimensions with $O\left( \kappa_D \right)$ corners, such that \UC, \CA\ and any other neighborhood-based algorithm only achieves approximation ratio $O\left( 1/ \kappa_D \right)$. It also means such algorithms have no hope of achieving a constant factor approximation that is independent of $D$, even with the density assumption.
\end{remark}

\section{Details of the \AM\ 2 Algorithm (CR2)}  \label{sec:appendix_am2}
In Section~\ref{sec:am2_high_level}, we briefly described the intuition behind the \AM\ 2 (CR2) algorithm, which improves upon CR1 using \textit{augmenting paths} to assign maximum potential audience to a creator based on the current matching. In this section, we present the CR2 algorithm in greater detail, including a motivating example, the formal definition of augmenting paths, explanations of how and why they can be used to assign users to a new creator, and finally, the CR2 algorithm itself with pseudocode.

\subsection{Motivation}
One drawback of CR1 is that it often assigns an unnecessarily large number of users to creators who are examined earlier in the algorithm. For example, consider an instance with $K=1$ that contains two creators $c_1$, $c_2$ and $2\abar$ users, who are split into four subsets with $\abar/2$ users each: $\userset_1$ and $\userset_2$, whose users are only happy with $c_1$ and $c_2$ respectively; and $\userset_{1,2}$ and $\userset_{2,1}$, whose users are all happy with both $c_1$ and $c_2$. Suppose the CR1 algorithm first examines $c_2$ (note that both creators have a potential audience size of $3\abar/2$). The algorithm will assign all $3\abar/2$ users in $\userset_1$,  $\userset_{1,2}$ and $\userset_{2,1}$ to $c_1$. Doing so reduces the potential audience size of $c_2$ to $\abar/2$, so she will be skipped in the next iteration as her constraint can no longer be satisfied. This means $c_2$ and all users in $\userset_{2}$ will leave the platform. However, \FL\ can retain both creators and all $2\abar$ users, by assigning $\userset_{1,2}$ to $c_1$ and $\userset_{2,1}$ to $c_2$.

The CR2 algorithm aims to resolve this issue by adding an ``error correction'' mechanism, which notices that when $c_2$ is examined, users in $\userset_{2,1}$ are unnecessary for $c_1$ (thus it was an ``error'' to assign excess audience to $c_1$ earlier in the algorithm), and these users can be reassigned to $c_2$ for her constraint to be satisfied. It implements this mechanism using augmenting paths. In this example, for any user $i\in \userset_{2,1}$, the recommendation or ``edge'' $\left( u_i, c_1 \right)$ is in the current matching but $\left( u_i, c_2 \right)$ is not. (Recall from Figure~\ref{fig:example_model_simple_bipartite} that our problem can also be seen as a many-to-many bipartite matching, where an edge from $u$ to $c$ exists in a bipartite graph if and only if $u$ is happy with $c$.) If we think of these two recommendations as a ``path'' of length 2, it is an augmenting path whose edges alternate between being in the current matching and not. The CR2 algorithm then ``flips'' all edges along the path, effectively reassigning $u_i$ from $c_1$ to $c_2$, allowing $c_2$ to gain a user, while $c_1$'s constraint is still satisfied even after losing a user. This can be done for all users in $\userset_{2,1}$, and combined with $\userset_2$ (where each user $u\in \userset_2$ forms an augmenting path $\left( u, c_2 \right)$ of length 1), they give $c_2$ enough audience.

\subsection{Augmenting Paths}
We first formalize the idea of viewing our problem as a bipartite matching:
\begin{itemize}
    \item Given an instance of our problem, its \textit{bipartite graph representation} is a bipartite graph $G = \left( \userset \cup \creatorset, E \right)$ whose vertices are all users and creators, and where there exists an undirected edge between user $i$ and creator $j$, denoted $\left( u_i, c_j \right)$, if and only if user $i$ is happy with creator $j$.
    \item Additionally, given a set of recommendations $\rec$ (a matching between some users and some creators that many not satisfy all participation constraints), we define the \textit{edge set representation} of $\rec$ as a set of edges $S\subseteq E$, such that $\left( u_i, c_j \right) \in S$ if and only if $j\in \rec\left(i\right)$.
\end{itemize}
We also define $a^{\rec}_j := \left\{ i\in \userset : j\in \rec(i) \right\}$ to be the audience of creator $j$ under matching $\rec$. Now, we give a formal definition of an augmenting path:

\begin{definition}[Augmenting path]
    Given a creator $j$ and a matching $\rec$, let $S$ be the edge set representation of $\rec$. Then, an \textit{augmenting path associated with creator $j$} is a path in the graph with the following characteristics:
    \begin{itemize}
        \item The path starts at $c_j$.
        \item The path ends at either a user $u_i$ such that $\left| \rec(i) \right| < K$ (i.e., the user has capacity for more recommendations), or a creator $c_k$ such that $\left| a^{\rec}_k \right| > \abar$ (i.e., the creator already receives strictly more than $\abar$ users, and can afford to lose one user).
        \item Each odd-numbered edge (from a creator to a user) is not in $S$, and each even-numbered edge (from a user to a creator) is in $S$.
        \item The path does not contain any cycles.
    \end{itemize}
\end{definition}
The following proposition characterizes the effects of ``adding'' an augmenting path to a matching:

\begin{proposition}[Effects of augmenting paths on audience size]  \label{prop:augmenting_path_audience_size}
    Let $\rec$ be an initial matching and $S$ be its edge set representation. %
    Given a creator $j$ and an augmenting path $P$ associated with $j$, let $\rec'$ and $S'$ be the resultant matching and edge sets after ``adding'' $P$ to $\rec$ and $S$. Formally, $S'$ is the XOR of $P$ and $S$, and $\rec'$ is the matching induced by $S'$. (We denote this as $S' = S \oplus P$ and $\rec' = \rec \oplus P$.)
   
    Then, the following properties hold:
    \begin{enumerate}
        \item Creator $c_j$ satisfies $a^{\rec'}_j = a^{\rec}_j + 1$. If $P$ ends at a creator $c_k$, then $a^{\rec'}_k = a^{\rec}_k - 1$. For all other creators $c_{j'}$, $a^{\rec'}_{j'} = a^{\rec}_{j'}$.
        \item If $P$ ends at a user $u_i$, then $\left| \rec'\left(i\right) \right| = \left| \rec\left(i\right) \right| + 1$. For all other users $u_{i'}$, $\left| \rec'\left(i'\right) \right| = \left| \rec\left(i'\right) \right|$.
        \item If $P$ ends at a user $u_i$ and $\rec'$ does not satisfy her participation constraints (due to insufficient number of recommendations), then $\rec$ also does not satisfy $u_i$'s constraints. For all other users, and for all creators other than $c_j$, $\rec'$ satisfies her constraints if and only if $\rec$ satisfies her constraints.
    \end{enumerate}
\end{proposition}
Proposition~\ref{prop:augmenting_path_audience_size} effectively states that each augmenting path is equivalent to creator $c_j$ getting a new user, while respecting all other participation constraints that have already been satisfied. The proof is trivial and left as an exercise.

\subsection{Description of the CR2 Algorithm}
At the core of CR2 algorithm is an improvement over CR1 that is motivated by Proposition~\ref{prop:augmenting_path_audience_size}: we can replace the step of ``assigning a user to creator $j$'' in CR1 with ``finding an augmenting path associated with $j$''. This achieves the same effect of increasing creator $j$'s audience size by 1, but unlike CR1 where the new audience is always assigned to $c_j$ directly (equivalent to an augmenting path of length 1), the augmenting path approach allows indirect assignments (path ending at a user that's not connected to $c_j$ directly) and redistributions of audiences among creators. 
This means the set of users that can be assigned to $c_j$ under the augmenting path approach is a superset of the CR1 approach, so using the former increases the likelihood of satisfying $c_j$'s audience constraint.\footnote{This ``improvement'' does not mean CR2 always returns a final matching with greater total engagement than CR1 on all instances. This is because due to the greedy nature of the \CRshort\ algorithms, sometimes there are ``wrong'' creators that should be avoided, i.e., not including this creator will result in a better matching. Nevertheless, in the simulations we present in Appendix~\ref{sec:appendix_sims}, CR2 shows a significant improvement over CR1 on average, which suggests CR2's other benefits generally outweigh the problem of picking the wrong creators occasionally.}

Therefore, CR2 uses the augmenting path approach when examining each new creator $c_j$ and providing her as much audience as possible. A pseudocode of CR2 is presented in Algorithm~\ref{alg:am2_high_level}.

\begin{algorithm}[h!]
\caption{\CR\ 2 (CR2) algorithm for time $t$}
 \label{alg:am2_high_level}
\begin{algorithmic}[1]
    \State $\userset_t = \{1, \ldots, U\} \gets$ current set of users, in arbitrary order
    \State $\creatorset_t = \{1, \ldots, C\} \gets$ current set of creators
    \State $\rect:\userset_t \mapsto P(\creatorset_t) \gets$ recommendations (initially empty, i.e., $\rect(i)=\emptyset \ \forall i\in \userset_t$)
    \While{$\creatorset_t \neq \emptyset$}
        \State $a(j) := \left\{ i \in \userset_t: u_i^T c_j \geq \ebar \text{ and } \left| \rect(i) \right| < K \right\}, \ \forall j \in \creatorset_t$  \Comment{Potential audience function for creator $j$}
        \State $j \gets $ creator in $\creatorset_t$ with smallest value of $\left| a(j) \right|$  \Comment{Smallest size of potential audience}
        \State $\creatorset_t \gets \creatorset_t \backslash \left\{j\right\}$
        \State $P \gets$ Augmenting path associated with creator $j$
        \Comment{Prioritize augmenting paths that end at users}  \label{algline:am2_augmenting_path_start}
        \While{$P \neq \emptyset$}  \Comment{Repeatedly BFS for augmenting paths until none can be found}
            \State $\rect \gets \rect \oplus P$  \Comment{Add $P$ to current matching}
            \State $P \gets$ Augmenting path associated with creator $j$
        \EndWhile
        \If{creator $j$ has fewer than $\abar$ users in $\rect$}  \Comment{Failed to add $j$ as she cannot be satisfied}
            \State $\rect \gets$ Restore original matching before examining $j$  \Comment{Undo all augmenting paths}
        \EndIf  \label{algline:am2_augmenting_path_restore_end}
    \EndWhile
    \State \Return{$\rect$}  \Comment{Recommendation for time $t$, note that some users may leave}
\end{algorithmic}
\end{algorithm}

The only change in CR2 is replacing Lines~\ref{algline:am1_creator_audience_compare}-\ref{algline:am1_creator_endif} in CR1 (Algorithm~\ref{alg:am1}) which simply try to add all potential audience of creator $j$ and check if they meet her audience constraint, with Lines~\ref{algline:am2_augmenting_path_start}-\ref{algline:am2_augmenting_path_restore_end} in Algorithm~\ref{alg:am2_high_level} which find as many augmenting paths associated with $j$ as possible.

In terms of implementation, we find augmenting paths one by one using a breadth-first search (BFS) that starts at $c_j$. We prefer augmenting paths that end at users over those that end at creators, since the former increases the total number of edges in the matching by one, and makes progress towards giving the user $K$ recommendations. Therefore, during the BFS, we record all completed augmenting paths once any possible end point is reached, but only terminate the search immediately when the end point is a user. If none are found, then a path that ends at a creator is used. %

Several properties of CR1 remain true in CR2. Both prioritize creators whose constraints are the hardest to satisfy. In both algorithms, once a creator is added to the returned set, it will never be removed (even though its audience size may decrease in CR2 while being at least $\abar$), and that the number of recommendations for each user can only increase during the execution. On the other hand, CR2 inherits several drawbacks of CR1, such as possibly choosing the ``wrong'' sets of users and creators, ignoring user engagement values, and not necessarily returning a stable set due to users with insufficient recommendations (so CR2 may need to be executed again at time $t+1$).

\section{Simulation Results}  \label{sec:appendix_sims} 
In this appendix, we present simulations on randomly generated instances with small parameter values. The main objectives of these simulations are to:
\begin{itemize}
    \item Evaluate the average performance of \IE\ relative to \FL, and identify choices of parameters in which \IE\ performs poorly. This generalizes Proposition~\ref{prop:increase_creators} and Theorem~\ref{thm:increase users} to less restrictive and non-asymptotic setups, while also uncovering additional scenarios that lack theoretical justifications.
    \item Evaluate the average performance of CR2 to \IE, and show that it achieves high approximation ratios relative to \FL, especially in the aforementioned situations where \IE\ performs poorly.
\end{itemize}
Empirically, most instances generated using uniform priors $\distribution$ for user and creator types do not satisfy the density Assumption~\ref{assumption:density}. As such, we did not run the \CA\ algorithm in simulations.

We first document the setup of our simulations, then list the main observations, and present plots of average approximation ratios to support each observation.

\subsection{Simulation Setup}
Our simulations are on simple randomized instances with small instance sizes, in terms of four main parameters $U$, $C$, $K$ and $\abar$. %
Typically, all four parameters are fixed in a single test case. We then compare different test cases and their trends under changes in parameters (e.g., increasing $U$). %

All users and creators have their types generated i.i.d. from the uniform distribution $\distribution$ over all non-negative unit vectors, following the random instances framework that we established in Section~\ref{sec:random_instances}. However, their constraint values are deterministic, with creator constraints given by $\abar$. 

We define the user constraint value $\ebar$ as the expected engagement from a random user to an independent random creator, multiplied by a parameter $\ebar_m$ which is set to $0.6$ in all plots presented below:
$$ \ebar = \ebar_m \mathbb{E}_{u,c\sim \distribution} \left[ u^T c \right]. $$
Intuitively, the multiplier $\ebar_m$ measures the strength of user constraints. A greater $\ebar_m$ means users are more demanding in the quality of recommendations, whereas a smaller $\ebar_m$ means they are happy with creators whose types are less similar to theirs. Note that $\ebar$ does not depend on $U, C$ and $K$, and only depends on $D$ and $\ebar_m$. In practice, we sample the value of $\mathbb{E}_{u,c\sim \distribution} \left[ u^T c \right]$ using Monte Carlo simulations, by generating 10,000 pairs of random user and creator types for each $D$ and finding their average dot products. While we only present plots with $\ebar_m=0.6$, we have verified that similar conclusions hold with $\ebar_m=0.8$ or $1$. %

Additionally, the plots below all have dimensionality $D=10$. However, similar observations can be drawn with $D=2$ and $D=5$.

For each test case with fixed $U$, $C$, $K$, $\abar$ (as well as $D$ and $\ebar$), we generate $N=1,000$ independent random instances, and run the following algorithms: \FL, \IE, and two versions of \AM\ (CR1 and CR2). Some of these runs may have $\longtermengagement\left(\FLseq\right)=0$ (no stable sets exist), especially with higher values of $\ebar_m$.\footnote{In the most extreme cases, with $\ebar_m=1$ and $K$ being close to $C-1$, most or all random instances have $\longtermengagement\left(\FLseq\right)=0$. This may be a consequence of our definition of $\ebar$: specifically, since it's defined as the expected distance from a user to a creator, each user will naturally have half of the creators below this threshold in expectation, but she requires to be assigned to all but one of them. Nevertheless, we decided to keep $\ebar$ constant when comparing different values of $K$ and other parameters to keep it as a control variable.} We measure the performance of \IE, CR1 and CR2 algorithms by computing the empirical averages of their approximation ratios in long-term engagement relative to \FL\ (on instances that satisfy $\longtermengagement\left(\FLseq\right)>0$).

\subsection{Main Observations}
Below is a list of key observations that emerge from simulations.
\begin{observation}[Increasing $U$]  \label{obs:increase_users}
    \IE's performance degrades when $U$ and $\abar$ both increase proportionally while $C$ and $K$ are fixed, subject to $UK=C\abar$ (the ``market balance'' condition).
\end{observation}

\begin{observation}[Increasing $C$]  \label{obs:increase_creators}
    \IE's performance degrades when $C$ increases, while $U$, $K$ and $\abar$ are fixed.
\end{observation}

\begin{observation}[Increasing $K$]  \label{obs:increase_K}
    \IE's performance is poor when the ratio of $K/C$ is high, subject to $UK=C\abar$ (market balance), and degrades when $K$ and $\abar$ increase proportionally while $U$ and $C$ are fixed.
\end{observation}

\begin{observation}  \label{obs:am2}
    Under conditions in Observations~\ref{obs:increase_users}, \ref{obs:increase_creators} and \ref{obs:increase_K}, CR2 achieves high approximation ratios, which are usually higher than \IE\ and are roughly constant following changes in the respective parameters.
\end{observation}

Note that Observation~\ref{obs:increase_users} generalizes Theorem~\ref{thm:increase users} to non-asymptotic settings, while reducing the dependence on $K>C/2$ and removing the requirement that $D=2$. Observation~\ref{obs:increase_creators} generalizes Proposition~\ref{prop:increase_creators} to non-asymptotic settings.

\subsection{Detailed Results}

\subsubsection{Increasing \texorpdfstring{$U$ and $\abar$}{U and a} (Observation~\ref{obs:increase_users})} 
 \label{sec:sims_increase_users}
\begin{figure}[h]
    \centering
    \includegraphics[width=\textwidth]{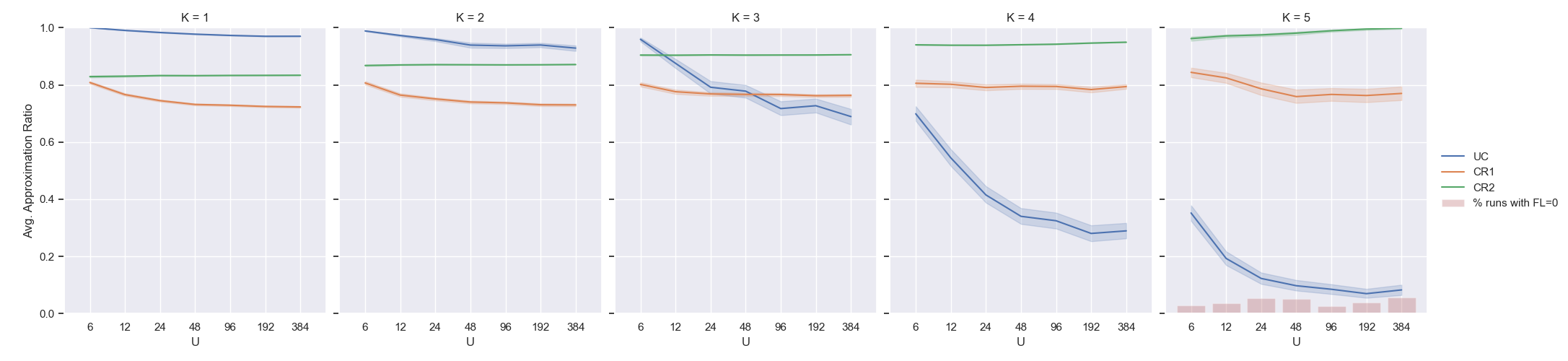}
    \caption[Approximation ratios of UC, CR1 and CR2 when $U$ and $\abar$ increase proportionally]{Approximation ratios of UC, CR1 and CR2 when $U$ and $\abar$ increase proportionally. In all plots, $C=6$, and $K$ increases across the subplots from $1$ (left) to $5$ (right). Within each subplot, $U$ increases exponentially on the $x$-axis, and $\abar = UK/C$. UC, CR1 and CR2 are in blue, red and green respectively, with 95\% confidence intervals.}
    \label{fig:sims_conj1_main}
\end{figure}

Figure~\ref{fig:sims_conj1_main}, which is a duplicate of Figure~\ref{fig:sims_conj1_maintext_am2} in Section~\ref{sec:am2_high_level}, shows average approximation ratios for test cases where each subplot captures the trend specified in Observation~\ref{obs:increase_users} and Theorem~\ref{thm:increase users}: both $U$ and $\abar$ increase proportionally while $C=6$ and $K$ remain constant, such that $UK=C\abar$.

In all subplots, the \IE\ algorithm shows a drop in performance relative to the \FL\ algorithm as $U$ and $\abar$ increase. The decrease is most notable when $K=4$ and $K=5$, both of which satisfy the assumption of $K>C/2$ in Theorem~\ref{thm:increase users}. However, a smaller decrease can still be observed when $K\leq 3$, which suggests the possibility that the result may still be generally applicable even for smaller values of $K/C$. We also observe that with a greater $\ebar_m$ where user constraints become stricter, \IE's decrease when $K\leq C/2$ can be more significant.

On the other hand, the CR2 algorithm consistently achieves at least 80\% of \FL's average long-term engagement in all test cases, and its performance is not significantly affected by the increase in $U$ and $\abar$. This is especially true in cases where \IE's degrade is the most notable.

\subsubsection{Increasing \texorpdfstring{$C$}{C} (Observation~\ref{obs:increase_creators})}  \label{sec:sims_increase_creators}

\begin{figure}[h]
    \centering
    \includegraphics[width=\textwidth]{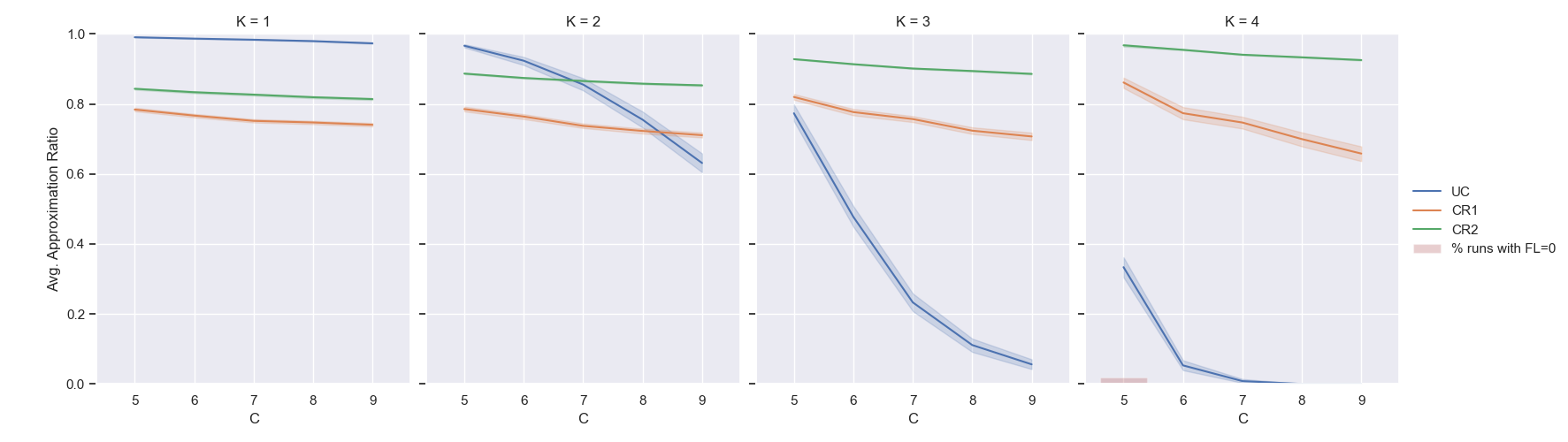}
    \caption[Approximation ratios of UC, CR1 and CR2 when $C$ increases]{Approximation ratios of UC, CR1 and CR2 when $C$ increases ($x$-axis in each subplot). Across subplots, $K$ increases from $1$ (left) to $4$ (right). All plots satisfy $U=10$ and $\abar = 2K$.}
    \label{fig:sims_conj2_main}
\end{figure}

Figure~\ref{fig:sims_conj2_main} shows average approximation ratios for test cases where each subplot captures the trend specified in Observation~\ref{obs:increase_creators} and Proposition~\ref{prop:increase_creators}: only $C$ increases, while $U$, $K$ and $\abar$ remain constant. We set $\abar=2K$, so that market balance is satisfied when $C=5$ (the leftmost point on the $x$-axis), while points further right have markets with disproportionately more creators than users.

The \IE\ algorithm sees a rapid decline in performance as $C$ increases, as long as $K>1$. This also suggests that Proposition~\ref{prop:increase_creators}'s asymptotic condition $C\rightarrow \infty$ may be unnecessary in practice, as the decrease in \IE's approximation ratio is already notable even with a small increase in $C$. The CR2 algorithm continues to achieve high and near constant approximation ratios in all cases.

\subsubsection{Increasing \texorpdfstring{$K$ and $\abar$}{K and a} (Observation~\ref{obs:increase_K})}  \label{sec:sims_increase_K}

\begin{figure}[h]
    \centering
    \includegraphics[width=\textwidth]{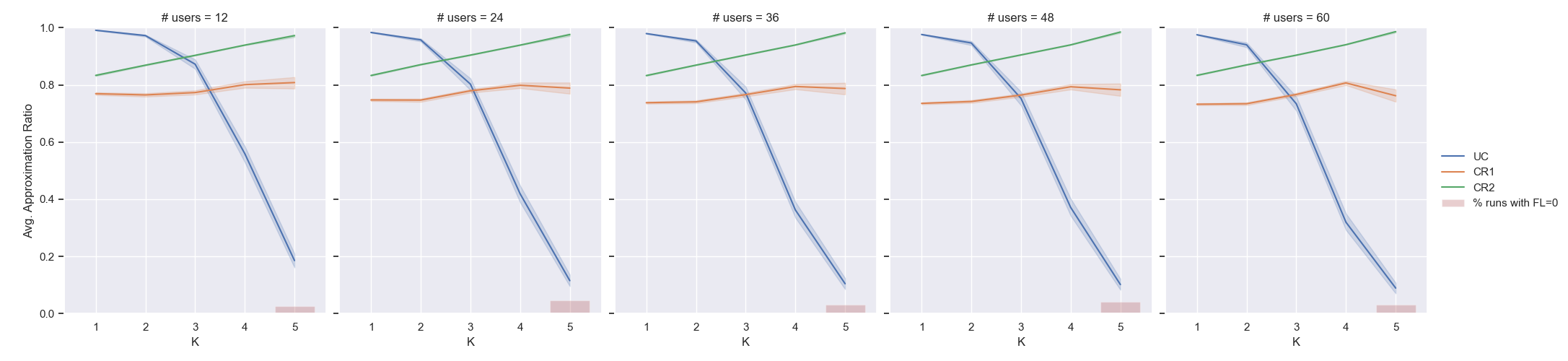}
    \caption[Approximation ratios of UC, CR1 and CR2 when $K$ and $\abar$ increase proportionally]{Approximation ratios of UC, CR1 and CR2 when $K$ ($x$-axis in each subplot) and $\abar$ increase proportionally. In all plots, $C=6$. Across subplots, $U$ increases from $12$ (left) to $60$ (right).}
    \label{fig:sims_conj3_main}
\end{figure}

Figure~\ref{fig:sims_conj3_main} shows average approximation ratios for test cases where each subplot captures the trend specified in Observation~\ref{obs:increase_K}: both $K$ and $\abar$ increase proportionally, while $U$ and $C=6$ remain constant with $UK=C\abar$.

While the \IE\ algorithm achieves high approximation ratios for small values of $K$ such as $1$ or $2$, its performance quickly declines for higher values of $K$, and becomes very low when $K=5$. In contrast, the CR2 algorithm sees an improvement as $K$ and $\abar$ increase, which is exactly when \IE\ degrades. This suggests that such instances with higher $K$, which are more challenging, are better suited for the creator-centric \AM\ algorithms.

Note that in this case, \IE's approximation ratio drops the most rapidly when $K\geq C/2$, which bears similarity with the condition in Theorem~\ref{thm:increase users} that $K>C/2$. Such a trend is also observed in other sets of simulations with different parameters; however, $K\geq C/2$ is not the only case where \IE\ performs poorly, as we briefly mentioned in Section~\ref{sec:sims_increase_users}. %

\clearpage
\bibliographystyle{informs4/informs2014}
\bibliography{refs/refs}

\end{document}